# PROBING FOR EXOPLANETS HIDING IN DUSTY DEBRIS DISKS: DISK IMAGING, CHARACTERIZATION, AND EXPLORATION WITH *HST*/STIS MULTI-ROLL CORONAGRAPHY

Short Title: PROBING FOR EXOPLANETS IN DUSTY DEBRIS DISKS

Glenn Schneider

Steward Observatory and the Department of Astronomy, The University of Arizona

933 North Cherry Avenue, Tucson, AZ 85721 USA

gschneider@as.arizona.edu

Carol A. Grady

Eureka Scientific

2452 Delmer, Suite 100, Oakland, CA 96002 USA

Dean C. Hines

Space Telescope Science Institute

3700 San Martin Drive, Baltimore, MD 21218 USA

Christopher C. Stark

NASA/Goddard Space Flight Center, Exoplanets & Stellar Astrophysics Laboratory

Code 667, Greenbelt, MD 20771

John H. Debes

Space Telescope Science Institute

3700 San Martin Drive, Baltimore, MD 21218 USA

Joe Carson

Department of Physics and Astronomy, College of Charleston

66 George Street, Charleston, SC 29424

Marc J. Kuchner

NASA/Goddard Space Flight Center, Exoplanets & Stellar Astrophysics Laboratory

Code 667, Greenbelt, MD 20771

Marshall D. Perrin

Space Telescope Science Institute

3700 San Martin Drive, Baltimore, MD 21218 USA

Alycia J. Weinberger

Department of Terrestrial Magnetism, Carnegie Institute of Washington

5241 Branch Road, NW, Washington DC 20015 USA

John P. Wisniewski

H. L. Dodge Department of Physics and Astronomy, University of Oklahoma

440 West Brooks Street, Norman, OK 73019 USA

Murray D. Silverstone

Department of Physics and Astronomy, University of Alabama

Box 870324, Tuscaloosa, AL 35487-0324

Hannah Jang-Condell

Department of Physics and Astronomy, University of Wyoming

Laramie, WY 82071

Thomas Henning

Max-Planck-Institut für Astronomie

Königstuhl 17, 69117, Heidelberg, Germany

Bruce E. Woodgate[1]

NASA/Goddard Space Flight Center

Code 667, Greenbelt, MD 20771

Eugene Serabyn

Jet Propulsion Laboratory, California Institute of Technology

4800 Oak Grove Drive, Pasadena, CA, 91109 USA

Amaya Moro-Martin

Space Telescope Science Institute

3700 San Martin Drive, Baltimore, MD 21218 USA

Motohide Tamura

The University of Tokyo, National Astronomical Observatory of Japan

2-21-1 Osawa, Mitaka, Tokyo 181-8588 JAPAN

Phillip M. Hinz

Steward Observatory and the Department of Astronomy, The University of Arizona

933 North Cherry Avenue, Tucson, AZ 85721 USA

Timothy J. Rodigas

Department of Terrestrial Magnetism, Carnegie Institute of Washington

5241 Branch Road, NW, Washington DC 20015 USA

---

[1] Deceased
[2] http://www.stsci.edu/hst/phase2-public/12228.pro
[2] http://www.stsci.edu/hst/phase2-public/12228.pdf
[4] http://www.stsci.edu/hst/stis/software/
[5] http://www.stsci.edu/hst/observatory/cdbs
[6] http://www.gemini.edu/node/12113
[7] http://www.eso.org/public/news/eso1417/
[8] FEPS: "Formation and Evolution of Planetary Systems" – M. Meyer, PI. This was the first of two light-scattering

# ABSTRACT

Spatially resolved scattered-light images of circumstellar debris in exoplanetary systems constrain the physical properties and orbits of the dust particles in these systems. They also inform on co-orbiting (but unseen) planets, the systemic architectures, and forces perturbing the starlight-scattering circumstellar material. Using *HST*/STIS broadband optical coronagraphy, we have completed the observational phase of a program to study the spatial distribution of dust in a sample of ten circumstellar debris systems, and one "mature" protoplanetrary disk all with *HST* pedigree, using PSF-subtracted multi-roll coronagraphy. These observations probe stellocentric distances $\geq$ 5 AU for the nearest systems, and simultaneously resolve disk substructures well beyond corresponding to the giant planet and Kuiper belt regions within our own Solar System. They also disclose diffuse very low-surface brightness dust at larger stellocentric distances. Herein we present new results inclusive of fainter disks such as HD92945 ($F_{disk}/F_{star} = 5\times10^{-5}$) confirming, and better revealing, the existence of a narrow inner debris ring within a larger diffuse dust disk. Other disks with ring-like sub-structures and significant asymmetries and complex morphologies include: HD181327 for which we posit a spray of ejecta from a recent massive collision in an exo-Kuiper belt; HD61005 suggested to be interacting with the local ISM; HD15115 and HD32297, discussed also in the context of putative environmental interactions. These disks, and HD15745, suggest that debris system evolution cannot be treated in isolation. For AU Mic's edge-on disk we find out-of-plane surface brightness asymmetries at $\geq$ 5 AU that may implicate the existence of one or more planetary perturbers. Time resolved images of the MP Mus proto-planetary disk provide spatially resolved temporal variability in the disk illumination. These and other new images from our *HST*/STIS GO/12228 program enable direct inter-comparison of the architectures of these exoplanetary debris systems in the context of our own Solar System.

# 1. INTRODUCTION

Understanding how planetary systems, including our own, form and evolve frame one of the most fundamentally compelling areas of inquiry in contemporary astrophysics (Beuther et al. 2014). Over the last few decades, optical, infrared and millimeter observations have shown that most stars are surrounded at birth by circumstellar (CS) accretion disks of primordial material. Separately, more than a thousand confirmed, and thousands more candidate, extrasolar planets detected to date mostly by photometric transit and radial velocity methods, are believed to originate in such disks with Jovian-mass planets forming before CS accretion disks become gas-poor over time. The evolutionary transformation of gas-rich disks to systems with small amounts of dust and little residual gas occurs over the first ~10 Myr of the host star's lifetime, with warm dust rare in systems after ~ 3 Myr (Haisch et al. 2001) and gas rare after 5-10 Myr and stellar mass accretion declining to ~ $10^{-11}$ $M_\odot$ yr$^{-1}$ over the same epoch (Fedele et al. 2010). Much of the dust may then be tied up in smaller rocky or icy bodies that can interact with co-orbiting giant planets or other small bodies, and collide, thus replenishing the CS environments (transformed into "debris disks") with second-generation material. Such collisional replenishment of dust-rich CS particles can continue throughout the lifetime of the system (Carpenter et al. 2005; Roccatagliata et al. 2009; Panić et al. 2013). In our own Solar System, even after 4.6 Gyr, this is evidenced by the presence of our own zodiacal dust cloud, asteroid and Kuiper belts, and population of comets. Indeed, this dust-production paradigm was reinforced with *Hubble Space Telescope* (*HST*) and *Rosetta* imaging of the aftermath of the apparent dynamical collision or rotational disruption of minor planet P/2010 A2, releasing a wake of trailing debris (Jewitt et al. 2010), even as competing forces (e.g., Poynting-Robertson drag and radiation pressure "blow out") attempt to clear the circumsolar environment of such particles. However, because these dynamical processes occur on much shorter timescales than a host star's lifetime, a quasi-equilibrium condition may exist where dusty debris is continuously generated through collisions and the net solid mass declines with age (Wyatt et al. 2007a; Wyatt et al. 2007b; Wyatt 2008). The outcomes of such interactions will analogously affect the compositional content and spatial distribution of dusty debris in exoplanetary systems on spatial scales large compared to planets.

Dynamical interactions of planets with residual planetesimals likely play a vital role in shaping the architectures of planetary systems. Recently formed terrestrial planets can generate copious quantities of dust in the early phases of planetary system formation (Kenyon & Bromley 2004). E.g., in the history of our Solar System, Jupiter is thought to have wrought considerable destruction in the early evolution of the asteroid belt (Bottke et al. 2005; Nesvorný, et al. 2013), scattering volatile-rich planetesimals towards the inner solar system that may have contributed significant water to, and affected the habitability of, the early Earth (Raymond et al. 2004; Morbidelli et al. 2012; Raymond et al. 2014; Albertsson et al. 2014). The Nice model (Tsiganis et al. 2005; Morbidelli et al. 2005) for the dynamical re-arrangement of the Solar System explains: (a) the migration of the outer planets; (b) the depletion of dust-producing planetesimals by factors of x10 in both the Asteroid and Kuiper Belts; and (c) evidence for a Late-Heavy Bombardment in the lunar cratering record (Strom et al. 2005). High spatial-resolution ($<\approx$ AU scale) and high fidelity maps of dusty CS debris in exoplanetary systems, in which the presence (or absence) of co-orbiting planetary-mass bodies may be imprinted or encoded on the imaged dust distribution, are needed to test the general applicability of such models around other stars.

CS dust disks (exo-Kuiper Belts) were first identified by thermal emission in excess of that expected from the stellar photosphere. This is how the majority of debris disks at d < 100 pc are known (Backman & Paresce 1993), and continue to be discovered (Eiroa et al. 2013). Most modeling of these disks has focused on the wavelength dependence of the infrared (IR) excess emission (i.e., their spectral energy distributions; SEDs), but is not unique. Different spatial distributions of dust and dust particle properties can produce similar SEDs. Breaking these degeneracies requires spatially-resolved imagery. With sufficient spatial resolution, we can potentially observe the effects of sculpting of the debris by dynamical interactions with planets and other phenomena. At long wavelengths, such imaging studies are currently limited to far-IR thermal emission from nearby (d < 50 pc) disks obtained with the 3.5m *Herschel Space Observatory*, and the few debris disks studied to date with the Atacama Large Millimeter Array (ALMA) sensitive to larger particles. Those disks that are bright enough to be imaged in scattered near-IR or optical light are observable with ground-based 8-10 m class adaptive optics (AO)-augmented telescopes with coronagraphs, or with *HST* coronagraphy. In particular, these CS debris systems are detectable and resolvable using the Space Telescope Imaging Spectrograph (STIS), with Point-Spread-Function (PSF) template-subtracted coronagraphy

(PSFTSC), and can uniquely provide detailed information about the systems and their architectures even when the individual planets remain elusive.

Initially, only ~15% of the thermally-emissive debris disks observed by *HST* were detected in scattered light, although additional disks continue to be discovered from archival data as post-processing techniques continue to improve, and as resolved detections from other wavelengths continue to accumulate. The discovery images of these light-scattering disks exhibit a diversity of radially "confined" and non-axisymmetric features in the stellocentric regions thus far explored, including: photocentric offsets (Schneider et al. 2009; Kalas et al. 2005a, 2008), warps (Heap et al. 2000), clumps (Holland et al. 1998; Greaves et al. 1998, 2012), non-coplanar features (Golimowski et al. 2006), spirality (Clampin et al. 2003), and brightness asymmetries (Telesco et al. 2000; Kalas et al. 2007a,b). These are exactly the morphological features predicted to arise in starlight-scattering CS debris perturbed by co-orbiting planets (Wyatt et al. 1999; Augereau et al. 2001; Kuchner & Holman 2003; Wyatt 2003, 2005; Wolf et al. 2007; Ertel et al. 2012). Some of these features, however, may also result from different evolutionary scenarios positing the production of belts of colliding planetesimals. The challenge then is to distinguish between disks arising from different mechanisms, e.g., between disks with planetesimals captured into mean motion resonances with planets, and disks where planetesimals are self-stirred (Wyatt 2008).

Distinguishing between these scenarios can best be achieved with images that reveal and spatially resolve most of the debris system sub-structures. Most of the "discovery" images, however, have been plagued with optical artifacts due to systematic effects from incomplete starlight suppression (i.e., residuals from limited-efficacy coronagraphy and PSF-subtraction), and are of insufficient (or questionable) fidelity. To overcome this problem, we have obtained new, much higher fidelity and sensitivity, visible-light observations of an *HST*-selected sample of CS disks using a multi-roll PSFTSC imaging technique with STIS. Our new STIS observations achieve inner working angles (IWAs) of $r \geq 0.3''$, equal to that previously provided by *HST*'s Near Infrared Camera and Multi-Object Spectrometer (NICMOS) coronagraph, but with a field of view and visible-light spatial resolution ($\approx$ 60 mas) comparable to that of *HST*'s Advanced Camera for Surveys (ACS) with its much larger effective IWAs. (Today, neither the NICMOS nor ACS coronagraphs are operational.) These new STIS images probe the interior regions of these debris systems, with inner working distances below $\approx$ 8 AU for about half the stars in this sample, corresponding to the giant planet and Kuiper belt regions within our own Solar System. This enables direct inter-comparison of the architectures of these exoplanetary debris systems in the context of our own Solar System, and also studies of disk dynamics, grain properties, and systemic evolution. These in turn can be used to deduce (or exclude) the existence of planets, and more importantly can set constraints on planetary masses, orbital distances, eccentricities and evolutionary history (Chaing et al. 2009; Rodigas et al. 2014a).

Our target sample is discussed in §2 of this paper. In § 3 we discuss the observational motivation and goals of the program predicated on the history of "lessons learned" from nearly two decades of *HST* coronagraphy and PSF-subtraction. In § 4 we detail our observational approach to greatly improve upon *a priori* anticipated sensitivity-limiting instrumental and extrinsic systematic effects that arise from incomplete or imperfect PSF subtractions (if observationally unconstrained or uncontrolled). Therein we elaborate upon the observations that define and enable our multi-roll PSFTSC by utilizing time-constrained multiple field (spacecraft roll) orientation ("N"-roll, with N = 6 by optimal design). In § 5 we present our image data reduction, and PSF subtraction methods applied across the target set. In § 6 we place the STIS multi-roll PSFTFC observations in the context of CS disk observations with other instruments, facilities, and at other wavelengths obtained in space and on the ground. In § 7 we present, at a high level across the target set, the observational results of the imaging program, with commentary in § 8. Finally, in Appendix A, we present and discuss the imaging results and derived properties and characteristics of each individual target system in detail. Detailed analysis and modeling of the image data for individual targets will be presented in follow-on papers, (e.g. Stark et al. 2014 for HD 181327, Hines et al. 2014 (in preparation) for HD 61005, Debes et al. 2014 (in preparation) for HD 32297, and others).

## 2. THE *HST*/GO 12228 TARGET SAMPLE

We observed an *HST*-selected ensemble of ten CS debris disks and one "mature" protoplanetary/transition disk of age comparable to the youngest debris disks in our sample (see Table 1). All targets were previously observed with *HST* coronagraphy in heterogeneously diverse discovery or survey programs. We re-visited disk targets whose surface brightnesses and geometries were sufficiently well determined (in the regions observed) from those previous ACS optical or NICMOS short wavelength near-IR images to ensure robust imaging with our multi-roll PSFTSC strategy.

**Table 1**
Survey Targets*

| Target | Vmag | B-V | Spec | Dist. (pc) | Age (Myr) | Disk $L_{IR}/L_{star}$ | Initial *HST* Imaging Instrument | Reference |
|---|---|---|---|---|---|---|---|---|
| HD 15115 | 7.150 | +0.35 | F2 | 45.2 | 12?[d] | 0.05% | ACS | Kalas et al. 2007a |
| HD 15745 | 7.824 | +0.32 | F2V | 63.5 | ~ 100?[g] | 0.12% | ACS | Kalas et al. 2007b |
| HD 32297 | 8.304 | +0.20 | A5V[+] | 112.4 | ~ 10[e] | 0.27% | NIC | Schneider et al. 2005 |
| HD 53143 | 6.820 | +0.80 | G9V | 18.3 | 1000 ± 300[j] | 0.025% | ACS | Kalas et al. 2006 |
| HD 61005 | 8.932 | +0.71 | G8V | 34.4 | 90 ± 40[f] | 0.25% | NIC | Hines et al. 2006 |
| HD 92945 | 8.592 | +0.89 | K1V | 21.4 | 80 – 150/300[h] | 0.076% | ACS | Golimowski et al. 2011 |
| HD 107146 | 7.622 | +0.62 | G2V | 27.5 | 80 – 200[b] | 0.12% | ACS | Ardila et al. 2004/05 |
| HD 139664 | 4.64 | +0.40 | F5V | 17.4 | 300(-200,+700)[i] | 0.009% | ACS | Kalas et al. 2006 |
| HD 181327 | 7.475 | +0.48 | F6V | 51.8 | ~ 12 – 20[a] | 0.25% | NIC/ACS | Schneider et al. 2006 |
| AU Mic | 10.277 | +1.44 | M1V | 9.9 | 6 – 20[c] | 0.44% | ACS[†] | Krist et al. 2005 |
| MP Mus | 10.44 | +0.94 | K1Ve | ~86 | 13 ± 5 | | NIC | Cortes et al. 2009 |

* For additional characteristics of the CS debris systems resulting from this study see Table 5.
Age estimation: [a] Schneider et al. 2006; [b] Williams et al. 2004; [c] Zuckerman et al. 2001; [d] Moór et al. 2006; [e] Schneider et al. 2005; [f] Hines et al. 2007; [g] Kalas et al. 2007b; [h] Lopez-Santiago / Plachan 2008, Mamajeck & Hillenbrand 2008; [i] Nordstrom et al. 2004; [j] Decin et al. 2000, Song et al. 2000, Nordstrom et al. 2004
- [†] AU Mic ground-based discovery imaging: Kalas et al. 2004
- [+] updated spectral type since Schneider et al. 2005 by Fitzgerald et al. 2007

Scattered-light images prior to this study, obtained with NICMOS, and/or ACS, were limited in the stellocentric regions probed as follows:

(*a*) only the outer regions (typically r > app 2") were probed due to the limiting effective IWA of the ACS coronagraph, or

(*b*) spatial resolution was limited (2 – 4x worse than optical coronagraphy) and PSF subtraction artifacts remain for those disks probed as closely as r = 0.3" with NICMOS, due to the longer (than optical) wavelength employed in the near-IR and two-roll only observations secured, or

(*c*) sensitivity to low-surface brightness (SB) dust (though detected either at low S/N or only where the dust was relatively brightest) was limited due to the relatively shallow exposure depths used in the imaging survey programs for many of the disk discovery images, or

(*d*) compositionally diagnostic "color" information (wavelength dependent spectral reflectivity of the grains), both in bulk and spatially resolved within the disks, is currently lacking because observations were conducted in only the optical or only in the near-IR (but not both) in the very limited regions (if any) of spatial overlap, and not at all in the inner regions of the disks.

This sample includes ten (42%) of the of 24 currently known light-scattering debris disks, and includes disks with host stars spanning spectral types from A0 to M1, distances from ~ 10 to 100 pc, and ages from ~ 10 – 1,000 Myr. These disks span a range of inclinations to the line-of-sight from edge-on ($i$ = 90˚; the most favorable geometry for discovery detection), to nearly face-on ($i$ = 0˚). These are the benchmark disks for thermal IR imaging and SED modeling (Lebreton et al. 2012; Donaldson et al. 2013), some of the earliest to be observed by ALMA (MacGregor et al. 2013; Walsh & Testi 2012), and anticipated for the Gemini Planet Imager (GPI) and the Spectro-Polarimetric High-contrast Exoplanet Research (SPHERE) coronagraph.

## 3. OBSERVATIONAL MOTIVATION/GOALS

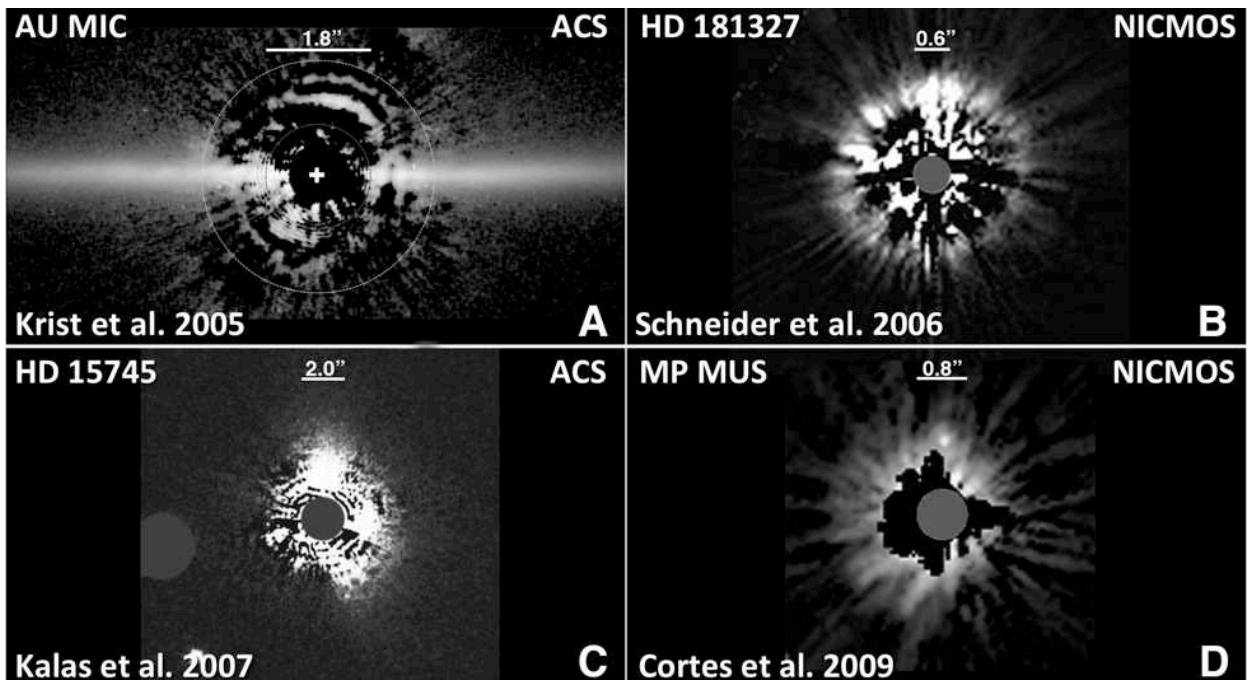

**Figure 1.** Prior *HST* PSF-subtracted optical (ACS) and near-IR (NICMOS) coronagraphic images of a representative sample of CS disks from prior programs are limited in image quality due to "pollution" by optical artifacts from PSF-subtraction residuals well beyond the physical limits of their image plane central coronagraphic obscurations.

Over the past 16 years, *HST* coronagraphy (with PSF subtraction) has provided some remarkable, even paradigm-changing, images of CS disks through the investment of many hundreds of *HST* orbits (see representative discovery images in Fig. 1). They are, however, derived from a diversity of heterogeneously planned and executed survey programs that typically have been limited in depth, inner (stellocentric) working angle, and are characteristically "polluted" with artifacts from non-optimal PSF subtractions; this impedes unambiguous interpretation and often provides only loose constraints for well-informed modeling, making comparison within the sample and with models challenging. However, studies of individual objects have demonstrated ways to improve upon the discovery imaging, both by using appropriate observational strategies (see §4) and advanced data reduction techniques (see §§ 5.4-5.8, and § 6.1).

"Raw" *HST* coronagraphy alone, with any of the telescope's coronagraphically augmented instruments, has insufficient starlight suppression to study even highly efficiently starlight-scattering CS dust without further "disk-to-star" contrast enhancement via PSF subtraction. PSF subtraction significantly enhances the imaging dynamic range (disk-to-star contrast) by 2 to 3 orders of magnitude, further removing residual starlight from the PSF halo that is incompletely suppressed by the *HST* coronagraphs alone. To enable quantitative metrical comparisons with other instrumental systems, in this paper, we specifically define "contrast" as the flux density contained within a resolution element at any stellocentric angular distance arising from the stellar PSF halo ratioed to the flux density within the central resolution element of stellar PSF core unsuppressed by coronagraphy and/or PSF-subtraction. *HST* PSF-subtracted coronagraphic images are often degraded or compromised by optical artifacts resulting from the subtraction of imperfectly "matched" stellar PSF templates due primarily to chromatism (differences in photopheric SEDs), particularly under the very broad STIS unfiltered bandpass (Grady et al. 2005), and by variable focus wavefront error caused by spacecraft thermal instabilities on orbital and longer timescales (a.k.a. "breathing" of the optical telescope assembly). Additional contributors to such artifacts that impact different spatial scales and correlation scale lengths in PSF-subtracted images are discussed by Schneider et al. (2010). To *a priori* minimize these in defining our GO 12228 observation plan, we adopted an approach with:

(1) Very close color-matching (in Δ[B-V], and Δ[V-R] when possible) of PSF template stars to the science targets, as indicated in Grady et al. (2005).

(2) Selection of PSF template stars to minimize target-to-template spacecraft slew distance (very similar sun angles with attitude changes)

(3) Interleaving of PSF observations with the science target in sequential, non-interruptible, orbits (as learned from experience with ACS; see Clampin et al. 2003)

(4) Maintaining similar target:PSF orbit orientation (thus similar β angles at nominal roll)

(5) Using 6 target and 2 PSF orbits per star to obtain data sampling of both inter/intra orbit

breathing phase.

(6) Obtaining observations at different spacecraft roll angles so that the final "analysis-quality", multi-roll combined, image covers as much of the 360° field about the star as possible; and that several images cover the same portions of the field of view, enabling robust separation of PSF and detector artifacts from the disk imagery.

(7) Obtain/combine data from different locations (and integrated exposure depths) along the STIS coronagraphic tapered wedge to sample the disk from an inner working angle from 0.3" $\leq$ r $\leq$ at least 12.5".

Details are given in Table 2.

**Table 2**
Planned, Contemporaneously Interleaved, PSF Template Stars

| Disk Target | PSF Star | PSF Spec | PSF V | PSF B-V | Δ(BV)[a] | PSF Slew[b] | PSF ΔONR[c] | |
|---|---|---|---|---|---|---|---|---|
| HD 15115 | HD 16647 | F3V | 6.26 | +0.37 | –0.02 | 3.5° | 1.7° | -8.6° |
| HD 15745 | HD 16199[d] | F2 | 7.47 | +0.34 | –0.02 | 28.3° | -38.3° | |
|  | HD 20499 | F2 | 7.27 | +0.35 | –0.04 | 9.2° |  | 33.3° |
| HD 32297 | HD 33403 | B9V | 8.31 | +0.01 | –0.19 | 4.9° | -9.8° | 5.6° |
| HD 53154 | HD 59780[e] | G8III | 6.52 | +0.92 | –0.12 | 4.8° | --- | 3.6° |
| HD 61005 | HD 56161 | G5IV | 6.94 | +0.75 | –0.04 | 4.8° | -4.5° | 7.1° |
| HD 92945 | HD 89585 | G5III | 7.41 | +0.88 | +0.01 | 9.3° | 0.8° | 0.8° |
| HD 107146 | HD 120066 | G0.5V | 6.30 | +0.66 | –0.04 | 23.8° | 17.5° | 11.1° |
| HD 139664 | HD 99353 | F7V | 5.17 | +0.50 | –0.10 | 39.7° | 48.4° | 68.55° |
| HD 181327 | HD 180134 | F7V | 6.37 | +0.45 | +0.03 | 1.4° | 0.5° | 0.7° |
| AU Mic | HD 191849 | M0V | 7.97 | +1.46 | –0.02 | 15.1° | 2.9° | 40.9° |
| MP Mus | CD-701017 | --- | 9.58 | +1.08 | –0.07 | 38.4° | 5.6° | 0.5° |
| Brt. Calib.[f] | Alpha Pic | A8V | 3.30 | +0.18 | N/A | N/A | N/A | N/A |

[a] Difference in B-V color index between target and PSF template stars
[b] *HST* attitude change in slewing between target and PSF template star
[c] ONR: "Off Nominal Roll"; ΔONR: roll angle change between 2$^{nd}$ target & PSF star visits in contemporaneous set
[d] Previously undetected close binary companion – unsuitable for PSF subtraction template
[e] PSF template obtained at in 2$^{nd}$ of two planned sets of observations – applied to both
[f] Bright PSF template test/calibration target (not directly used for science target PSF subtractions)

## 4. OBSERVATIONS – PLANNING & EXECUTION

### *4.1. Strategy*

We obtained multiple orientation STIS coronagraphic imaging of the disks listed in Table 1 and single orientation observations otherwise obtained in the same manner, of a color-matched set of PSF template stars (Table 2), one for each science target. Our goals were to obtain high spatial resolution images of the light-scattering CS material in these disk systems over a wide stellocentric radial and azimuthal range. We imaged as close to the star as possible in stellocentric angle with the STIS occulting Wedge A (limited by the 0.3" half-width at the smallest, standard, position of WedgeA-0.6), and also exposed deeply, commensurate with a requisite "multiple roll" strategy to minimize PSF subtraction residuals, to image tenuous low-SB dust in the outermost reaches of these CS debris systems (at WedgeA-1.0 position) previously unimaged with ACS. The multiple-orientation observing approach, importantly and simultaneously, provides both rejection (by anti-coincidence through median filtering) of optical artifacts that could otherwise be mis-construed as intrinsic sub-structures in CS disks, and completely (or nearly so) azimuthally samples "around" the otherwise obscuring STIS coronagraphic wedge and unapodized *HST* diffraction spikes. The effectiveness of this approach for debris disk imaging into sub-arcsecond IWA was demonstrated previously to lower efficacy with only two spacecraft roll angles (e.g., Schneider et al. 2009). In this study we planned 6 spacecraft roll angles, to provide complete azimuthal coverage at r > 0.3".

### *4.2. STIS Coronagraphic Exposures*

For optimal S/N at the edge of the r = 0.3" occulting wedgeA-0.6, our individual short exposures were designed on a target by target basis to reach (but not exceed) 90% of the full-well depth of the STIS CCD from the stellar light PSF-halo that dominates significantly over starlight scattered by CS dust before PSF subtraction. For this we scaled (based on stellar V and B band magnitudes) from previous observations of STIS coronagraphic PSF template stars as well as prior observations of protoplanetary and debris disk hosts. Multiple "short" exposure images (target specific integration times of a few to < tens of seconds depending upon stellar brightness) were obtained to improve the S/N as demonstrated previously for HR 4796A per Schneider et al. (2009). We supplemented these "shortest" exposures with much deeper exposures (typically ~ 10x each, multiply CR-split for cosmic ray rejection and compensation) to map the mid and outer regions of the disks where the CS disk SB is (in general) much lower (due to the $r^{-2}$ dilution of the stellar radiation field), and imaging is otherwise photon, rather than contrast, limited. We additionally employ the 0.5" half-width position of WedgeA-1.0, advantageously using the larger unvignetted coronagraphic FOV, to further investigate external perturbations to the CS dust in the form of "tidal tails" and other structures from posited companions and interactions with the local

ISM, and blow-out of small scattering particles at large stellocentric distance.

The WedgeA-0.6 images were designed to: (a) image as close to the central star as possible without saturating at the edges of the wedge in the pre-PSF subtracted images; and (b) provide full (or nearly so) azimuthal coverage unaffected by the wedge and *HST* diffraction spikes in multiple orientation combination, to r ≤ 3.3". Deeper images (of necessity saturating the regions close to the star) extending to much larger angular distances (using a commensurately larger sub-array read out) were separately obtained using WedgeA-1.0. These WedgeA-1.0 images were designed to simultaneously provide: (a) deep images of the outer portions of the target disks to stellocentric distances of ≥ 6"; (b) unsaturated imaging overlapping into the smaller stellocentric regions (to r > ~ 2") separately covered in the non-overlapping domain by the WedgeA-0.6 images to the physical limit of the occulting wedge edges. See Table 2 for target-specific details with additional information available in the as-executed *HST* Phase 2 observation plan[2].

All data (following standard mode-2 coronagraphic target acquisitions) were obtained with STIS instrumental CCDGAIN = 4, and the longer (deeper) exposures with CR-SPLIT = 3 to 5, optimized for the length of target (declination) specific visibility windows. We employ sub-array readouts tuned to the maximum anticipated angular extent of the outer regions of the light scattering disks detectable to the anticipated levels of the background noise to minimize dead time and maximize observational efficiency. For details see Table 3.

### *4.3. STIS Coronagraphic Orbits*

For each science target we obtained a sequence of short and long coronagraphic exposures (see above), in a total of 6 orbits to observe at 6 different spacecraft roll (celestial field orientation) angles for spatial coverage, and to minimize PSF-subtraction residuals that are only partially correlated with different rolls (Schneider et. al. 2009). We "broke" the six orbits into two sets of three sequentially executed orbits with, ideally (but seldom realized) relative orientations -30°, 0°, and +30° (with loose tolerance) from off-nominal roll. The two sets of three orbits were each scheduled (at different epochs) at nominal roll orientations differing by 90°–150° to fully sample, (or nearly so) azimuthally beyond the r = 0.3" edge of the coronagraphic wedge, and probe beyond the r = 3.4" edge of the STIS FOV on the narrow side of the wedge's taper in the WedgeA-0.6 position. This also had the advantage, by observing at two epochs and thus different Sun and β angles (stochastically, not by requirement), of guarding against the possibility of an anomalous "breathing" phase in a single set of observations that otherwise can significantly degrade PSF subtraction. We imposed no absolute orientation constraints on the three-orbit sets, even in cases where the celestial orientation of disk major axes are known. This eased already constrained schedulability of the individual visits with no penalty or degradation for meeting our science goals. Along with each three-orientation visit set, we also identically observed, but only at nominal roll for one orbit, a bright target-color matched PSF template star to be used for contemporaneous PSF subtraction. Thus, a total of eight orbits were required for each disk-host/PSF-star pair. Hence, we consumed a total of 88 orbits to observe our eleven CS disk targets and matched PSF subtraction template stars.

---

[2] http://www.stsci.edu/hst/phase2-public/12228.pro

### *4.4. PSF Star Selection and Coronagraphic Performance*

Extensive experience with *HST* broadband coronagraphy (with NICMOS R = $\Delta\lambda/\lambda_{cent}$ = 0.53 with F110W and even more extreme with STIS unfiltered R = 0.75), had clearly demonstrated the need for PSF stars very well matched in color indices under the instrumental bandpass to mitigate differential chromatic effects in PSF structure that otherwise hinder efficacious PSF subtraction (Grady et al. 2005; Schneider et al. 2010). Hence, whenever possible given other constraints, we selected PSF template stars of very similar photospheric SEDs (and of sufficient brightness) well-matched to each of our CS disk hosting target stars (with [B-V] = +0.02 to +1.44, see Table 2) differing by $\Delta|B-V| < 0.1$ for each target/PSF-template star pairs.

A second major contributor to optical artifacts in PSF-subtracted images arises from secular changes in the fine structure of the coronagraphic PSF over time (spacecraft orbit and orbital phase) and with spacecraft repointing (attitude and roll). Both differential chromatic and PSF instability effects have hindered ACS coronagraphic observations, very often having prevented even the much larger limiting IWAs of the ACS coronagraph's r=0.9" and 1.8" masks to have been reached as those observations had been conducted. To mitigate these issues to the greatest extent possible, along with each of our CS disk-hosting target stars we interleaved (as the 3$_{rd}$ of 4 sequential orbits), at nominal roll, a target-specific color matched PSF subtraction template star in each of the 90°–150° rotationally offset visit sets. Whenever possible, so as not to excite

"breathing" modes with thermal instabilities in the *HST* Optical Telescope Assembly (OTA), we constrained the cross-sky target/template slew distance to < 10° (to minimize spacecraft attitude changes), and kept roll angle differentials from target/template nominal roll as close to zero as possible. See Table 2 for details.

### *4.5. Details of the As-Executed Observations*

Table 3 provides an as-executed visit-level observation summary for each disk target and its contemporaneously observed PSF star, each observed at two epochs in four contiguous *HST* orbits. In this program, each visit is comprised of a target acquisition exposure (not shown in Table 3) followed by multiple WedgeA-0.6 and WedgeA-1.0 exposures in a single orbit. At each epoch, in chronological order, the 1$^{st}$, 2$^{nd}$, and 4$^{th}$ visits are of the disk target with celestial field orientation angles scheduled as close to -30°, 0°, +30° from off-nominal roll as possible, given *HST* guide star selection and other scheduling constraints. The Table 3 "Orient" angles give the field angle measured from the 50CCD aperture's image +Y axis to celestial north counter-clockwise (through east). In the 3$^{rd}$ visit at each epoch the PSF star is imaged at its nominal roll angle, thus interleaving PSF observations between the 2$^{nd}$ and 3$^{rd}$ disk target orbits (orbits 2 and 4 in each set of target+template images). In most cases the absolute orientation angle of the spacecraft for the PSF orbit, at its nominal roll, is similar to the 2$^{nd}$ disk target orbit to avoid driving breathing modes with large changes in spacecraft β angle. For each visit, columns 4 & 6 in Table 3 indicate the total number of WedgeA-0.6 & WedgeA-1.0 exposures taken in the three same-epoch disk visits (or single PSF visit) followed in columns 5 and 7 by the corresponding total integration time. In detail, from visit-to-visit and between the two epochs the exposure times in particular for the WedgeA-1.0 images are not always identical for the same targets due to variances in target visibility intervals at different times of the year and at different celestial orientations. A detailed "Phase 2" exposure-level observation plan, and orbit layout, with a higher level of detail than summarized in Table 3 is available from STScI[3].

Each visit has a unique identifier (Table 3, column 8), monotonic for epochal target/template pair, with #'s 1 – 4 or #'s 5 – 8 corresponding to the 1$^{st}$ – 4$^{th}$ visits at that epoch. Throughout this paper we refer to visit numbers as V## in accordance with the column 8 visit data id. The raw and pipeline processed exposure level data archived in the Mikulski Archive for Space Telescopes (MAST) are entered as program OBIW## followed by these visit-level data identifiers.

---

[2] http://www.stsci.edu/hst/phase2-public/12228.pdf

**Table 3**
*HST* GO 12228 Observation and Data Log

| Target (Disk/PSF) | UT Date Obs. Start | Orient ° | W0.6 # Exp | W0.6 $T_{EXPT}$ *all visits* sec | W1.0 # Exp | W1.0 $T_{EXPT}$ *all visits* sec | Visit Data ID[g] |
|---|---|---|---|---|---|---|---|
| HD 15115 | 08OCT11 | 195.1, 220.1, 245.1 | 60 | 1026 | 12 | 4566.8 | 41, 42, 44 |
| HD 16647 | 08OCT11 | 221.7 | 24 | 261.6 | 7 | 1439.9 | 43 |
| HD 15115 | 19DEC11 | 0.1, 27.6, 30.1 | 60 | 1026 | 12 | 4866 | 45, 46, 48 |
| HD 16647 | 19DEC11 | 19.0 | 24 | 261.6 | 7 | 1439.9 | 47 |
| HD 15745 | 02JAN12 | 20.1, 40.1, 60.1 | 42 | 1436.4 | 9 | 5022.7 | 71, 72, 74 |
| *HD 16199*[a] | *02JAN12* | *1.7* | *14* | *521.4* | *3* | *1639.8* | *73* |
| HD 15745 | 03NOV12 | 105.1, 127.1, 154.1 | 42 | 1436.4 | 9 | 4199.4 | B5, B6, B8 |
| HD 20499 | 03NOV12 | 160.4 | 11 | 304.7 | 4 | 1644.8 | B7 |
| HD 32297 | 22DEC11 | 325.1, 355.1, 25.1 | 27 | 1652.4 | 9 | 4961.7 | 05, 06, 08 |
| HD 33403 | 22DEC11 | 345.3 | 8 | 576 | 3 | 1647.9 | 07 |
| HD 32297 | 06NOV11 | 222.1, 242.1, 262.1 | 27 | 1652.4 | 9 | 4961.7 | 01, 02, 04 |
| HD 33403 | 06NOV11 | 247.6 | 8 | 576 | 3 | 1647.9 | 03 |
| HD 53154 | 10JAN11 | ~~301.1~~[c], 318.6, 336.1 | 40[c] | 688[c] | 12[c] | 3600[c] | ~~81~~[c], 82, 84 |
| HD 53154 | 04APR11 | 30.8, 48.2, 65.7 | 60 | 1032 | 18 | 5400 | 85, 86, 88 |
| HD 59780[d] | 04APR11 | 45.7 | 24 | 333.6 | 8 | 1668 | 87 |
| HD 61005 | 17FEB11 | 340.1, 10.1, 20.1 | 30 | 1893 | 9 | 4748.4 | 55, 56, 58 |
| HD 56161 | 18FEB11 | 5.5 | 16 | 313.6 | 3 | 1773 | 57 |
| HD 61005 | 06DEC11 | 250.1, 270.1, 290.1 | 30 | 1893.6 | 9 | 4748.4 | 51, 52, 54 |
| HD 56161 | 06DEC11 | 277 | 16 | 313.6 | 3 | 1773 | 55 |
| HD 92945 | 29FEB2012 | 25.7, 47.7, 67.7 | 41[b] | 1514.4[b] | 9 | 4098.3[b] | 61, 62, 64 |
| HD 89585 | 29FEB2012 | 48.5 | 16 | 475.2 | 3 | 1614 | 63 |
| HD 92945 | 06MAY2011 | 25.7, 46.7, 67.7 | 42 | 1575 | 9 | 4761 | 65, 66, 68 |
| HD 89585 | 06MAY2011 | 48 | 16 | 475.2 | 3 | 1614 | 67 |
| HD 107146 | 22FEB11 | 192.1, 209.6, 227.1 | 48 | 1065.6 | 12 | 4764 | 25, 26, 28 |
| HD120066 | 22FEB11 | 227.1 | 24 | 264 | 7 | 1421 | 27 |
| HD 107146 | 03MAY11 | 66.7, 81.7, 96.7 | 48 | 1060.8 | 12 | 4988.4 | 21, 22, 24 |
| HD120066 | 03MAY11 | 86.7 | 24 | 264 s | 7 | 1445 | 23 |
| HD 139664 | 31JUL11 | 24.1, 48.1, 66.1 | 48 / 24 | 115.2 / 1152 | 9 | 4469.4 | 91, 92, 94 |
| HD 99353 | 31JUL11 | 96.4 | 16 / 8 | 62.4 / 624 | 3 | 1410.0 | 93 |
| HD 139664 | 23MAY11 | 303.1, 327.1, 351.1 | 48 / 24 | 115.2 / 1152 | 9 | 4469.4 | 95, 96, 98 |
| HD 99353 | 23MAY11 | 35.6 | 16 / 8 | 62.4 / 624 | 3 | 1410.1 | 97 |
| HD 181327 | 20MAY11 | 222.7, 242.7, 262.7 | 60 | 1314 | 24 | 5124 | 11, 12, 14 |
| HD 180134 | 20MAY11 | 243.2 | 24 | 262.8 | 8 | 1656 | 13 |
| HD 181327 | 10JUL11 | 293.1, 313.6, 334.1 | 60 | 1314 | 24 | 5060 | 15, 16, 18 |
| HD 180134 | 10JUL11 | 314.2 | 24 | 285.6 | 8 | 1656 | 17 |
| AU Mic | 09AUG10 | 311.1, 341.1, 10.1 | 36 | 1638 | 9 | 4887 | 35, 36, 38 |
| HD 191849 | 09AUG10 | 343.9 | 16 | 489.6 | 3 | 1713 | 37 |
| AU Mic | 16JUL11 | 228.1, 257.1, 287.1 | 36 | 1638 | 9 | 4887 | 31, 32, 34 |
| HD 191849 | 16JUL11 | 298.0 | 16 | 489.6 | 3 | 1657.8 | 33 |
| MP Mus | 03MAR11 | 278.1, 300.1, 321.1 | 21 | 8158.5 | 0[e] | 0[e] | A1, A2, A4 |
| CD-701017 | 03MAR11 | 305.7 | 11 | 2681.1 | 0[e] | 0[e] | A3 |
| MP Mus | 06JUN11 | 359.1, 15.6, 36.1 | 21 | 8158.5 | 0[e] | 0[e] | A5, A6, A8 |
| CD-701017 | 06JUN11 | 18.0 | 11 | 2681.1 | 0[e] | 0[e] | A7 |
| Alpha Pic[f] | 10JAN2011 | 324.0 | 16/8 | 11.2 / 112 | 16 / 6 | 224/1530 | 83 |

[a] V73: Close angular proximity background star and/or companion revealed – ill-suited as PSF subtraction template.
[b] V62: Interrupted by recoverable guide star loss-of-lock. #EXP, $T_{expt}$ tabulated as executed, not as planned.
[c] V63: Target Acquisition Failure – no data from this visit. #EXP, $T_{expt}$ tabulated as executed, not as planned.
[d] Contemporaneous (interleaved) PSF template observation only for HD 53143 only for V85-V88.
[e] VA1-VA8: Observations planned for only WedgeA-0.6 for this PMS star disk target and its PSF template.
[f] V83: Bright, blue, program calibration target. Obtained contemporaneously with V81-V84.
[g] Visit level dataset id as assigned by MAST. GO 12228 data archived as obiw + datset_id + *.

## 5. IMAGE DATA CALIBRATION, REDUCTION & PSF SUBTRACTION

### 5.1. Exposure-Level Basic Instrumental Calibration

The GO 12228 observations executed between May 2011 and November 2012. All raw image data were retrieved soon after the execution of each target/epoch visit set from MAST. Basic image data calibration of each individual exposure in the raw data frames, including bias correction, dark subtraction, flat-fielding, linearity correction, and identification of instrumentally deficient and saturated pixels, was performed using the STSDAS *calstisa* software[4] as routinely used in the *HST*/OPUS pipeline. Initial basic instrumental pipeline processing is usually done by STScI immediately following the epoch of observation with bias and dark reference files that often are "stale" and do not best reflect the observationally contemporaneous state of the instrument. Thus, prior to final reduction and PSF subtraction, all raw images were re-calibrated using later-obtained (by STScI as part of the Cycle 19/20 calibration program) bias and dark reference data closer to the dates of actual observations subsequently provided to us through STScI's Calibration Database System[5].

### 5.2. Visit-Level Reduction

All instrumentally calibrated frames were visually inspected to check for (compensate if possible, or vet if necessary) any significant imaging performance anomalies such as exceptionally high cosmic-ray event hit rates (e.g., in V45 3$^{rd}$ WedgeA-1.0 exposure) and loss of two-FGS fine-lock pointing control (e.g., in V62 loss of two WedgeA-0.6 exposures and curtailment of one WedgeA-1.0 exposure). Sky backgrounds were estimated from manually selected regions in the periphery of the field far from the target star/disk and not coincident with any stars/galaxies in the field, and subtracted. Within each visit, separately, the multiple-exposure data obtained at each of the two Wedge A occulter positions were median combined for cosmic ray rejection and diminishment of random instrumental and photon noise compared to stellar/disk signal, and then converted to count rate images based on shutter open times. The visit-level median combined count rate images were then screened for background biases statistically unequal to zero within the pixel-to-pixel noise in the field periphery (and corrected in rare cases where outliers were detected). These, and subsequent steps in image processing and later analysis were carried out using the IDL-based IDP3 software (Stobie et al. 2006).

---

[4] http://www.stsci.edu/hst/stis/software/
[5] http://www.stsci.edu/hst/observatory/cdbs

### 5.3. Absolute Photometric Calibration

Throughout this paper we adopt STScI's absolute photometric calibration of the STIS unfiltered 50CCD instrumental response, applicable for all coronagraphic imaging, as codified through the STSDAS synphot/calcphot software package and calibration database reference files. We summarize in Table 4.

**Table 4**
STIS Coronagraphic Mode Absolute Photometric Calibration

| Full Spectral Passband | 2000 - 10500 Å |
|---|---|
| Pivot Wavelength ($\lambda_p$) | 5752 Å ; $F_\lambda = g(F_\nu * c) / \lambda_p^2$ |
| FWHM of Unfiltered Passband | 4330 Å |
| Instrumental Sensitivity* | 1 count (ADU gain = 4.096) s$^{-1}$ = 4.55E-7 Jy or = AB mag 26.386 (with gain = 4.096 e$^-$ ADU$^{-1}$) |
| (AB) Zero-point magnitude | 3671 Jy |

* for a spectrally-flat source.

In terms of SB, then, for an HRC astrometric ("plate") scale of 0.05077 arcsec pixel$^{-1}$ then:
1 count s$^{-1}$ pixel$^{-1}$ = 0.1765 mJy arcsec$^{-2}$.

### 5.4. Locating the Target Star

Because of systematic limitations in the *HST* on-board autonomous target acquisition process, the final targeting slew to place the target star at the desired location in the STIS 50CCD field can deviate by ~ ± 1 pixel (50 mas) from a fixed fiducial position w.r.t. the grid of detector pixels, and typically by ~ ± 0.25 pixels from the deployable focal plane coronagraphic mask nominal target location. Thus, prior to later two-Wedge position and multi-roll image combination, and subsequent star/disk relative astrometric measures, the location of the coronagraphically occulted target star in visit-level median-combined science image aperture frame (SIAF) must be determined to astrometrically high precision. No direct (unocculted) image of the star itself

(other than a target acquisition image at a different field position and with an attenuating filter in place, both insufficient for this purpose) was possible at its post-slew position used for science imaging due to saturation limits from the bright target stars. Thus, we used the previously demonstrated "X marks the spot" method (c.f., Schneider et al. 2009 § 4.2) to determine the location of the coronagraphically occulted star. We use the bright (but unsaturated) orthogonal *HST* diffraction spikes, unsuppressed by the STIS coronagraph, that are rotated ~ ± 45° from the occulting wedge axis (see Fig. 2A) and with respect to the STIS 50CCD SIAF x/y pixel grid.

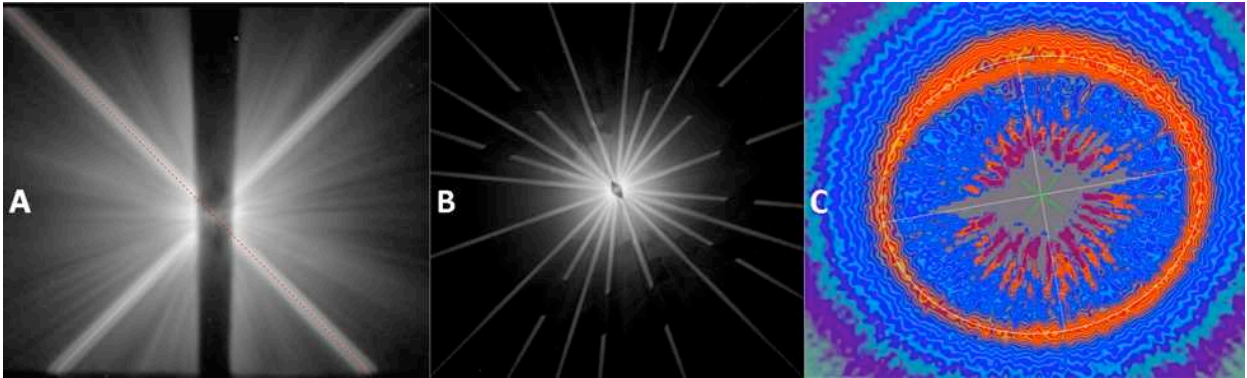

**Figure 2.** A - Representative STIS coronagraphic stellar PSF (WedgeA-0.6 location) showing the best-fit to the ridge of pixels along one of the two sets of orthogonal diffraction spikes that "point back" to the star (the other spikes unannotated to show better their morphology). The intersection of the two linear fits locates the star to a precision typically of ~ 4 - 5 mas for the depth of integration systematically used in individual images (differently for WedgeA 0.6 and 1.0) throughout the GO 12228 program. B – In two-wedge combined multi-roll imaging (shown here before PSF subtraction for illustrative purposes) the multiple diffraction spikes in co-aligned images allow maintaining a systematic precision in co-registration at the same 4 – 5 mas level. C – Precise determination of the star location (green marker) enables high-precision differential astrometric measures and evaluating the significance of the measured positions of disk sub-structures and features w.r.t. the disk-hosting star, such as photocentrically determined offsets in CS debris rings (in this case for HD 181327 by elliptical isophote fitting with the best fit to the elliptical annulus of peak radial SB and corresponding minor/major axes overlaid; see § A.1).

The target star position in each visit-level frame is typically located with an experimentally determined precision w.r.t. the SIAF frame of ~ 4 - 5 mas (~ 0.1 pixels) RMS. These diffraction spike determined stellar centroids are subsequently used to initially co-register intra-visit target images acquired at their differing WedgeA-0.6 and 1.0 positions. They are also used (following PSF-subtraction) for inter-visit (multiple roll) image alignment at a common-origin, astrometrically-determined, stellocentric reference position (see Fig. 2B) against which disk sub-structures can be located w.r.t. the position of the host star (e.g., Fig. 2C) with high precision.

*5.5. Visit-Level PSF-Subtractions*

The subtraction of PSF template images from those of the disk-hosting stars follows the process described in detail by Schneider et al. (2009). The iterative global PSF-subtraction process, treating template brightness and X/Y positions as free parameters, is initially done with target and template images in the detector frame orientation (so disk features rotate in the PSF-subtracted frames from visit to visit at different spacecraft roll angles). A representative full set of visit-level PSF subtracted images (optimized to globally minimize residuals from PSF-subtraction) is shown for HD 181327 WedgeA-0.6 and 1.0 in Figs. 3 and 4, respectively.

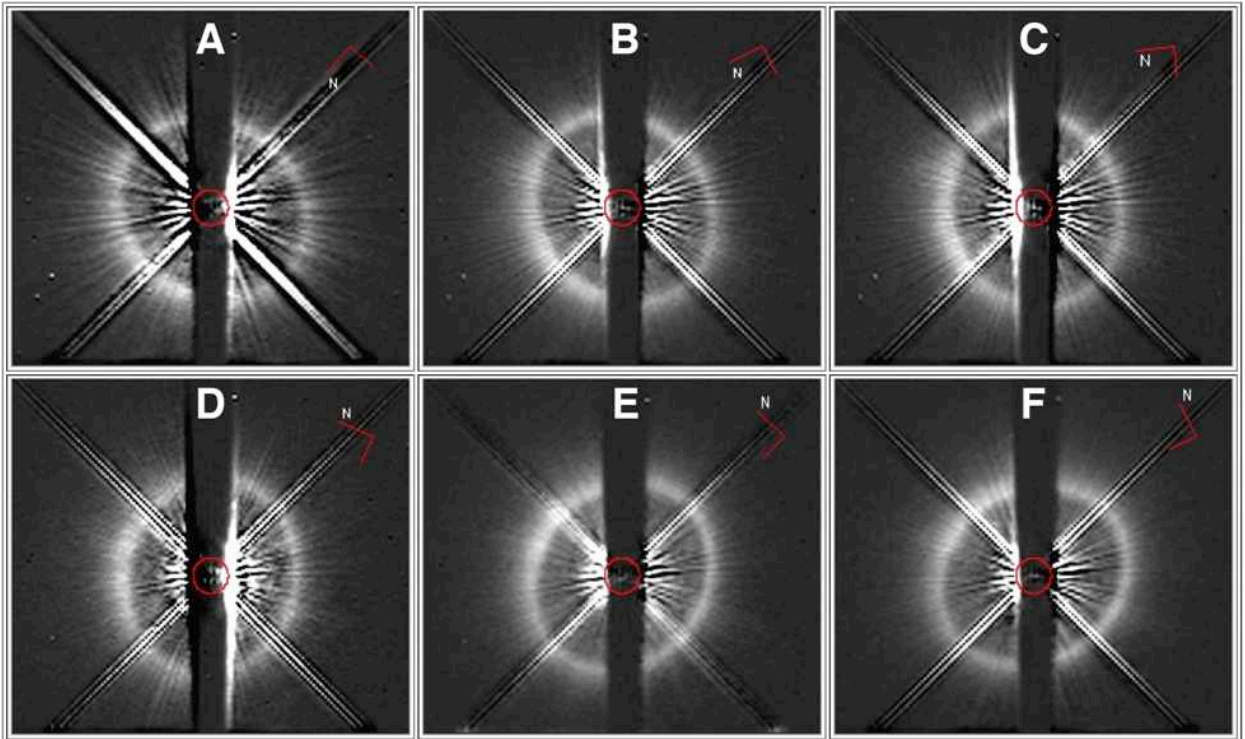

**Figure 3.** WedgeA-0.6 visit-level PSF subtractions - HD 181327. Top (A-C): Visits 11, 12, 14. Bottom (D-F): Visits 15, 16, 18. Top and bottom images respectively using contemporaneously interleaved Visit 13 and 17 HD 180134 PSF templates (see Table 3 for details). 6.2" x 6.0" sub-array regions shown with a linear display from -1.0 (black) to +7.0 (white) cts s$^{-1}$ pixel$^{-1}$. Central red circle overlaid is r = 0.3" centered on the occulted position of the star.

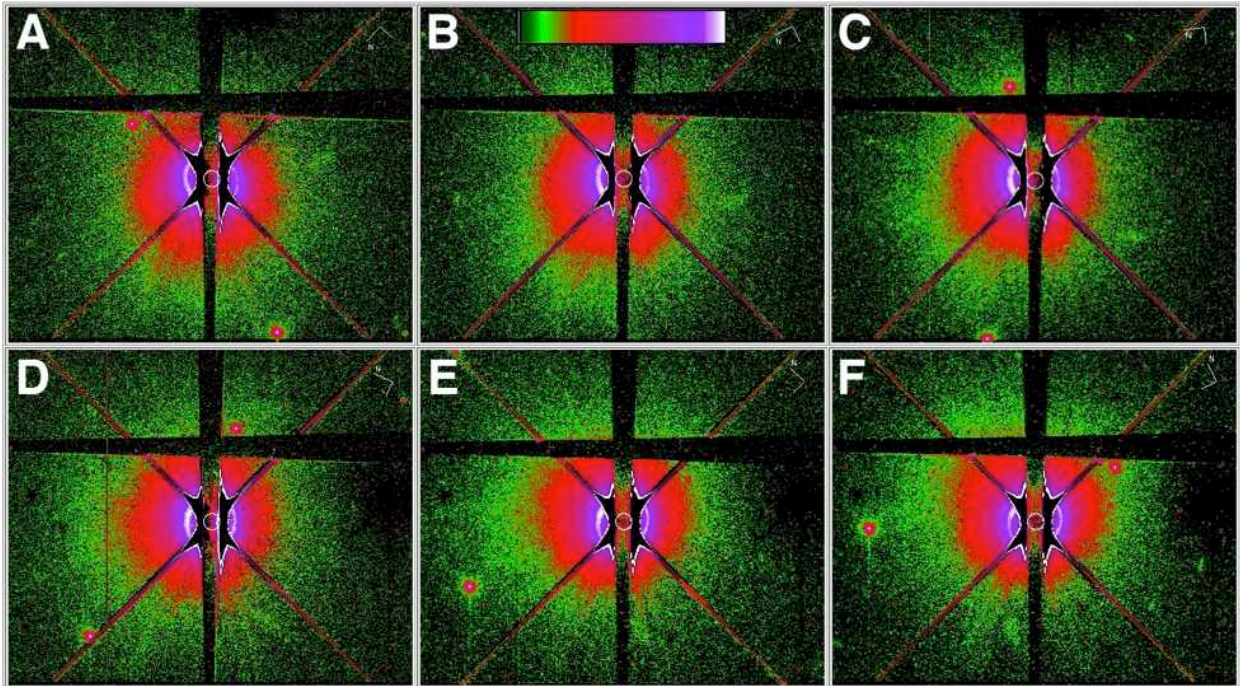

**Figure 4.** WedgeA-1.0 visit-level PSF subtractions - HD 181327. Top (A-C): Visits 11, 12, 14. Bottom (D-F): Visits 15, 16, 17. Top and bottom images respectively using contemporaneously interleaved Visit 13 and 17 HD 180134 PSF template (see Table 3 for details). 25.3" x 21.0" sub-array regions shown with a $\log_{10}$ display from [-2.0] to [+0.75] cts s$^{-1}$ pixel$^{-1}$ {dex} (inset color bar). Central white circle r = 0.5" at the occulted position of the star. The flanking saturated regions (black) extend beyond the edge of the wedge.

Not all visits for a given target result in PSF-subtracted images with the same image quality. For example, in Fig. 3 the visit 16 (panel E) PSF-subtracted image of the HD 181327 debris ring is most free of the radial "tendril"-like artifacts that typify differences in *HST* "breathing" (driving differential wavefront focus errors), while the visit 11 (panel A) image is most affected by a breathing phase differential. It is not unusual for the first visit in an imaging set to have larger breathing residuals, in particular if the thermal state of the telescope is significantly changing following a slew from a different location in the sky from a preceding observation. The relatively low level of dispersion seen amongst the two sets of six HD 181327 images (WedgeA-0.6 and 1.0) is typical. However, significantly degraded results, despite all efforts for "PSF matching", can occur with episodes of large "breathing" excursions.

Apart from "breathing" residuals, the on-board autonomous coronagraphic target acquisition process attempts to place targets on the mid-line of the occulting wedge. Because of repeatability limitations in the STIS slit wheel that deploys a retractable focal plane mask used for coronagraphic observations, target placement "behind" the wedge can vary by typically ~ ¼ pixel on initial placement in each visit, as we see in the GO 12228 data and as previously reported by Kriss et al. 2013. As a result, there is often a brightness differential from the PSF halo at the opposite side wedge edges that can change both in brightness and parity from acquisition to acquisition. With PSF-subtraction this can result in positive and negative edge gradients that can extend over several pixels thus increasing the innermost working angle otherwise possible beyond the physical (0.3" from the mid-line) limit of the WedgeA-0.6 mask, and the reach of the image saturation region with WedgeA-1.0. This can be seen, for example, in the visit 14 imaging of the HD 181327 debris ring (Fig. 3, panel C) where wedge-edge brightness gradients saturate the regions immediately exterior to the wedge edges a few pixels beyond the r = 0.3" physical limit of the wedge. In the visit 15 imaging (Fig. 3, panel D) exactly the opposite parity (but with similar reach) is seen – indicative of a relative displacement of the target and template stars w.r.t. the mid-line of the wedge. Finally, in this regard, the visit 16 imaging (Fig. 3, panel E) had no perceptible difference in target/template star relative decentering. Serendipitously, the visit 16 image as previously noted, was also least subject to breathing effects, so in combination this represents a near best-case for PSF-subtraction performance in the face of instrumentally induced residuals due to both breathing and decentering.

### *5.6. Chromatic Correction and Self-Referential PSF Templates*

Despite best efforts in the selection of PSF template stars with "color matching" of optical color indices from catalog photometry, chromatic residuals in PSF subtraction can, sometimes, still appear. In such cases the dominant stellocentric residual patterns manifest themselves as intensity modulated (bright and dark) circular "rings" that are readily apparent in debris systems viewed edge-on, or nearly so where there is little or no disk flux in the chromatically affected off-disk background. These annular-like chromatic residuals are invariant in the frame of the detector, as the (spacecraft) orientation angle on the sky is changed (i.e., where the residuals are revealed in azimuthal angle relative to the position angle of the disk on the sky). By using disk-masked image combinations at multiple rolls to map and characterize these residuals in median combination, an empirical chromatic correction to apply after PSF-subtraction can be obtained. We have done this in the cases of HD 32297 (§ A.3) and AU Mic (§ A.10) where we discuss this in detail. In the case of edge-on AU Mic disk, we investigated and compared this approach to a "similar" one alternatively using the off-disk portions of the disk-masked AU Mic halo (including one with the disk fully "hidden" behind the occulting wedge) as its own, composited, PSF template – thus fully eliminating any differential chromatic effects. We discuss the efficacy of this approach in light of potential concerns of "self-subtraction" of the disk if incompletely or imperfectly masked in the creation of a composite PSF template also in section § A.10.

### *5.7. WedgeA-0.6 and 1.0, and Multiple Roll Image Combination*

For each target, and separately its matched PSF template, whenever possible, we fully combine all WedgeA-0.6 and 1.0 visit-level count-rate images from all successfully executed visits. On an individual target basis, however, we reject images (or visits) that suffer performance degradation due to anomalous breathing excursions or other operational issues. Such cases are noted in the Table 3 footnotes, and discussed for individual targets in Appendix A.

With the optimization of the PSF-subtraction free parameters ($\Delta X$, $\Delta Y$, and template brightness scaling) established at the visit-level in the SIAF image orientation, all WedgeA-0.6 and 1.0 images are translated to a common stellocentric origin based upon target image "X marks the spot" centroids, and rotated to a common celestial ("north up") orientation (based upon down-linked engineering telemetry derived from the *HST* pointing control system). Image translation and rotation is accomplished by remapping the original images onto an output pixel grid via bi-cubic pixel interpolation apodized by a sinc-function of kernel width commensurate with the ~60 mas FWHM of the STIS/50CCD PSF (to prevent "ringing" with less than critical sampling in some pixels). Once all images are co-aligned and in a common celestial frame and PSF-subtracted, image-specific digital masks were created ("manually" crafted for each PSF-subtracted image using the IDP3 software "mask builder") to individually exclude "bad" pixels from each image that were: (a) obscured by the STIS coronagraphic wedges A or B, (b) corrupted by OTA diffraction spikes, (c) saturated close to the star (typically as planned in WedgeA-1.0 imaging, or due to non-ideal target centering), or (d) suspect in calibratability at or very near the physical edge of occulting wedge. (For an idealized illustration, of (a) and (b) only using only a single wedge and optimal differential rolls, see, *c.f.* Fig. 2 of Debes, Perrin and Schneider, 2013). These artifact-masked images are then median combined into an Analysis

Quality (AQ) data image comprised of either just the WedgeA-0.6 or WedgeA-1.0 input images, or both utilizing all unrejected data.

In final AQ image combination using both Wedge A 0.6 and 1.0 data, we typically truncate the radial extent of the WedgeA-0.6 images at a (target dependent) radius of ~45 – 60 pixels (r ≈ 2.5″ – 3″) beyond which the short exposure time WedgeA-0.6 images typically become photon starved and would contribute only noise in comparison to 10-20 x more deeply exposed WedgeA-1.0 images. With less complete spatial coverage, we also separately create "epoch 1" and "epoch 2" AQ images to enable comparative analysis for temporal variations between the two epochs within the disk (e.g., see MP Mus, § A.11) or for identification of background objects with non-common proper motions.

In addition to these AQ images themselves, for analysis purposes we also propagate and keep in ancillary images for every pixel: (a) the number of images contributing to every output pixel, (b) the standard deviation about the masked-median median, and (with the matched data masks) (c) the total exposure time in each pixel contributing to the final AQ images.

### 5.8. *Mitigation of Differential Residuals from Quasi-Static Wavefront Errors*

In this paper, all images presented and results discussed are derived from the AQ images processed as described in §§ 5.1 - 5.7. In some cases, some additional improvement can be realized by the identification and further rejection of still-remaining PSF-subtraction residuals arising from identifiable quasi-static wavefront errors (i.e., from "breathing" of the *HST* optical telescope assembly). A Multi-roll Residual Removal Routine (MRRR) for doing so is presented, discussed, and demonstrated in detail by Stark et al. (2014). MRRR is enabled by the multi-roll correlation of the imprinted quasi-static PSF-residual patterns in the instrument (detector) frame oriented images. The applicability and efficacy of the MRRR approach is dependent upon the target scene and the roll-angle overlapping of unmasked field coverage at each roll. MRRR is best for disks where such imprinted patterns are independently detectable with significance in the relative absence of disk-flux in the instrument-frame images (before rotating to a common celestial orientation). In such images while the disk-flux is re-oriented on the detector in the images acquired at different spacecraft rolls, the quasi-static PSF-structure remains rotationally invariant. In the context of the well-suited observations of the HD 181327 debris ring, following the application of MRRR, Stark et al, *ibid*, found a reduction in the photometric measurement uncertainties along its debris ring of ~ 5%.

## 6. STIS MULTI-ROLL PSFTFC OBSERVATIONS - COMPARATIVE CONTEXT

Herein we comparatively summarize some of the key (and unique) attributes of the STIS multi-roll PSFTSC observations in the context of prior *HST*, ground-based, and thermal IR observations of the CS disks observed in the GO 12228 sample.

### *6.1. Comparison with HST/NICMOS Reprocessed Archival Data*

Recent advances in post-processing techniques such as: the application of Locally Optimized Combinations of Images (LOCI); adaptive filtering via principal value decomposition; and recombination, e.g., with the Karhunen-Loeve Eigenimage Projection (KLIP) method (e.g., see Soummer, Puey & Larkin 2012), have demonstrated the ability to achieve incremental improvements in image quality through the partial identification and rejection of PSF-subtraction residuals even over more aggressive approaches in "classical" (global) PSF-subtraction such as have been enabled by the NICMOS Legacy Archive PSF Library (Schneider et al. 2011). We illustrate in Fig. 5 with a comparison of the discovery image of the HD 181327 debris ring (panel A), subsequent image processing improvements for the NICMOS data with increasingly aggressive post-processing methods, (panels B-D) and our STIS six-roll PSFTSC image (panel E). In the STIS image, for the first time, we clearly resolve: (1) a crisp inner edge to the debris ring, (2) demonstrate its difference from the diffuse outer edge, (3) detect scattered light interior to the ring, and (4) measure brightness asymmetries around the ring that are not preserved in the LOCI or KLIP re-processed data; see § A.9. Local optimization methods enhance the visibility of the ring, however, they: (a) do not conserve low-SB flux density in spatially extended regions, and (b) require a large, observationally diverse, "library" of template coronagraphic PSFs that simply do not yet exist for *HST* optical images.

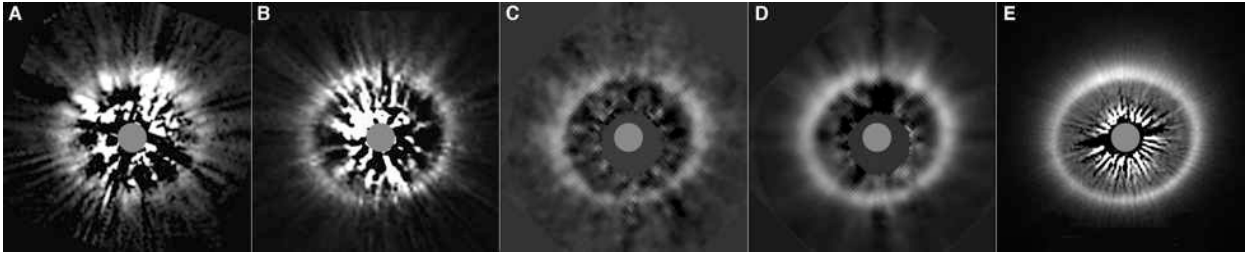

**Figure 5.** Comparison of PSF-subtraction methods revealing the HD 181327 debris ring in *HST* coronagraphic images with (A–D) derived from the same raw data (NICMOS 1.1 μm imagery) and (E) STIS 6 roll (6R) contemporaneous observationally matched-PSF template subtracted coronagraphy (PSFTSC). A: NICMOS discovery image using two (of ten) non-contemporaneously observed PSF template stars (Schneider et al, 2006; *HST* GO program 10177). B: "LAPLACE" (*HST* AR program 11279) re-processing and globally optimized re-reduction with PSF-matching from a down-selected 53 template ensemble (Schneider et al., 2010). C: LOCI re-processing (without regularization) with a 232 LAPLACE recalibrated PSF template library. D: KLIP re-processing (35 coefficients) using same PSF template library as (C) with regularization. C and D from Soummer, Pueyo, and Larkin 2012 (*HST* AR program 12652). E: STIS 6R-PSFTSC (result from this program discussed in this paper). Gray circle indicates the location and size of the NICMOS r = 0.3" coronagraphic circular obscuration.

## 6.2. Comparison with HST/ACS Observations

While the large angular extent of the ACS coronagraphic masks (r = 0.9" and r = 1.8") preclude CS observations at small IWA's the instrument (unlike NICMOS) does provide a coronagraphic FOV comparable to STIS. This has been used advantageously for (angularly) large CS debris systems, but is significantly less efficient than STIS. For example, the full extent of the HD 181327 debris system (unseen with NICMOS) is revealed with STIS six-roll PSFTCS and compared in Fig. 6, over the full dynamic range of imaging sensitivity, to a discovery epoch PSF-subtracted ACS image at very similar central wavelength (from Schneider et al 2006). Nebulosity in the STIS image is traced to stellocentric distances of 9.5" with more complete sampling about the star, better image fidelity, and higher sensitivity to low surface-brightness light-scattering material in the outermost, photon-limited, portions of the HD 181327 debris system. The STIS instrument's near full-throughput pupil (compared to ~ 50% for ACS in its coronagraphic mode), and unfiltered spectral sensitivity ($\Delta\lambda/\lambda$ = 75%, compared to 25% for ACS/F606W) together provide an ≈ 6x gain in exposure depth per unit integration time. For these STIS observations, with the additional investment in exposure time of a factor ≈ 4.5x over the ACS images, an improvement in exposure depth by a factor of ≈ x27 and photon-limited S/N of ≈ x5.2 by was realized in the outer-disk photon-limited regime (as closely predicted by the STSDAS/SYNPHOT synthetic photometry instrument models). STIS, however, having only an unfiltered coronagraphic mode provides no diagnostic "color" information as with ACS multi-band imaging.

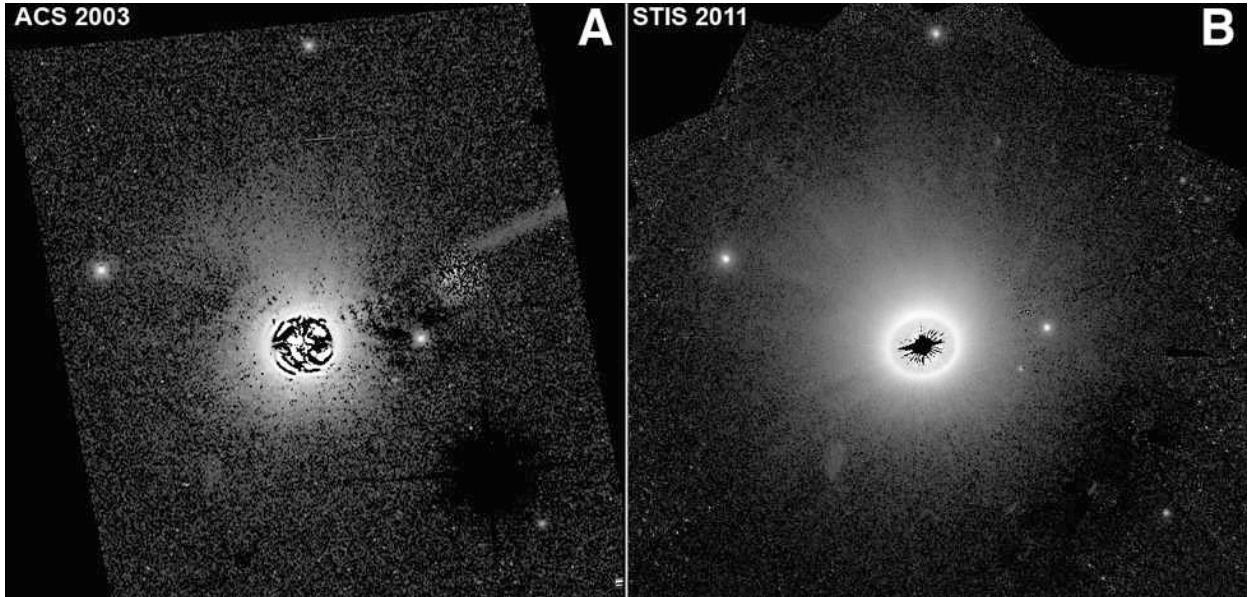

**Figure 6.** The smallest of the ACS coronagraphic masks (used in the HD 181327 scattered-light disk co-discovery imaging, panel A) provides r ≤ 0.9" IWA obscuration. However, ACS PSF-subtraction residuals completely dominate the disk light interior to the radius of the peak SB of the ring (r = 1.7") revealed by NICMOS PSFTSC (see Fig. 1B) using various methods of different aggressiveness. The residuals seen in the ACS image are largely suppressed to much smaller stellocentric angles with STIS 6R/PSFTSC beyond the effective 6-roll combined, WedgeA-0.6 limiting, $IWA_{effective}$ of r=0.3". Here, deeper (and better sampled) imaging of the outer portions of the disk with STIS 6R/PSFTSC (panel B) simultaneously provides high-fidelity, high S/N imaging of the outer portions low SB of the disk, the bright debris ring, and the largely cleared region in the ring interior. Both images: full FOV = 25" x 25", north up, east left.

The improvement in image quality and gains in S/N with STIS multi-roll PSFTFC, enable higher precision measurements of the spatial and SB distributions of the CS materials in, and possibly escaping from, these debris system with more comprehensive, and better observationally constrained, analysis and modeling than possible with the original discovery images alone. For example, in the above case of HD 181327, differences now observable in scattering phase functions in the ring and in the outer halo further suggests the presence of an unseen, planetary-mass perturber that may lie exterior to the ring. Additionally, these improved data have resulted in the putative discovery of a debris "spiral", or spray of ejecta, from a posited recent massive collision in the exo-Kuiper Belt of HD 181327 debris system. Both of these findings are discussed in detail in Stark et al. (2014), with the observational data presented for this, and the other disks in this sample, herein. With these data, we are now at the point of directly imaging phenomena postulated as important in the evolution of our Solar System, bringing the analysis of exoplanetary systems with *HST* into a new regime.

### *6.3. Comparison with Ground-based Observations*

Ground-based coronagraphs on large, AO-augmented, telescopes are currently capable of revealing the brightest features of CS debris systems that possess strong local contrast gradients (rather than diffuse, extended flux). This is well exampled with the HD 61005 debris system. The HD 61005 ("the Moth") debris system was first imaged with *HST*/NICMOS (Hines et al. 2007), and its brightest features with strong SB gradients have now been imaged from the ground. The NICMOS discovery image showed a bright, but diffuse, skirt of material apparently "blown" off an inner disk viewed nearly edge-on. The skirt of escaping material could be traced 6" from the star, with a brightness (inferred surface density) enhancement posited from a bow-shock on the "front" edge of the debris system in the direction of its relative motion running into an ISM-wind. ACS optical coronagraphic polarimetry better revealed the outer portions of the debris disk, the fan of material, and the "limb-brightened" edge of the blow-out fan. Maness et al. (2009) found the optical-to-nIR disk color blue, and the polarization properties of the outer parts of the disk and skirt were different; in particular, with the skirt dominated by small, compact, light-scattering particles. Subsequently, Buenzli et al. (2010) imaged the brightest (lowest contrast) features of the HD 61005 debris system using VLT/NaCo with ADI and LOCI image processing, resolving the disk as a highly inclined debris ring, and measured its apparent eccentricity, and thus inclination, as $84.3°± 1.0°$ and suggested a 2.75"±0.85" AU photocentric offset. A crisp inner edge to the ring was inferred, but the skirt of blow-out material was not observed. Our STIS 6-roll PSFTSC imagery (see § A.5): fully discloses the skirt and the forward (higher optical-depth) limb-brightened bow-shock; resolves both the "forward" and "back" sides of the debris ring (similar on its bright side to that seen in Buenzli et al. 2010); the ring inner clearing; recovers the photocentric offset; and reveals, for the first time, that the disk is a three component system. In addition to the ring-like disk and fan, there is an additional disk component of material beyond the launch point for the fan, which is interpreted an analog to the scattered Kuiper Belt (Hines et al. 2014, in preparation). We also image interior to the ring, finding a largely cleared zone down to 10 AU from the star. The debris ring, as seen in the STIS imagery, is coincident with the planetesimal belt resolved at sub-mm wavelengths (Ricarte et al. 2013).

### *6.4. "Comparison" with Thermal IR Excess as a Predictor of Optical Brightness*

IR excess emission above stellar photospheric levels is the signature of re-radiating CS dust attributable to orbiting exoplanetary debris for the majority of systems. Thus, over the past 16 years, *HST* coronagraphic imaging surveys designed to discover optical/near-IR light-scattering counterparts to (mostly) thermally emissive debris disks have targeted nearby stars with the brightest 12-100 μm thermal IR excesses. These candidate systems were identified from space-based surveys (e.g., ISO, IRAS, and *Spitzer* and *Herschel* photometry and/or spectral energy distributions). In nearly all cases, only IR excess sources with $L_{IR}/L_{star} \geq 10^{-4}$, as a presumed order of magnitude proxy to *a priori* unknown optical/near-IR scattering fractions ($F_{disk}/F_{star}$), were targeted with anticipation of high survey yields. This was a "conservative" selection criterion given then anticipated (but initially not well quantified) contrast-limited coronagraphic imaging sensitivity to disk-scattered starlight with *HST*'s coronagraphically augmented instruments. Collectively, however, (only) ~ 15% of candidate debris systems surveyed were found to possess CS dust with high enough SB to be detected with *HST* coronagraphy as conducted in a diversity disk-finding survey programs. Only two-dozen spatially-resolved starlight-scattering debris disks emerged from 1998 – 2013; most recently with a few previously observed, initially undetected, but newly discovered with advanced re-processing techniques. This relatively low success rate underscores that thermal IR excess *alone* is not a good predictor of the fraction of starlight scattered by a CS debris disk into the observers line-of-sight (e.g., see Table 1 and Fig 7.). IR excess data also provide no insight into how that total flux is actually spatially distributed (Booth et al. 2013) and thus visible in the face of instrumental sensitivities and systematics; i.e., the scattered-light component of most disks with $L_{IR}/L_{star} \geq 10^{-4}$ remained elusive.

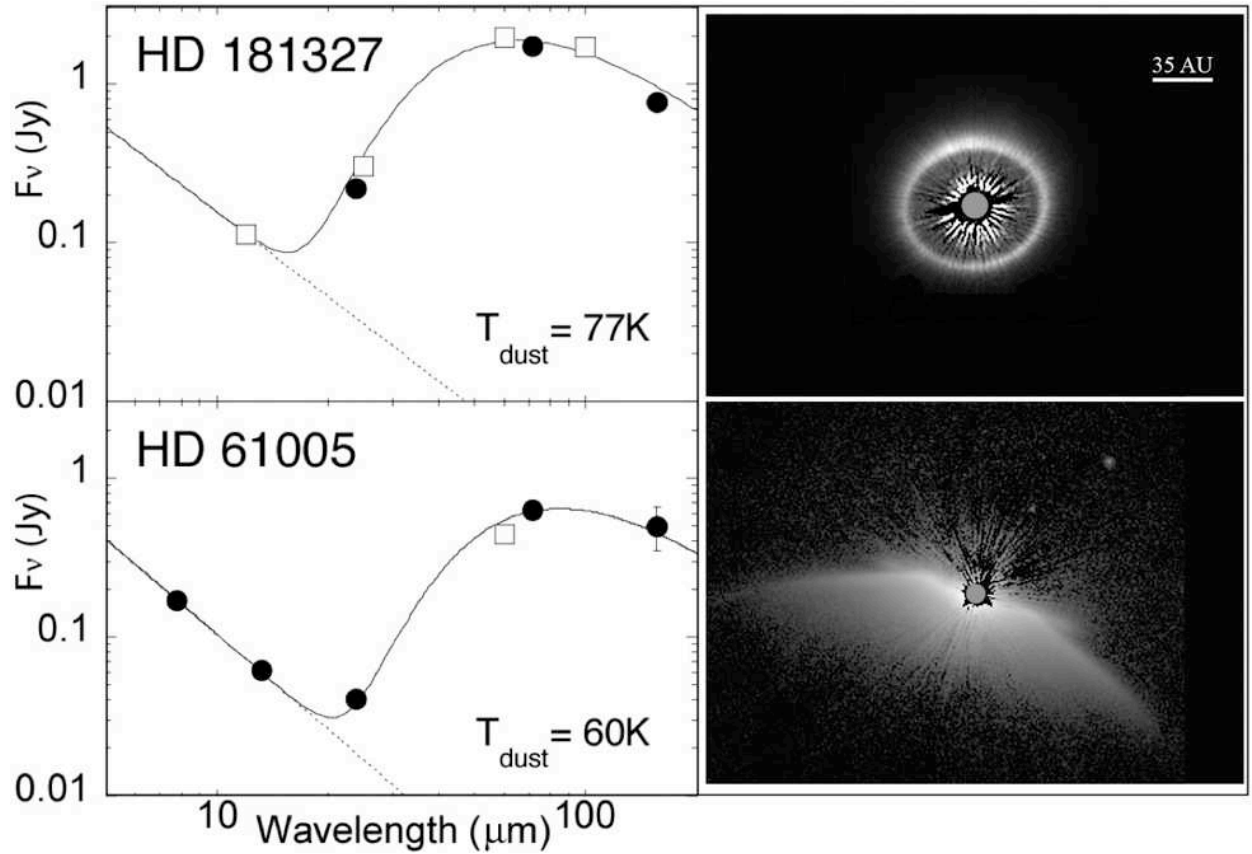

**Figure 7.** Photometric measures of thermal IR excesses (fit to single temperature blackbodies) from the HD 181327 (top) and HD 61005 (bottom) debris systems exhibit close similarity in their SEDs (left) though very different morphologies are revealed with STIS 6R/PSFTSC scattered-light imaging (right).

For detectable disks, a loose correlation between IR excess and optical/near-IR scattering fraction exists (e.g., see Fig. 8 for the debris systems observed in this sample). This is particularly the case for systems with ages $\leq$ 100 Myr, and excluding specific systems, like AU Mic, where radiation pressure blowout of small dust particles is negligible. On an individual target basis, with additional dependences, $L_{IR}/L_{star}$ is only a weak predictor of SB and detectability in spatially resolvable light-scattering disks. Grain properties (optical constants), the dust density distribution, the viewing geometry of any disk, and systemic age, can lower disk visibility (contrast) globally, or for sub-structures within, in a complex manner unpredictable without a foreknowledge of those systemic parameters. Additionally, CS debris systems with very similar spectral energy distributions can have vastly different systemic morphologies, geometries, and dust density distributions uniquely informed by spatially resolved imaging.

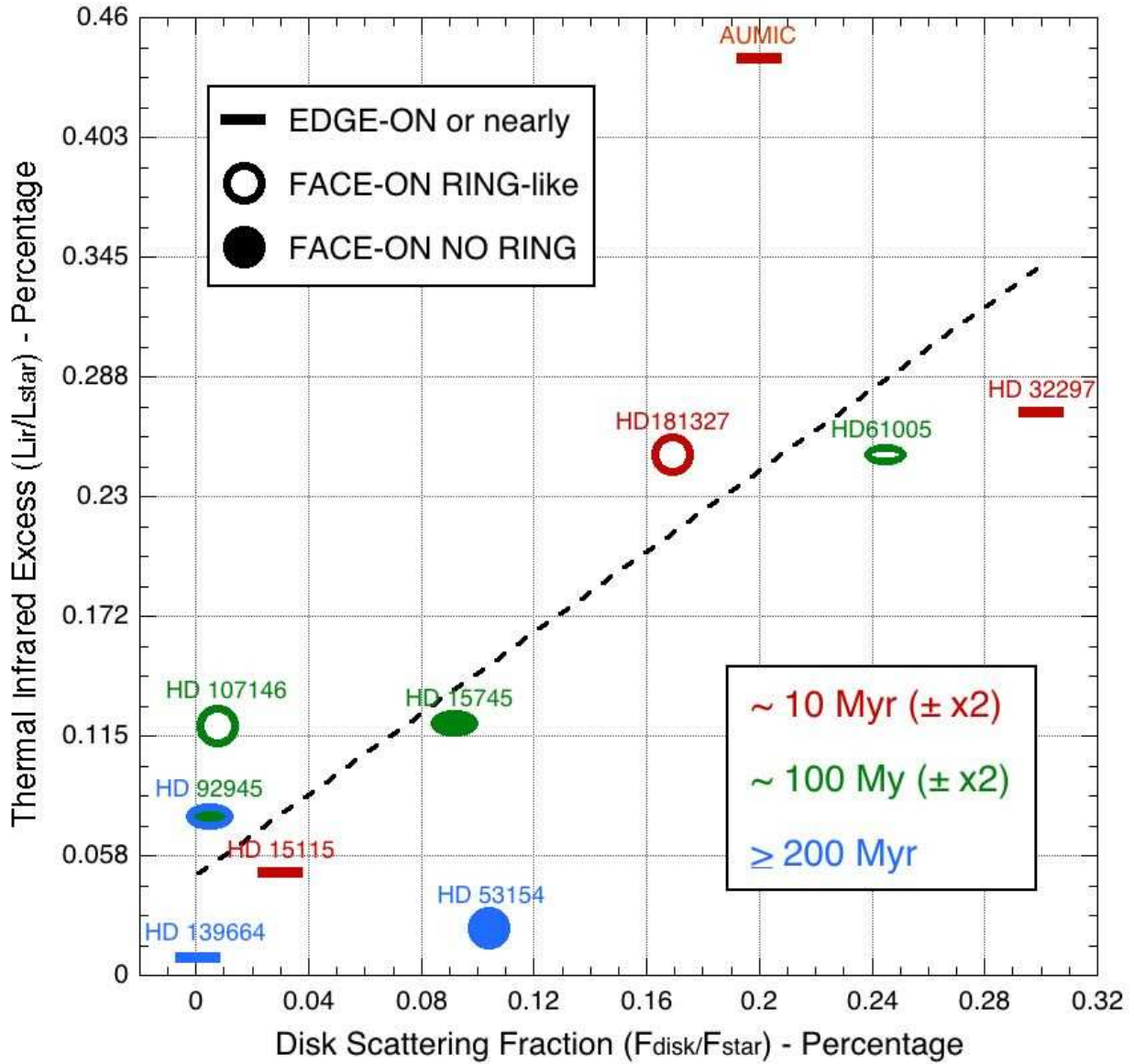

**Figure 8.** Overall, debris disk scattering fractions and thermal excesses are cross-correlated with age, but with multi-parametric dependencies, $L_{ir}/L_{star}$ may be only a weak predictor of optical brightness for individual systems (e.g., HD 15115 in the GO 12228 sample).

## 7. Observational Results

The GO 12228 targets constitute a follow-up sample of previously discovered light-scattering CS debris systems and thus excludes "non-detections". A gallery of disk images (discussed in detail in Appendix A) that highlight their morphological diversities is presented in Fig. 9. The ability to discern with clarity sub-structures within these disks, depends not only upon global or large spatial scale disk-to-star contrasts, but also on localized disk signal gradients in the presence of any remaining PSF-subtraction artifacts. I.e., lower contrast substructures (e.g., rings, gap edges, and perceptively edge-on mid-plane regions), are typically more readily detectible (with sufficient integration time) than diffuse, extended features. The latter remain elusive in ground-based images reliant on aggressively advanced locally optimized PSF subtraction/calibration methods such as LOCI, KLIP, and (observationally) ADI that do not preserve disk flux on spatial scales larger than the dominant spatial frequencies of the detectable, lower contrast and spatially localized, disk features. Simply, these methods, though of utility for disk imaging, do not preserve low SB extended flux and typically render such features undetectable. Conversely, the gallery of GO 12228 debris disk images presented in Fig. 9, derived from STIS multi-roll PSFTSC, are immune to these post-processing effects. These images may be used directly not only to characterize and quantify disk, and disk grain, properties but as "truth" (within their own levels of efficacy and uncertainties) in assessing independent results from ground-based coronagraphs with advanced processing techniques that intrinsically spatially filter disk images. The disk images in Fig. 9 are presented at differing angular scales due to the differences in target distances with our goal of representing well the debris system morphologies both individually and comparatively. Each disk, and its observationally derived characterizing parameters, is discussed in more detail in Appendix A.

The primary morphological, geometric, and photometric characteristics of the disks, derived from these images, are summarized in Table 5. Disk inclinations (i), except for those very close to edge or face on, and major-axis celestial position angles (PA), are estimated photometrically from stellocentric elliptical isophote fitting. Inner working distances (IWD) are the closest stellocentric distances to which the CS brightness could be measured. The radial extent of the debris systems with large halos are expressed to where the disk flux declines to +3 $\sigma$ resel$^{-1}$ of the background far from the disk; Inner (i) and outer (o) radii are separately estimated for disks with concentric components. Total (disk integrated) flux density measurements ($F_{disk}$) and scattering fractions ($F_{disk}/F_{star}$) at the STIS 50CCD pivot wavelength of 0.57 μm fully enclose the disks, but exclude the small areas unsampled at and interior to the IWD (e.g., see Fig. 41). For $F_{star}$ we assume catalog V-band magnitudes, as given in Table 1, transformed into the STIS 50CCD spectral band (Table 4). Note that $F_{disk}/F_{star}$ is simply an observationally derived parameter, along the line-of-sight to the observer. For moderately to highly inclined systems, the

IWA could conceal a significant fraction of the potentially (otherwise) observable scattered flux, in particular if the dust is highly forward scattering.

**Table 5**
Disk Morphology, Geometry, Extent, Brightness

| Target | Morphology | Est. i° | PA° | IWD AU | Inner Clearing? | Radial Extent AU | Brightness Asymmetries | $F_{disk}$ mJy | $F_{disk}/F_{star}$ percent |
|---|---|---|---|---|---|---|---|---|---|
| HD 15115 | Nearly Edge-On* One side bifurcated ("The Needle") | 80* | 99.1 ±0.6 | 18 | Obscured by geometry ?? | E:320 W:570 | E/W Extent E. side bifurcation W. side above-plane | 2.04 | 0.030% ± 0.004% |
| HD 15745 | Featureless (?), bi-lateral asymmetry | 52 | 22.5 ±1.5 | 19 | None to IWD limit = 19 AU | i: 255 o: 430 | Along Minor Axis | 3.26 | ~0.092% |
| HD 32297 | Edge-On Bow Concavity | near edge on | 47.5 ±1.0 | 34 | Obscured by geometry | ~ 1560 asymmetric | ⊥ to disk plane. Disk SB asymmetry w.r.t. minor axis. | 5.98 | 0.30% ± 0.03% |
| HD 53154 | symmetric inner, outer asymmetric faint "arcs" | < 5 | -- | 5.5 | None to IWD limit = 5.5 AU | i: 37 o: 70 | No @ < 3" > 4" NW gap/arc | 6.92 | ~ 0.104% |
| HD 61005 | Low SB Skirt ("The Moth") | 85 ± ~ 1 | 70.6 ±1.2 | 14 | Yes. Inner edge @ ~ 45 AU | ~ 330 | ⊥ to disk plane Ring SB asymmetry w.r.t. minor axis. | 4.46 | 0.245% ± 0.03% |
| HD 92945 | Broad Halo + Narrow inner ring | 28 | 100 ±2 | 6.4 | Yes. Inside inner of inner ring | ~ 118 ring: 59 | Inner ring better seen on east side Outer W brighter | 0.143 | ~0.0051% |
| HD 107146 | Nearly Face-On Broad Ring | 18.5 ± ~ 2 | 38 ±3.1 | 11 | Yes. Shallow edge @ ~ 60 AU | ~ 220 symmetric | H-G scattering phase angle only | 0.404 | 0.0077% ± 0.0004% |
| HD 139664 | Near Edge-On | ~ 90 | 75.5 ±0.5 | 20 | Not observable | ~ 100 | E/W Peak ~ x 1.6 (S/N limited?) | 0.260 | ~0.0005% |
| HD 181327 | Inclined Narrow Ring + Diffuse Outer Halo | 30.1 ± 1.2 | 102 ±4 | 18 | Yes. Sharp inner edge @ ~ 80 AU | ~ 460 asymmetric | Non H-G azimuthal. Inner/outer skew. 25% ansal ΔSB | 7.81 | 0.17% ± 0.015% |
| AU Mic | Edge-On with one-side "bump" and warp | 90 | 37.8 ±0.2 | 5 | Obscured by Geometry | ~ 130 symmetric | Out-of-plane. Warp. Sub-structures. | 2.51 | 0.20% ± 0.02% |
| MP Mus | Asymmetric, "Featureless" | 27.3 ±3.3 | 10 ±2 | 30 | None to IWD limit = 30 AU | Disk: 155 Halo: > 340 | H-G. Front/back Temporal variability | 1.74 | ~ 0.68% |

\* inner "half-ring" → ~ 80°, extended un-bifurcated disk → edge-on; see Fig. 12.

In Fig. 10 we plot the disk-to-star signal contrast as a function of stellocentric angular distance along the disk major axis where the disks are brightest (or nearly so) in CS azimuthal angle on both sides of the star. Fig. 10 is presented in dimensionless contrast units (to facilitate comparison and scaling in other instrumental systems), per the definition of image contrast in §4. To the extent the CS light-scattering material may be spectrally neutral across the STIS passband (that cannot be ascertained from these data alone, but in general deviations will be only small), these contrast curves may be closely transformed to photometric SBs at the 50CCD pivot wavelength of 0.57 μm by multiplying by 0.178 mJy arcsec$^{-2}$.

For most disks, their SBs decline radially from IWAs along the major axis of $\geq 0.3"$ to the outermost disk regions ~ 10" (or where the noise level declines to ~ 3 σ resel$^{-1}$ over the sky background) by typically ~ 4 dex. At common stellocentric distances, the diversity in debris disk SBs amongst the sample varies at the extrema from the highest to lowest contrast disks by ~ 3 dex. The highest contrast disks, e.g. HD 139664 and HD 92945, prove a contrast challenge to STIS PSFTSC at the smallest stellocentric distances and in some cases, in particular with < 6 rolls, a mask-limited IWA$_{minimum}$ = 0.3" is not reached.

Fig. 10 is of particular utility to those considering future coronagraph designs, and disk imaging programs, where instrumental performance is a prime consideration, hence it is cast in stellocentric angular distance, although each of the debris systems are at different physical distances from the Earth. Such curves may be applied directly to instrument performance models to assess disk visibilities and to derive integration time requirements for fainter disks with scalable S/N. Table 1 gives the GO 12228 target distances in parsecs, to readily allow transformation of stellocentric angle into astronomical units for each of the disks in Fig. 10.

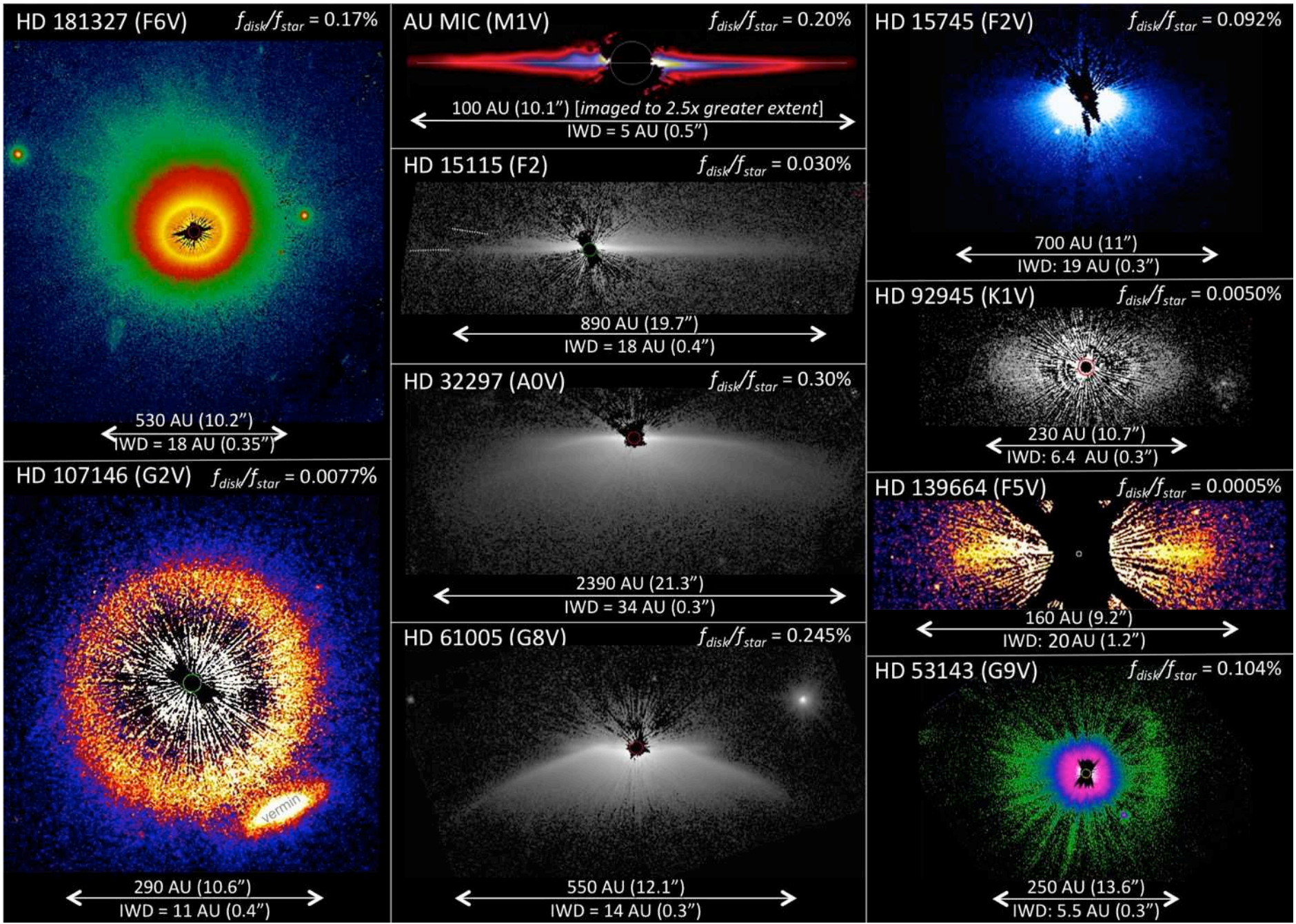

**Figure 9.** Analysis Quality scattered-light images of the GO 12228 debris disks discussed in Appendix A. Arrows indicate the full physical and angular extent of the disks (except AU Mic) in AU and arcseconds (scaled differently for each disk), and below the inner working distances realized (though for all disks not at all azimuth angles) with PSFTSC imaging.

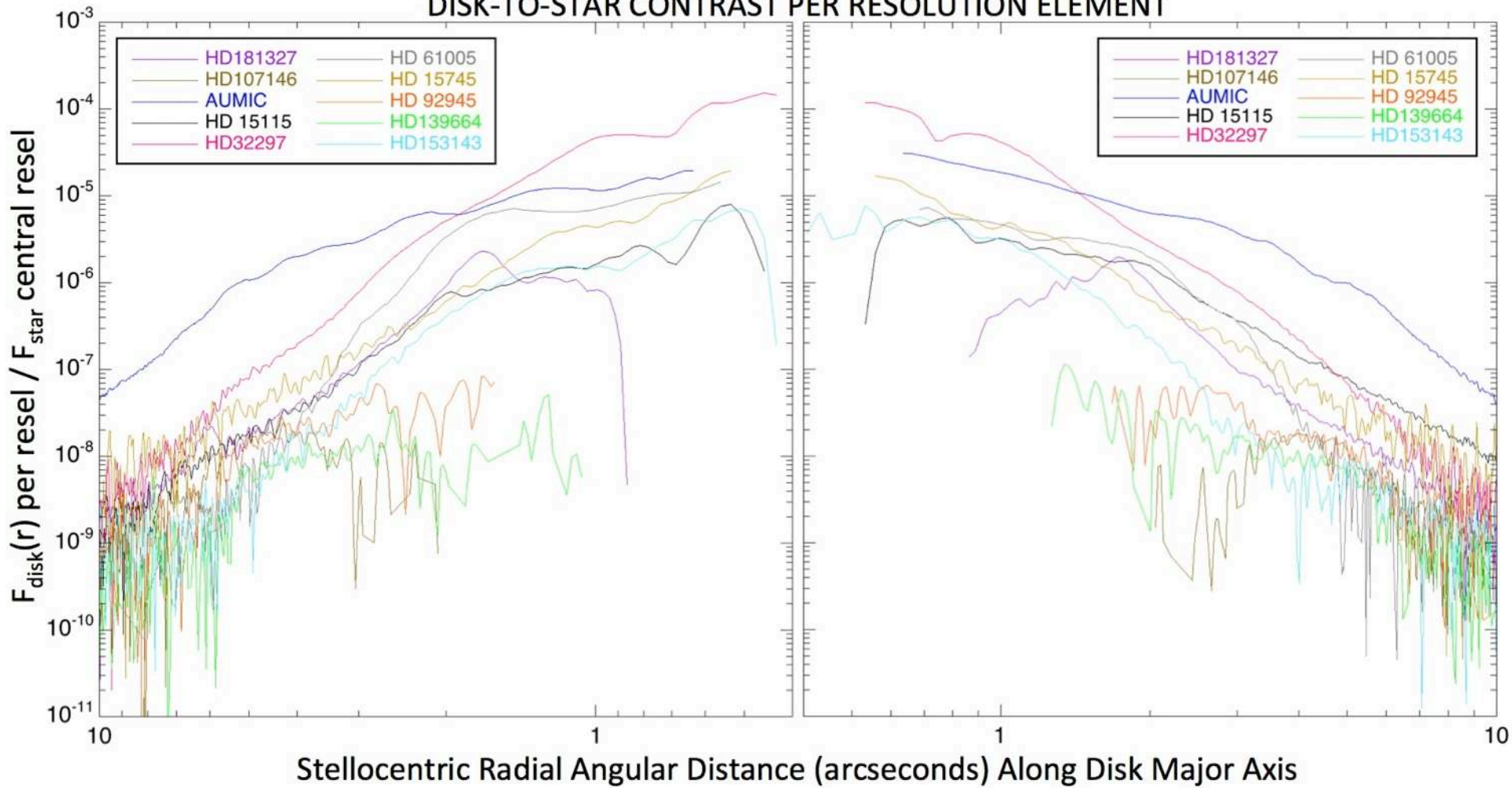

**Figure 10.** Radial SBs for the GO 12228 debris disks measured along the disk minor axes as a function of stellocentric distance expressed as resel$^{-1}$ image contrasts, derived from AQ images shown in Fig. 9. In detail, the minor axis may not be coincident with the smallest, asymmetric IWA realized, so not in all cases the smallest stellocentric distance from the star observed.

# 8. DISCUSSION

## 8.1. Placing HST Optical Coronagraphic Imagery in Context

High-fidelity ground-based imaging of CS material, in general, remains a difficult challenge even as the newest generation of "extreme" adaptive-optics augmented coronagraphic imagers are coming on line. Historically, most ground-based CS disk imaging has relied on coronagraphy with polarimetric differential techniques (Hinkley et al. 2009), where light from the disk is simultaneously imaged in the orthogonal (ordinary and extraordinary) linear polarization components, with the unpolarized light of the star used as the PSF to achieve contrast augmentation; e.g., see Perrin et al. (2014a) for a comprehensive review. Images obtained in this mode can reveal the "polarized intensity" of the CS material imaged, i.e., the "p*i" product of the polarization fraction and "total intensity", a useful diagnostic, but from which these fundamental quantities are inseparable.

Because of the nearly ideal properties (brightness, contrast, size, morphology) of the HR 4796A debris ring for high contrast imaging, it has become somewhat of a *de facto* "truth test" target for ground-based AO systems (both "extreme" and less aggressive) that can be compared against either the HST/NICMOS nIR discovery imaging (Schneider et. al 1999) or the subsequent HST/STIS optical follow-up imaging (Schneider et al. 2009). E.g., this debris ring has been spatially resolved from the ground with the Subaru HIgh Contrast Instrument for Adaptive Optics (HICIAO; Thalmann et al. 2011), the Very Large Telescope Naysmith Adaptive Optics System (NAOS) Near-Infrared Imager and Spectrograph (CONICA) coronagraph (NaCo; Lagrange et al. 2012), and Gemini South Near-Infrared Coronagraphic Imager (NICI; Wahhaj et al. 2014), none extreme AO systems, and also very recently with the Clay Magellan AO (MagAO; Rodigas et al. 2014b, in preparation) instrument, with varying levels of fidelity.

A recent demonstration observation by GPI produced both total light and polarized intensity images of the exceptionally bright ($F_{disk}/F_{star}$ = 1.6x10$^{-3}$) HR 4796A CS debris ring (Perrin et al. 2014b, in preparation) fully resolving the apparently (p*i) brighter side of the ring beyond the instrument's IWA at high (but uncalibrated) contrast and spatial resolution[6]. These may be qualitatively compared with another first-light total intensity HR 4796A debris ring image from SPHERE[7].

---
[6] http://www.gemini.edu/node/12113

Ground-based "total intensity" imaging of CS debris disks, today, however, generally requires observing, and/or post-processing methods, such as ADI, or LOCI/KLIP for contrast enhancement that do not preserve the flux (polarized, or in technically feasible cases, total intensity) in spatially extended regions and with shallow contrast gradients. Thus, the highest fidelity images with recoverable total-light photometry remains, currently, in the domain of space-based coronagraphy, and in particular the highest efficacy yet demonstrated with STIS multi-roll PSFTSC.

With *HST*, PSF-subtraction residuals from chromatism and wavefront error instability (breathing) effects can be minimized with judicious PSF template star selection, and observation scheduling and design allowing the "effective" IWA to approach the physical limit of the occulting mask, i.e., r = 0.3" for STIS WedgeA-0.6. Thus, resolving angularly small inner debris rings, centrally cleared regions in CS disks, and constraining their inclinations may be done also with photometric robustness in total light and without need for advanced processing techniques that do not spatially conserve the target flux. In comparison to techniques such as principal component analysis, our variant on classical (global) PSF subtraction does not require an existing *a priori* geometrical model for the debris system. This is likely to be particularly valuable for systems that have only been resolved at longer wavelengths or those whose debris disks are inferred only from IR excesses; e.g., the many newly-identified debris systems discovered from WISE observations (Patel et al. 2014), or those which are spatially resolved by *Herschel*, but for which optical imagery does not exist.

Imaging the outer, faint and diffuse, regions of CS disks, where small particles may be escaping the system or interacting with the ISM is also difficult (if not problematic in most cases) from the ground where Angular Differential Imaging (ADI) facilitates observations. ADI data are obtained as suites of exposures where the rotation of the system on the sky enables the high-contrast detection of sources. ADI relies on azimuthal median filtering where the angular size of the source is less then the system rotation in individual images. Exposure times are limited by the need to avoid rotational smearing, which increases in tangential scale at larger stellocentric distances. ADI is very effective for point sources, i.e., exoplanets in close angular proximity to their host stars (~ 10s to 100s AU in physical scale in these disks). However, ADI-assisted detection of CS debris suffers for systems lacking small spatial scale sub-structures and without locally strong brightness gradients (e.g., face-on disks and those with large, diffuse, outer disk halos), and in particular for stellocentric regions beyond about 2". These issues do not arise with *HST*. With *HST*, "maximum" exposure times are set by Earth-occultation interruptions, or splitting long integrations for cosmic-ray detection and correction, and are on the order of a ks or longer. Total integration times for 6-roll observations are typically >10 ks and thus with sufficient S/N in the outer disks to map faint outer halos, tracing either loosely bound particles, or those which are being actively ejected from the system.

---

[7] http://www.eso.org/public/news/eso1417/

### *8.2. All Debris Systems are NOT the Same*

The imagery presented in this paper shows systems with debris rings of different widths, morphological and SB asymmetries, sub-structures, broader debris disks, and debris systems with blow-out structures that do not imprint conspicuous signatures in their SEDs, in sub-millimeter observations, or in FIR thermal emission.

Detailed dynamical and compositional modeling of these and augmenting multi-wavelength data, beyond the intended scope of this paper, is required to fully exploit the new images obtained in this study. The detection of radial differences in the scattering phase functions in inner debris rings and outer disk halos, with (or without) pericenter offsets of debris belts from their host stars, offer new constraints to such models. Modeling efforts incorporating these data are still in their early phases, but have already yielded indirect signatures suggesting the presence of wide exoplanets in HD 181327 (Stark et al. 2014) and HD 61005 (Hines et al. 2014). The detection of spiral-structured debris from a posited major collision originating in a well-revealed exo-Kuiper Belt 86 AU from HD 181327 (Stark et al., *ibid*) challenges models for when larger bodies are present in such structures, and highlights debris disks as highly dynamical in nature. These data further suggest that environmental interactions of the outer portions of debris disks with the ISM also appear to be common (Debes et al. 2014, in preparation).

Posited implications for other debris system features and sub-structures, revealed by the images we have obtained, are discussed on a target-by-target basis in Appendix A. Therein we demonstrate the analytic and interpretive potential of such high-fidelity optical coronagraphic imaging of CS material uniquely enabled with STIS multiroll PSFTSC. These images highlight the fact that debris disks are highly dynamic and complex environments. The picture that emerges, even from this small sample of previously imaged debris disks, is one of system-to-system diversity, perhaps matching the diversity of exoplanets and exoplanetary systems known from RV and transit studies. Fig. 9 clearly illustrates the diversity in debris system global morphologies and architectures (separate from those expected from simple line-of-sight inclination effects) as well as asymmetries and sub-structures. We comment on several below.

*(a) Ring-Like Disks:* Ring-like features in CS debris systems are readily, and unambiguously, observable in CS disks with favorable (intermediate to face-on) viewing geometries when they exist, such as in HD 92945 (§ A.6), HD 107146 (§ A.7), and HD 181327 (§ A.9). The improved inner working angle of our data compared to previous *HST* observations, has resulted in imaging rings also in the nearly edge-on systems HD 61005 (§ A.5) and (partially detected in) HD 15115 (§ A.1). Even in our small sample of systems, the ring architectures are diverse and range from: broad and low SB in HD 107146; a narrow ring with an outer diffuse and very low SB component in HD 92945; an extremely bright ring with a extended and highly asymmetric outer component in HD 61005; a narrow and very bright ring seen in HD 181327; and an (apparently) extreme "front/back" brightness asymmetry in HD 15115. These five disks are also extremely diverse in scattering fractions from $F_{disk} / F_{star}$ = 0.25% for HD 61005 to 0.005% for HD 92945, though these two disks in particular have rings of approximately the same physical size (r ~ 60 AU) and are of similar estimated ages (~ 100 – 200 My). Conversely the broad ring of sun-like star HD 107146, though with a similar spectral type host star as HD 61005, is ~ 2x as large. Despite their specific differences, the prevalence of such structures of sufficient brightness to be imaged (seen in other debris disk as well, e.g., HR 4796A and Fomalhaut), suggest they are common features in debris systems.

*(b) Highly Asymmetric Disks*: The HD 61005 debris ring was not resolved in the lower spatial resolution scattered-light discovery imaging in the near-IR with NICMOS (Hines et al. 2007), due to its nearly edge-on inclination (nor with ACS due to its large IWA). Those observations, however, revealed a large "skirt" of likely small-grain material (earning it the name of "the Moth" by its morphology), attributed to material escaping the system posited by ISM ram pressure ejection (Hines et. al 2007; Mannes et al 2009). This suggested that prior modeling and interpretation of debris disk dynamics and evolution, as isolated systems, was deficient.

Environmental interactions could be important and must considered, e.g.: chemical evolution of the disk material by ISM "pollution"; compositional evolution and differentiation by stripping of small particles from the disk; and thus mass loss (disk erosion) over time by systemically extrinsic (as well as prior considered intrinsic) forces. The GO 12228 AQ images better revealed the HD 61005 debris system structures, but also three other examples of disks, two at an nearly edge on viewing geometry, HD 15115 and HD 32297 (§ A.3), and the prior described "fan-like" disk of HD 15745 (§ A.2), where ISM/environmental interactions may be responsible for their morphologies. Low S/N detection of diffuse material also beyond the outer edges of the HR 4796 A (Thalmann et al. 2011), and Fomalhaut (Kalas et al. 2013) debris rings, suggest that such structures may be common, and potentially important in understanding debris disk evolution and dynamics.

*(c) Out-of-Plane features/sub-structures in edge-on Disks*: Disks viewed edge-on (or nearly so) while (at least partially) self-obscuring the in-plane inner disk regions, fortuitously provide a most favorable geometry for readily studying "out of plane" CS material without ambiguities. The prototypical β Pictoris edge-on debris system sports a strong "warp" (Heap et al. 2000), or inclined (out-of-plane) secondary disk (Golimowski et al. 2006; Apai et al. 2014). Its origin is posited as possibly arising from a forcing interaction with its interior, imaged, co-orbiting giant planet, β Pic b and/or yet undetected additional planets in the system. In AU Mic (§ A.10), we detect an out-of-plane "bump" (SB and presumed dust density enhancement) on just one side of the disk at ~ 13 AU, but with an asymmetric warp about the disk minor axis that may be difficult to explain in the absence of one or more co-orbiting planets. HD 15115 is viewed nearly-edge on (but not quite so by a few degrees). While no such features are seen in the inner disk, the outer disk on one side only, appears intriguingly bifurcated at large stellocentric distance whose origin remains (at this time) only speculative.

### *8.3. Novel Structures and Behaviors in Older Protoplanetary Disks*

While small-particle material originating in debris disks can be seen ejected to large stellocentric distances (with some likely escaping from some systems), distant material associated with protoplanetary disks informs on the evolutionary history, and particularly the star-forming environment, of the host-star system. Large, remnant proto-stellar envelopes have previously been observed at stellar ages of 4 Myr (AB Aur, Grady et al. 1999) and at 10 Myr (HD 100546, Grady et al. 2001), in association with intermediate-mass stars. The scale of these structures suggests system formation in isolation, and without a history of dynamical perturbations, and again highlights structures inconspicuous at longer wavelengths, and that do not leave a footprint in the disk IR excess. The data for MP Mus (§ A.11) obtained in this survey provide not only a superb image of the disk of this system ready for detailed modeling that can be compared to co-eval systems in the TW Hya association, but also provides evidence for a remnant envelope, indicating that at least some T Tauri stars form in isolation as profound as some of the nearest Herbig Ae stars. Our observation of MP Mus also demonstrates the utility of the observational approach for studying spatially resolved temporal variability in disk-surface illumination.

### 8.4 Symbiosis with Other Wavelength/Instrumental System Data

Our STIS six-roll PSFTSC images directly reveal disk features and substructures unseen in precursor *HST* discovery data, and components that had been inferred from SED modeling, but not previously observed directly. These visible-light observations provide imagery with spatial resolution comparable to that which ALMA should soon provide, and an order of magnitude better than delivered by the *Herschel Space Observatory*. Optical (*HST*) and sub-mm (ALMA) imaging at similar spatial resolutions are highly symbiotic, with *HST* tracing the small grain structures in the debris systems, and ALMA tracing the population of larger particles (see Boley et al. 2012 and Moor et al. 2013), and E-VLA sensitive to cm size material.

Extreme A/O systems are now coming on-line for ground-based 6.5 – 8 m class telescopes (e.g., SCExAO, MagAO, GPI, SPHERE), and offer higher ground-based Strehl ratio imagery than the previous generation of AO systems, though with more restricted fields of view than offered by *HST*/STIS. They none-the-less offer symbiotic capabilities for imaging moderate contrast features in the bright regions of angularly small disks, enhanced when limited to polarized flux, in particular in the nIR where Strehl ratios and imaging performance is improved (and with spatial resolution, e.g., in K-band, comparable to STIS in the optical and thus providing, in combination, significant color diagnostics).

In addition to the remainder of the approximately dozen *HST*-discovery images of debris disks not followed up in this study, including a few recently found in NICMOS LAPLACE (Schneider et al. 2011) reprocessed data with KLIP (Elodie et al. 2014), there are now known debris disks at < 50 pc that are resolved at long wavelength by the *Herschel Space Observatory*. E.g., HD 139664 (reported in this paper with STIS), HD 127821 and HD 13337 all $L_{IR}/L_{star} \sim 10^{-4}$ (Su et al. 2013), HD 207129 (Löhne et al. 2012) and Eta Crv (Duchêne et al. 2014). These *Herschel*-resolved debris disks are typically too large in angular extent to image with extreme AO systems (and in some cases too large for complete imaging by the *James Web Space Telescope* (*JWST*)), but can be fully imaged with *HST*/STIS. Some of these newly-discovered ring-like disks show IR pericenter glow in *Herschel* PACS imagery, but the low (order arcsecond) angular resolution of the *Herschel* data preclude a quantitative measurement of pericenter offsets that are classic signatures of exo-planets on eccentric orbits. Optical coronagraphy with *HST*/STIS, with spatial outer working angle coverage of the full exo-Kuiper belt regions of such disks, provides an order of magnitude improvement in angular resolution, compared to *Herschel* data. This also enables measurements of pericenter offsets < 3 AU at d=27.5 pc, and of outer belts inclinations to ± ~ 1° that are an order of magnitude improved over those achievable from the thermal IR imagery (Greaves et al. 2014). The availability of such inclination constraints, in systems with known exoplanets with orbital data, would result in more robust estimates of the fraction of exoplanets with orbits that are close to coplanar with their host star Kuiper belts, and the fraction that have been perturbed into higher inclination orbits (Dawson & Dodson-Robinson 2011). Moreover, the frequency of pericenter offsets and their quantitative measures allow us to evaluate similar data for eccentric planets (see Beust et al. 2014 for application to Fomalhaut). High-quality inclination measures for systems further allow selection of disks to search for transiting exo-comets (Roberge et al. 2014).

## *8.5. Conclusion*

High-fidelity scattered-light images of circumstellar disks are needed to: (a) constrain the modeling of individual systems; (b) provide reference data for the interpretation of observations obtained in thermal emission; or (c) directly reveal material which is inconspicuous at longer wavelengths (Lebreton et al. 2012). In some cases, the discovery images of the disks that were targeted in this study were sufficient to establish the presence of CS material, or to constrain models of the IR excess. However, the discovery images were typically not suited for dynamical modeling of individual systems, or for identifying indirect signatures of planet presence: such studies require both the image fidelity and S/N provided by our multi-roll observing strategy. Such data enables us, for the first time, to obtain holistic views of exoplanetary systems that have yet to be achieved for our own Solar System.

The combined image quality, photometric efficacy, and inner to large outer working angle depth (sensitivity) offered by *HST*/STIS multi-roll PSFTFC for CS disk imaging is unique, and currently problematic with any other existing instrumental systems. We argue, that as a foundational basis and legacy for both future multi-wavelength (for example in combination with ALMA), and follow-on in-development or conceived high-performance ground, or space-based disk-imaging coronagraphic facilities (e.g., *JWST*, NASA's *Astrophysically Focused Telescope Assets*, the *Exoplanetary Circumstellar Environments and Disk Explorer* mission concept), new observations of as many disks as feasible, should be carried out while *HST* and STIS are still available.

## ACKNOWLEDGEMENTS

The HST/GO 12228 investigation team (co-authors of this paper) dedicate this paper to the memory of our late colleague and co-investigator, Bruce E. Woodgate (1939 - 2014) whose influence and insights in advancing astronomical instrumentation and astrophysics has touched so many. As Principal Investigator and architect of Space Telescope Imaging Spectrograph, our investigation, and so many others would not have been possible without his vision and foresight. He will be sorely missed by us all.

This study is based on observations made with the NASA/ESA Hubble Space Telescope, obtained at the Space Telescope Science Institute, which is operated by the Association of Universities for Research in Astronomy, Inc., under NASA contract NAS 5-26555. These observations are associated with program # 12228. Support for program # 12228 was provided by NASA through a grant from the Space Telescope Science Institute, which is operated by the Association of Universities for Research in Astronomy, Inc., under NASA contract NAS 5-26555. JC was supported by the Research Corporation for Science Advancement through a

Cottrell College Science Award. CCS acknowledges the support of a Carnegie Fellowship and an appointment to the NASA Postdoctoral Program at NASA Goddard Space Flight Center, administered by Oak Ridge Associated Universities through a contract with NASA.

# APPENDIX A: INDIVIDUAL OBJECTS

## A.1 – HD 15115

*Introductory Notes.* A starlight scattering debris disk with an "extreme" asymmetry about its F2 main sequence star, a posited member of the ~ 12 Myr old β Pic moving group, was discovered by Kalas et al. (2007) from ACS F606W coronagraphic imaging data obtained in GO program 10896 (reproduced in Fig. 11A). Though contrast-limited by PSF-subtraction residuals at stellocentric angular distances ≤ 2.5", the ACS imagery revealed the outer disk's edge-on geometry, and highly asymmetric extent to ~ 315 AU on the east side of the star, and ~ 550 AU on the west side. Follow-up J, H, and K' adaptive optics imaging of the disk with the NIRC2 instrument on the Keck-II telescope confirmed the existence of the disk asymmetry beyond ~ 2.5″, but suggested a more symmetric structure from 0.7" – 2.5". With combined *HST* and Keck photometry in commonly sampled stellocentric regions, Kalas et al. (2007), suggested the optical-to-NIR color as blue, with small dust particles postulated similar to those presumed to be in the AU Mic disk. The morphology and V-H color suggested an informal designation of the HD 15115 disk as the "blue needle" by Kalas, Graham & Fitzgerald (2007).

With subsequent *HST*/NICMOS F110W (1.1 μm) coronagraphic imaging, Debes et al. (2008) found, in combination with the prior V and H band photometry, asymmetric radial color gradients in the disk, with the redder dust particles closer to the star, and suggested a "warp" on the western side of the disk (only). Kalas et al. (2007) earlier posited the extreme asymmetry might originate from external dynamical perturbations to the debris dust from HIP 12545, a nearby co-moving M-star member also, likely, a coeval member of the β Pic moving group. Debes et al. (2008) alternatively suggested that a recent collision in the disk might explain the asymmetry, color gradients and the suggested "warp".

More recently, Rodigas et al. (2012) obtained diffraction-limited imaging of the inner (r < 2.5") portions of the HD 15115 disk with the PICES instrument on the Large Binocular Telescope (LBT) in Ks band from 1.0" – 2.5", and LBTI/LMIRcam adaptive optics imaging at L' from 1.2" – 2.1". They found the disk more symmetrical at the longer wavelength, and a spectrally neutral Ks – L' color in the ~ 2" region on both sides of the disk. In combination with the prior shorter-wavelength derived colors, they suggest the scattered light on the west side originates from a dust particle population with smaller grains, suggesting the disk dynamics are perturbed by a local ISM wind. Rodigas et al. (2012), also suggest a "bow like" appearance of the inner disk in both Ks and L' bands offset by a few AU to the north of the star.

*Observations and PSF Subtraction.* The STIS 6R/PSFTFC observations of HD 15115 and its PSF template star executed with no anomalies or degradations of significance with reduction and template PSF subtractions. At most celestial position angles (inclusive of the direction along the disk major axis), an inner working angle of r = 0.4" was achieved (green circle in Fig. 11 C). The relative orientation constraints due to *HST* guide star availabilities were a bit more restrictive than typical, so spatial coverage between 0.4" and 0.8" due to multi-roll combined WedgeA obscuration in the sectors flanking approximately celestial PA 30° degrees (and affecting he visibility of the disk minor axis) is lacking. This is the "hard black" region of "no data" in the center of the image in panels B and C, in this stellocentric radial zone.

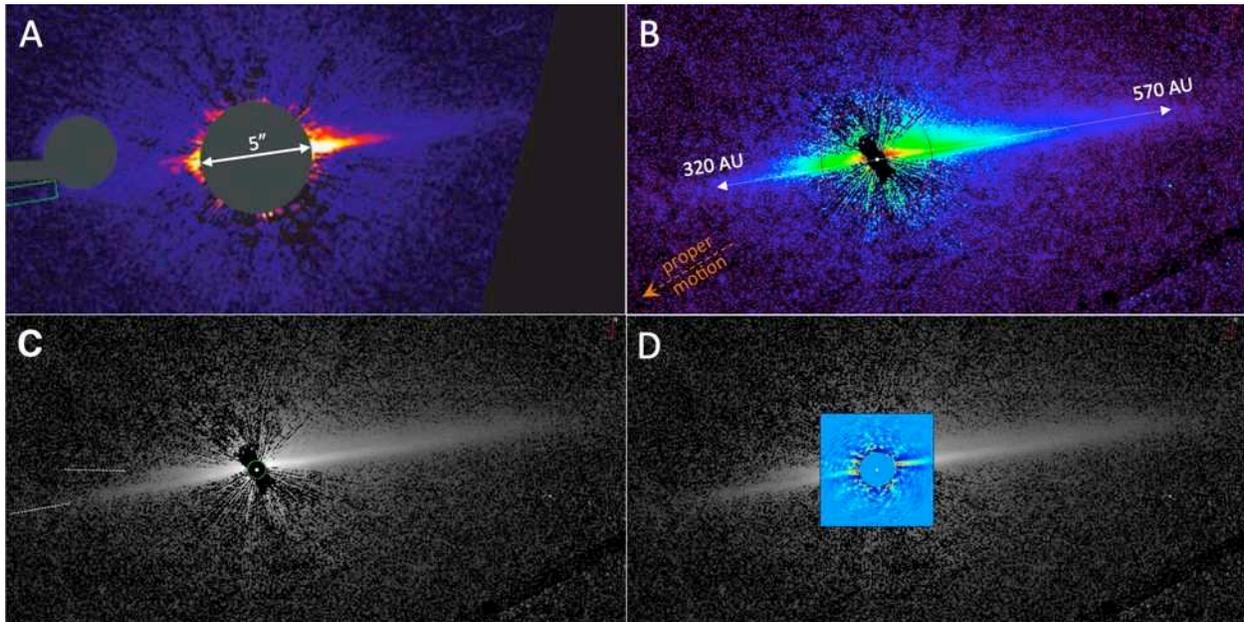

**Figure 11.** A: HD 15115 scattered-light ACS discovery image from reproduced from Kalas et al. 2007. B. STIS 6R/PSFTSC image, $\log_{10}$ display with x20,000 imaging dynamic range from [-3.0] to [+1.0] cts s$^{-1}$ pixel$^{-1}$ {dex} (0.177 to 3532 μJy arcsec$^{-2}$). The radial extent of the disk along the mid-plane to sensitivity-limited stellocentric distances at 3-σ resel$^{-1}$ over the background are indicated. Note the preponderance of starlight-scattering material preferentially "above" the disk mid-plane to the NW of the star (only). The region interior to the r = 2.5" red circle was unimaged with ACS. C: Same data as panel B shown in grayscale with a $\log_{10}$ stretch from [-3.5] to [+1.0] cts s$^{-1}$ pixel$^{-1}$ {dex} (0.056 to 3532 μJy arcsec$^{-2}$). The green circle indicates the r = 0.4" IWA along the disk mid-plane. The mid-plane diverging, bifurcated, linear light-scattering structures on the east side of the disk are indicated. D: Disk mid-plane from ~ 1.0" to 2.5" (inset FOV 5" x 5"; linear display stretch from -1 to + 2.5 mJy arcsec$^{-2}$) resolved

by Rodigas et al. 2012 at Ks band with LBT/PICES overlaid on the STIS image. All images are north up, east left.

*Principal Results*. The STIS 6R/PSTFSC AQ image of the nearly edge-on HD 15115 CS disk has (within a factor of a few) similar photon-limited sensitivity to the dust in the outer portions of the debris system as was revealed by ACS, and is seen to a similar angular extent (see Fig. 11B). The inner disk, however, revealed at similar optical wavelengths enabling spatially resolved image interpretation and photometry between 0.4" and 2.5" (red circle in panel B), is not accessible in the ACS image (panel A). The primary geometrical and photometric results derived from the AQ image are as follows. The total flux density from the disk at r ≥ 0.4" ( ≥ 18 AU projected on the sky) is ≈2.04 mJy. The 0.58 μm disk scattering fraction ($F_{disk}$ / $F_{star}$) @ r ≥ 0.4" is 0.030% (approximately an order of magnitude less than HD 32297 (see § A.3) and HD 61005 (discussed in context in §A.5). Light scattering material in the plane of the disk is seen to at least 320 AU to the east, and 570 AU to the west, of the star (panel C). The celestial orientation of the disk major axis appears to be ~ 99.1° (E of N) as determined from photometric isophote fitting along the apparent mid-plane of the disk at r > 1.2" on both sides of the star.

The asymmetric "needle"-like morphology of the HD 15115 disk, as originally suggested by Kalas et al. (2007), is confirmed also into the inner regions unseen previously in the optical. However, the suggestion of a more azimuthally symmetric disk, as inferred from the H-band NIRC2 image, is not confirmed, and instead suggests that the symmetrical structure appearing in the ground-based Keck AO image may be an artifact from imperfect wavefront error control and starlight rejection. In addition to the E/W asymmetry in disk extent along the major axis as previously noted, three other very significant disk features are seen in the STIS AQ image:

(1) The "bowing", or concavity, of the disk along the minor axis to the north of the star suggested by the Rodigas et al. (2012) Ks band image (reproduced and astrometrically overlaid on the optical image in Fig. 11C), is seen to continue inward to the STIS optical IWA$_{effective}$ r = 0.4″ (18 AU) along the disk major axis. This suggests that we are seeing the brighter (front?) scattering surface of an inner debris ring, viewed in nearly edge-on projection, that may be at least partially cleared; See schematic suggestion of a posited inner ring component in Fig. 12 (inset). The "invisibility" of the opposing (southern) side suggests a very high Henyey-Greenstein scattering asymmetry parameter *or* dust obscuration through an incompletely cleared inner disk component. Detailed modeling will be needed to arbitrate the possibilities. A similar structure, with a more well-defined central clearing and lesser "front/back" scattering asymmetry, was revealed for the nearly-edge on debris system of HD 61005 (§ A.5).

(2) On the east side of the debris structure, the light scattered from the dust appears to be bifurcated away from the inferred disk major axis direction. This is indicated in Fig. 11C by the two dashed lines annotating these disk features. The origin of this feature is uncertain, but could be due to a line-of-site opacity enhancement of the dust in that direction. We note more well-defined bifurcated structures associated with the outer disk of HD 61005 as shown in Fig. 23C.

(3) There is a significant out-of-plane (vertical) extension in the scattered light distribution at r ~ 2.5" above (to the north of) the disk plane *only* on the west side of the star, i.e. the region in green as stretched in Fig. 11 B to the NW of the star.

All three of these asymmetries must be modeled (and reproduced) prior to ascertaining which of the prior posited asymmetry mechanisms (stellar perturbation, recent episodic collisions, or ISM interaction) may be dominantly causal, though this is beyond the scope this paper.

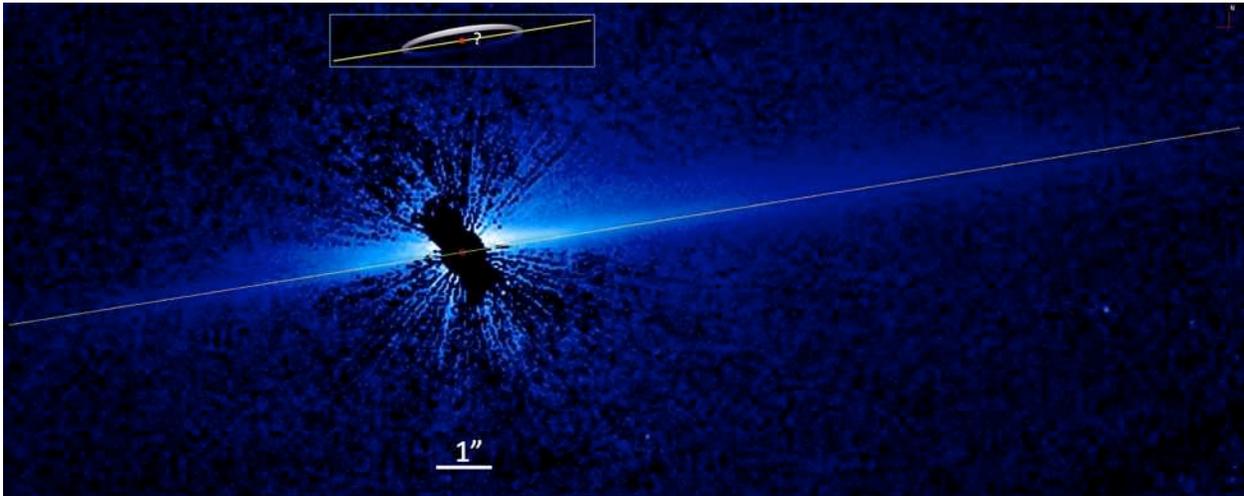

**Figure 12.** The mid-plane of the asymmetrically extended HD 15115 CS debris disk, found from fitting the locus of points defining the peak SB across the apparent disk major axis at r > 1.5" on both sides of the star, is over plotted on the disk image. The morphology, and deviation, of the flux density excess at smaller stellocentric distances is to the north of the star only. This, qualitatively, suggests a nearly edge-on inner debris ring, (schematically represented in the image inset) reminiscent of HD 61005 (see § A.6) that may be only partially cleared in the interior. The western side of the posited ring in this image, "above" the plane of the disk, is more readily seen (compare to the cartoon inset). The eastern of the debris ring may be distorted (or partially obscured) by more complex disk features, e.g., as evidenced by the bifurcated structure highlighted in Fig. 11C, arising from extrinsic forces noted in the main text. This image north up, east left with a $\log_{10}$ display stretch from [-3.5] to [+1.0] cts s$^{-1}$ pixel$^{-1}$ {dex} (0.056 to 3532 mJy arcsec$^{-2}$) and 25.4″ x 10.4″ FOV.

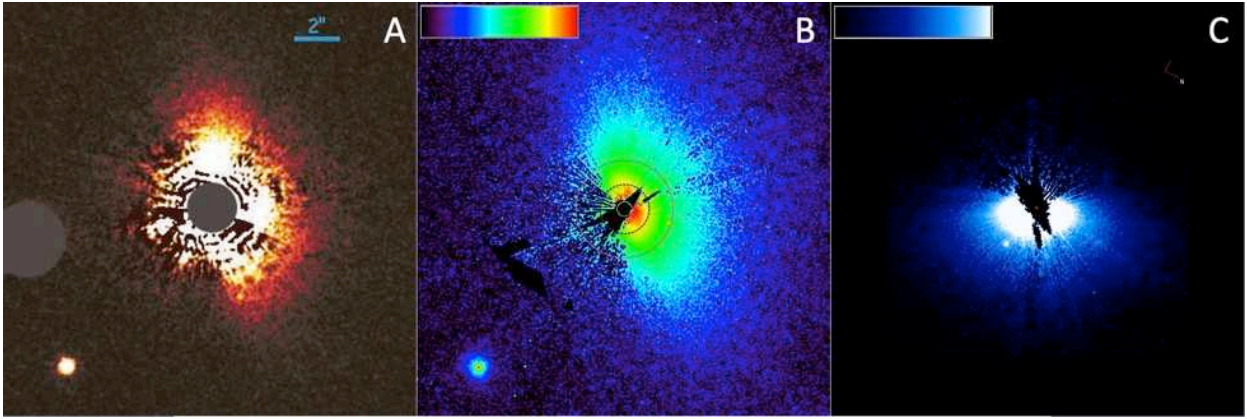

**Figure 13.** A: HD 15175 ACS discovery image (from Kalas et al. 2006; *c.f.,* Fig. 1). Central masked region r =1" but "*the presence of quasi-static speckles produces significant PSF-subtraction residuals, limiting the detection radius to ~ 2 arcseconds.*" B: STIS *three*-roll combined image (Visits B5 – B8 only) improves over ACS IWA$_{effective}$ = 2" undegraded from residuals by ~ x7 to IWA$_{effective}$ = 0.3" (19 AU projected; central yellow circle). Dotted black circle corresponds to the r = 1" digital mask imposed on ACS image in panel A. Solid gray circle corresponds to the larger r = 2" effective IWA in the ACS. Log$_{10}$ display [-2.2] to [+2.5] {dex} cps/pixel. C: Same STIS data as B, but with disk major axis rotated to image horizontal and log$_{10}$ monochrome display [-1.5] to [+1.5] {dex} cps/pixel. Field shown: 350 x 350 STIS pixels (17.75" x 17.75"). N.B.: Point-object (revealed as binary with STIS) is non-common proper motion background star superimposed on the disk, as determined with NICMOS (2004)/STIS(2012) two-epoch data.

*Introductory Notes*. HD 15745 is an ~ 100 My old F2V star for which an extended scattered light disk with a "fan"-shaped morphology was reported by Kalas et al. (2007) based on ACS coronagraphic imaging (reproduced in Fig. 13A).

*Observations and PSF Subtraction – Tortured History*. The first of the two initially planned sets of observations of HD 15745 and its PSF star executed 2011-Nov-05. The three disk-target orbits (V75, V76 & V78) executed nominally. However, the guide star acquisition for V77—the contemporaneous interleaved PSF template observation—failed and no PSF data were obtained. This full set of observations was then rescheduled as Visits B5 – B8 with a different set of guide stars on 2012-Nov-03. The second set of observations (Visits 71 – 74) executed nominally on 2012-Jan-02. However, the PSF star was then found to be an *a priori* unknown close binary with the companion sufficiently bright to render its unsuitable as a PSF template. Thus, prior to the repeat of the originally failed set of observations, six orientations of target imaging data had been obtained, but without a suitable PSF template star. The 2012-Nov-03 earlier-failed observations re-executed nominally, with a PSF template (VB7) that proved to be an excellent match for its flanking contemporaneous visits. An attempt was made to use the VB7 PSF template for both the non-contemporaneous disk imaging data obtained in Visits 71 – 74 and 75 – 78. In combination with visits B5, B6, & B8, visits 71, 72, & 74 would provide complete roll coverage. Visit 75, 76, & 78 were respectively redundant with the repeated visits B5, B6, & B8, but could in principle improve the S/N with replicated observations. Unfortunately (though reductions were attempted), the VB7 PSF was badly mis-matched in breathing phase relative to the visit 71, 72, & 74 data. Though after its application, a scattered-light disk could be seen in these data, the PSF subtracted images were riddled with artifacts not suitable for combination with the much more successful Visit (B5, B6, B8) PSF-subtracted images. PSF subtractions from the visit 75, 76, 78 data were somewhat better qualitatively, but still significantly inferior to those obtained at exactly the same roll angles in visits B5, B6, & B8, and could not be used to further improve the data in combination. Other PSF templates obtained in the GO 12228 program were tested as well, but none produced visit 7* PSF-subtracted images of sufficient quality to combine for additional roll coverage with the successful visit B* images. Hence, the fully reduced AQ data images of the HD 15745 disk—unlike the other AQ images in this program—are combined from only *three* differently oriented visits (see Table 3 for details). Nevertheless, the data quality and *spatial* coverage, though not fully complete closest to the star, are both very good and the resulting 3R/PSFTSC combined image shown in Fig. 13A/B was used to ascertain the principal metrical results discussed below.

*Principal Observational Results*. The STIS 3R/PSFTSC AQ image of the HD 15745 CS debris disk, to first order, appears have a "featureless," bi-laterally symmetric, morphology about the disk axes (see Fig. 14, left). There are no visually apparent sub-structures evident in the disk, which appears smoothly contiguous, radially and azimuthally, with no turn-over or "central clearing" of the disk radial SB to a smallest $IWA_{effective}$ of appx r = 0.3", but with a posited break in the NE side radial SB profile at r ~ 1.2" (see Fig. 14) under investigation (Debes et. al 2014). The celestial orientation of the major axis of the disk appears ~ 22.5° (E of N), with a bright-minor to semi-major axial ratio of the disk of ≈ 0.615 as determined from fitting photometric isophotes (e.g., see illustration of best-fit 1 ct s$^{-1}$ pixel$^{-1}$ SB contour in Fig. 14, left panel). For an intrinsically circularly symmetric disk, this axial ratio implies a disk inclination of ≈ 52° from face-on. The estimated center of the disk, also determined from isophotal ellipse fitting, appears offset from the position of the star by 0.258" (16.4 AU projected) along the disk minor axis toward the SW—indicative of strong directionally preferential scattering by the disk material. This significant offset and front/back scattering phase asymmetry in SB, in combination with longer wavelength data, can be modeled to better constrain the physical properties of the light-scattering debris.

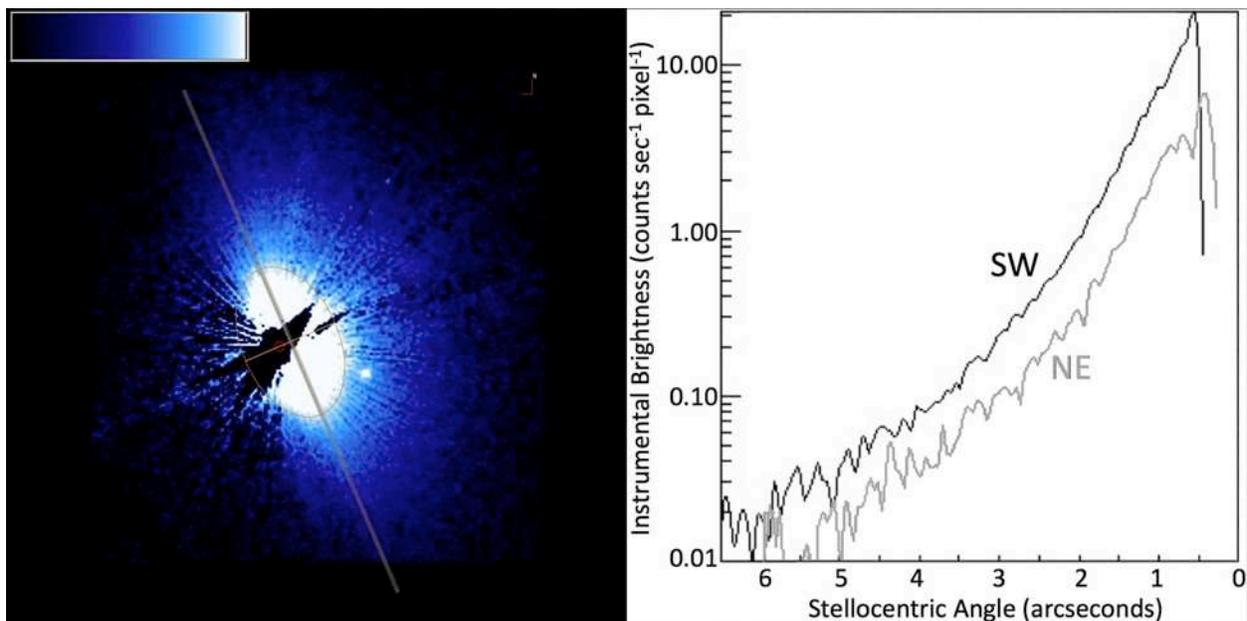

**Figure 14.** Left: Geometrical/isophote elliptical fit to the HD 15745 CS disk, in comparison to the location of the star (red circle) offset along the disk minor axis is indicative of very strong directionally preferential scattering by the disk material. The ellipse-fit is to the 0.8 cts s$^{-1}$ pixel$^{-1}$ SB isophote (1 cts s$^{-1}$ pixel$^{-1}$ = 0.177 mJy arcsec$^{-2}$), which is at r = 34.0 pixels (1.73") from the disk center along the major axis (with minor:major axial ratio 0.615). Right: Radial SB profiles along the disk major axis are of common slope, but differ throughout in brightness suggesting a higher dust density distribution on the SW side of the star.

We measured the total flux density from the HD 15745 CS disk to a stellocentric angular distance of r ≤ 10", excluding the small unsampled area close to the star (shown in black in Fig. 14A) with IWD ≤ 19 AU as projected on the sky, as $F_{disk}$ = 3.26 mJy with an uncertainty of a few percent. The 0.57 μm disk scattering fraction ($F_{disk}$ / $F_{star}$) at r ≥ 0.3" is ≈ 0.092%. Ninety percent of the dust-scattered starlight is contained within r < 3.55" (255 AU projected) of the star. Diffuse material is seen with a SB 3 σ pixel$^{-2}$ brighter than the background noise far from the disk to a distance of 6.8" along the disk major axis.

With the geometry of the light scattering disk established from the AQ image, we measured the disk major axis SB profile on both sides of the star (Fig. 14, right panel). Measurements were made in single pixel increments, in a 0.1" (2 pixels ≈ 1 resel) wide photometric aperture, ± ≈ 6" in length *w.r.t.* the disk minor axis as illustrated in Fig. 14 (left panel). The SW and NE side SB profiles have identical slopes. At all stellocentric angular distances, however, the SW side of the disk along the minor axis is brighter at the NE side by factors ≈ 3x at r = 0.6" to ~ 2x at r = 6.0". This global (large-scale) "scalable" brightness asymmetry suggests, possibly, a higher surface density of scattering grains on the SW side of the disk—perhaps from a recent collision, or stirred by planet(s) though other morphological signatures of potentially co-orbiting planets are not apparent.

The spatially unresolved point-object that appears superimposed upon the disk was undetected in the ACS discovery image due to the presence of strong PSF subtraction residuals in that region. The object was found to be a non-comoving background star based on differential proper motion measures using epoch 2004 NICMOS imaging (GO 10177) in which the inner most portion of the disk (only) was only marginally detected.

### A.3 – HD 32297

*Introductory Notes.* The light-scattering debris disk about the 112 pc distant, ~10 My old, A0 main sequence star HD 32297 was discovered from relatively shallow depth, 2-roll PSF-subtracted 1.1 μm coronagraphy with *HST*/NICMOS (Schneider et al. 2005). The NICMOS image revealed a nearly edge-on disk with a linear morphology extending to a sensitivity limited stellocentric distance of r = 3.3", and was traced inward to the NICMOS r = 0.3" IWA. The disk was noted as asymmetrically brighter to the SE along the disk major axis at r ~ 0.6" (50 AU), closely coincident with a peak in the millimeter brightness later reported by Mannes et al. (2008). Schneider et al. (2005) noted a minor-axis mirrored asymmetric break in the radial 1.1 μm SB profile along the plane of the disk ~180 AU from the star, seen only on the NE side of the disk. A subsequent re-examination of the NICMOS image revealed a slight concavity of the disk with the ansae of the major axis slightly deflected to the NW as shown in Fig. 15A. Kalas et al. (2005) obtained a ground-based R-band image of a much larger nebular structure enveloping the *HST* resolved debris disk with the U. Hawaii 2.2 m telescope and r = 3.25" coronagraphic mask, but with a larger IWA of r ~ 5". I.e., the ground and *HST* images explored mutually exclusive domains in stellocentric distance. The asymmetric morphology and SB maps of the wide-field ground-based, and narrow-angle *HST* images, however, by extrapolation were strongly spatially correlated (Fig. 15B), and suggested an unseen extended debris structure linking the inner disk seen by NICMOS with the enveloping nebulosity seen from the ground. Kalas et al. (2005) suggested that the swept-back morphology of the high SB regions in the nebulosity, deviating in celestial PA from the linear extension of the *HST*-imaged inner disk, was due to a "sandblasting" interaction of the disk dust particles by the ISM or from perturbations from two nearby stars. The millimeter appearance of this system is discussed in Maness et al. (2008).

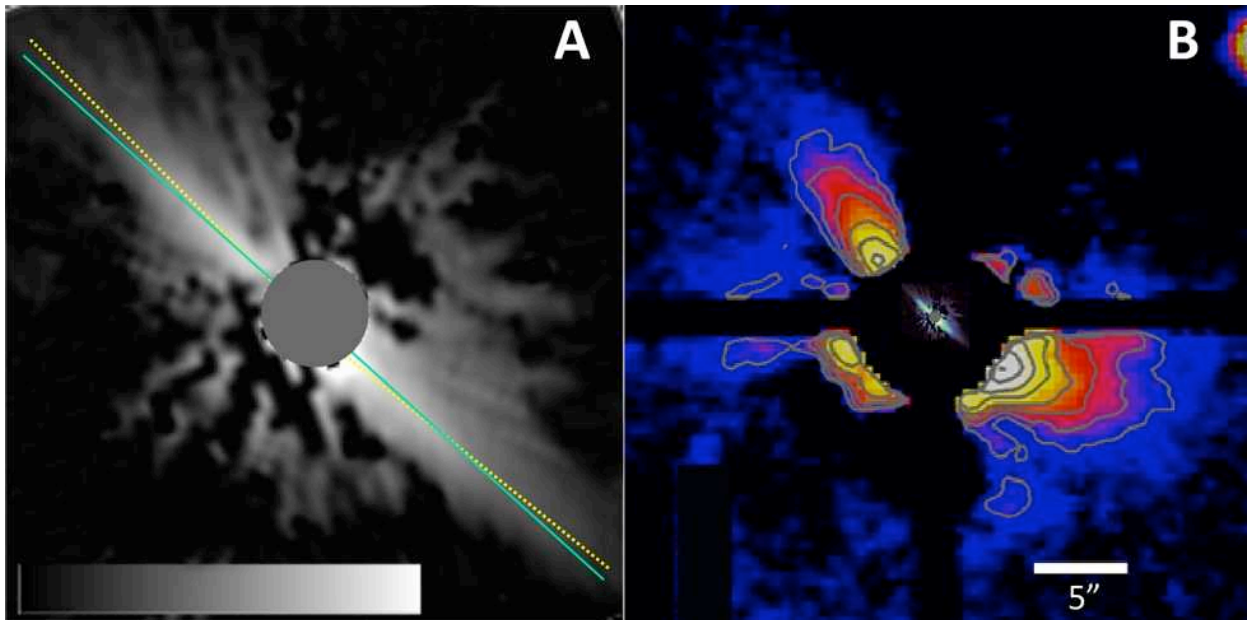

**Figure 15.** A: NICMOS image of the r ≤ 3.3" region of the HD 32297 disk (replicated from Schneider et al. 2005; gray circle overlays the r = 0.3" coronagraphic obscuration) separately fitting image isophotes on opposite sides of the star (yellow dotted lines) indicates a departure from a linear fit to both sides of the disk (green line). B: NICMOS image astrometrically inset with the surrounding large-scale asymmetric nebulosity seen with wider-field ground-based imaging (r ≥ 5", reproduced from Kalas et al. 2005), suggested a then unseen contiguous debris structure undergoing possibly extrinsic perturbations. In both panels residual diffraction spikes in the NICMOS image, not co-incident with the edge-on disk, have been digitally masked.

Follow-on *HST*/NICMOS imaging by Debes et al. (2009) at 1.6 μm (H-band) and 2.05 μm, providing color constraints on the disk dust particles, found both color and morphological asymmetries (including warps) that supported the conjecture by Kalas et al. (2005) of the HD 32297 system encountering a dense ISM gas cloud. Subsequent ground-based AO imaging confirmed several aspects of the *HST* observations. Ks-band imaging with the PALAO system and a phase mask coronagraph, detected the diametrically opposed "lobes" of the disk, affirmed the r ~ 50 AU E/W brightness asymmetry, and provided additional color constraints on the CS grains (Mawet et al. 2009). Later AO imaging with the Keck 2 telescope using NIRC2 with LOCI processing techniques, better informed on the inner disk at Ks band in the 0.3" ≤ r ≤ 2.5" explored at shorter wavelengths with *HST*/NICMOS (Currie at al 2012). Boccaletti et al. (2012) used VLT/NaCo and constrain the disk inclination to 88°, and detected an inner, cleared cavity at H and Ks. Donaldson et al. (2013) modeled the SED, including new *Herschel* Space Observatory data, and found that the best fit to the stellar properties, including *HST*/STIS UV spectra is for an A7V star ($T_{eff}$=7750 K, $A_V$=0 with L=5.6 $L_\odot$). They fit the SED with two dust belts, one a ring centered near 110 AU and an inner dust belt at 1.1 AU with an unconstrained outer edge. Their best-fit model for the outer disk was for a minimum particle size of 2.1 μm, and a 1:2:3 ratio of silicates, carbonaceous material, and water ice with 90% porosity, although Rodigas et al. (2014) argue that this particle mixture may not be consistent with L' imagery of the disk.

*Observations and PSF Subtraction - Details*. Due to a proposal implementation error, the contemporaneous PSF template star scheduled with our HD 32297 orbits was B-V redder, Δ(B-V) = +0.19, than tolerable for unbiased PSF-subtraction. Fortuitously, off-rolled PSF self-subtraction, using HD 32297 observed at different celestial roll angles as its own PSF template, revealed that the out-of-plane scattered light excess from the disk itself was asymmetrical originating with significance from nearly only one "half" of the disk, to the NW of the disk major axis (see Fig. 16). While the flux from the disk in such a self-subtraction may be partially entangled, and not fully photometrically preserved, this suggested an alternate strategy not unlike that adopted for AU Mic, that proved successful.

As anticipated with a PSF template with |Δ(B-V)| > 0.1, PSF-template subtraction resulted in a very strong chromatic artifact resulting in significant over-subtraction of the disk flux close to the star. This is illustrated in the six panels of PSF-subtracted images in Fig. 17 row A, of the HD 32297 disk from the six visits executed using WedgeA-0.6. (Very similar results, in terms of a repeatable chromatic residual, arises also from the WedgeA-1.0 PSF-subtracted imaging). Note that as the asymmetric disk rotates in the frame of the detector, the chromatic under-subtraction artifact (across row A) exposed at changing celestial PA's remains invariant.

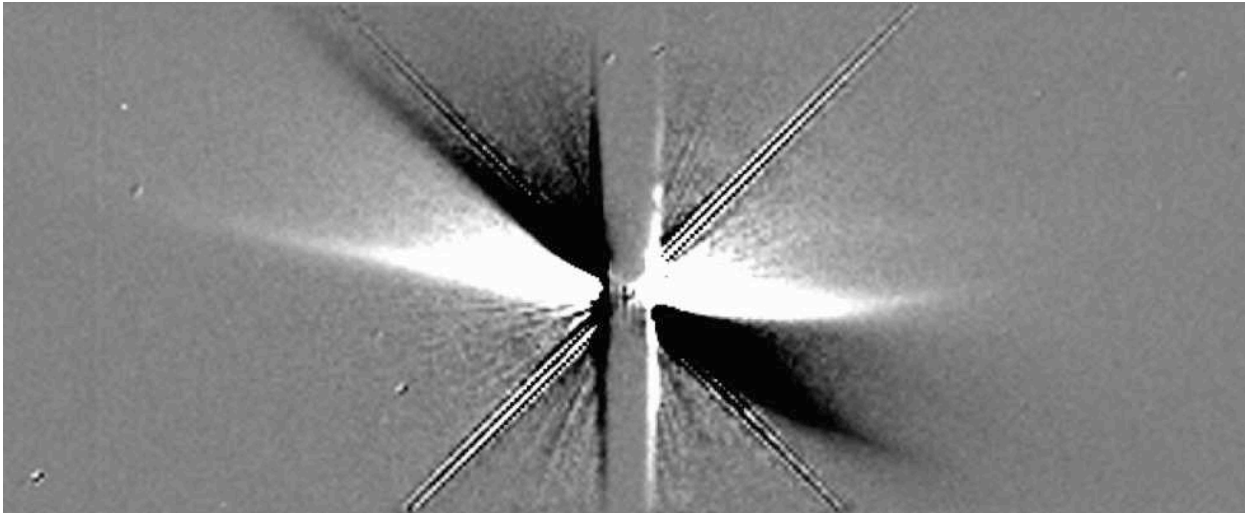

**Figure 16.** V05 minus V06 difference image of the HD 32297 CS disk, differentially rolled by the planned, nominal, 30° with the star occulted at the 0.6" wide WedgeA occulter position in both orientations. In the regions where the positive and negative imprints of the disk flux do not overlap, this (and similar other paired) images inform that the sensible disk flux all originates only to the NW side of the star and no more than within 0.2" (narrower than the half-width of the WedgeA-0.6 occulter) to the SW of the plane of the edge-on disk.

We thus created digital masks for each of the PSF-subtracted images (Fig. 17 row B), to obscure the disk itself in each frame and enable a "build up", in multi-image combination, of the chromatic over-subtraction signal. A median combination (with missing-data exclusion) of these six images, that individually do not fully sample the space around the star, results in a nearly complete image of the over-subtraction component as shown in panel C1. While there is rotationally invariant azimuthal sub-structure at high cyclical frequencies, at low spatial frequencies the over-subtraction pattern is globally symmetric radially about the star. Hence, we create a radial median image as shown in panel C2, to "fill in" by approximation (without additional higher spatial frequency information) the "missing data" regions (panel C3). We then produce a composite C1 + C3 image as a *correction template* to apply to each of the images in row A to remove (most of) the chromatic over-subtractions. The PSF-subtracted disk images, corrected in this fashion, are shown in row D. We validated the efficacy of this approach by separately performing PSF-subtractions with a non-contemporaneously observed PSF calibration star, α Pictoris, who's B-V color is a near (but not exactly) perfect match to HD 32297. Comparing the α Pic subtracted images in row E to the chromatically corrected images using our mis-begotten PSF star in row D, and in contrast to the uncorrected images in row A, clearly shows the efficacy of this approach. Note that we do not simply (alternatively) use α Pic as a

PSF-template to build a multiple-roll AQ image, as in those subtractions (row E) breathing residuals dominate (i.e., the radial streaks in those images) that are very largely mitigated with the use of the (color mismatched, but corrected) PSF template star. The same is seen with all the WedgeA-1.0 imaging, and we proceed identically with both sets of images to produce the full WedgeA-0.6 and WedgeA-1.0 combined AQ data image.

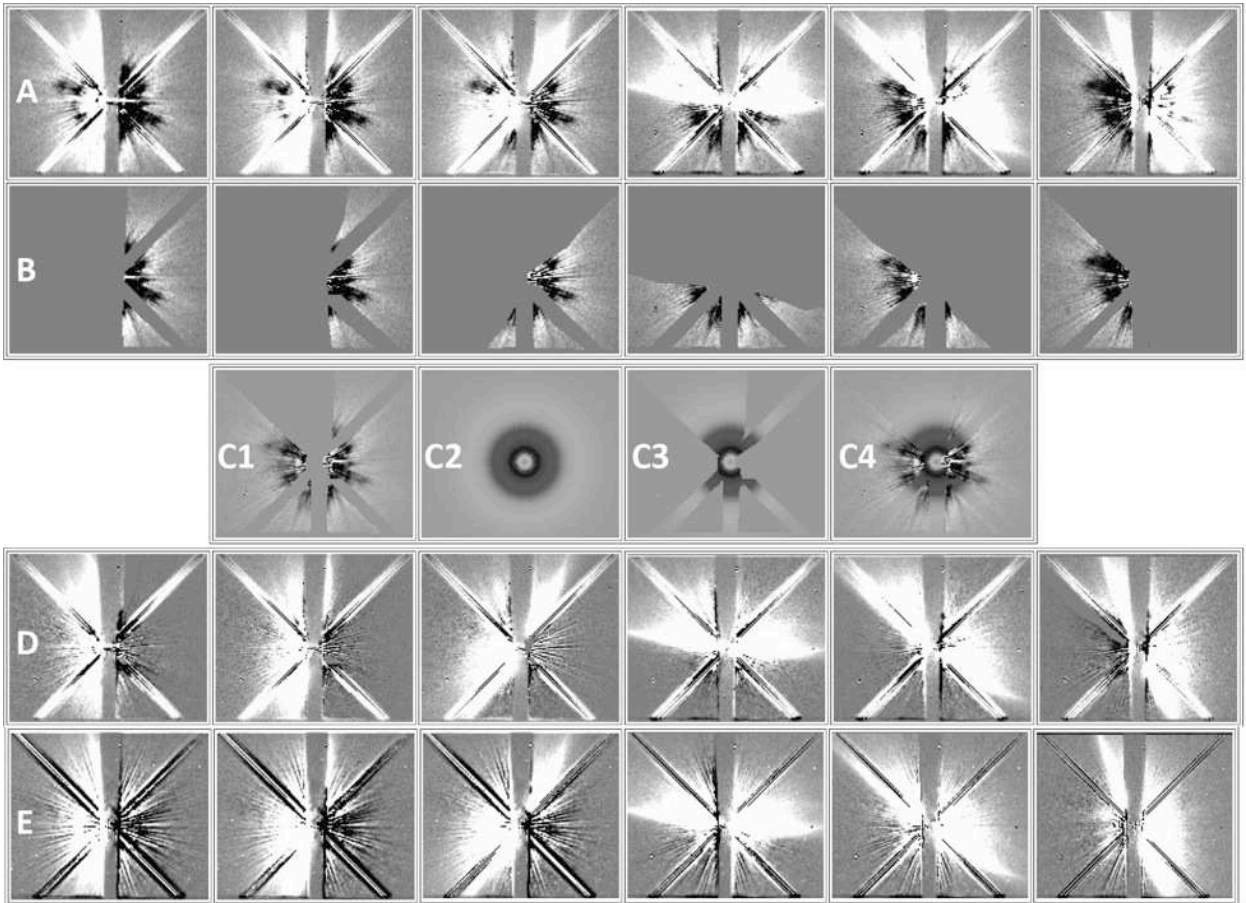

**Figure 17.** A: PSF-subtracted WedgeA-0.6 images of the HD 32297 CS disk using unintentionally color-mismatched PSF star, left to right V05, V06, V08, V01, V02, V04. B: Masking out the disk signal from the corresponding images. C1: Masked-median combination of all six images in row B. C2: Radial profile image from image C1 excluding unsampled regions. C3: Using **C2** as a proxy to estimate the over-subtraction in unsampled regions. C4: The final chromatic correction template to apply to the images in row A. D: PSF-subtracted images of the HD 32297 disk after chromatic correction (Row A images minus C4). E. Comparison to PSF-subtractions using non-contemporaneous calibration PSF star, a Pic with a closely color-matched Δ(B-V) = -0.02.

*Principal Results*. The STIS 6R/PSFTSC AQ image of the HD 32297 CS debris system, utilizing all data obtained from both Wedge occulter positions, is illustrated over its full spatial extent in Fig. 18. The STIS image indeed reveals a contiguous debris structure "connecting" the inner disk seen with *HST*/NICMOS and ground-based AO imaging, to the outer nebulosity observable with conventional ground-based coronagraphy, with unprecedented clarity.

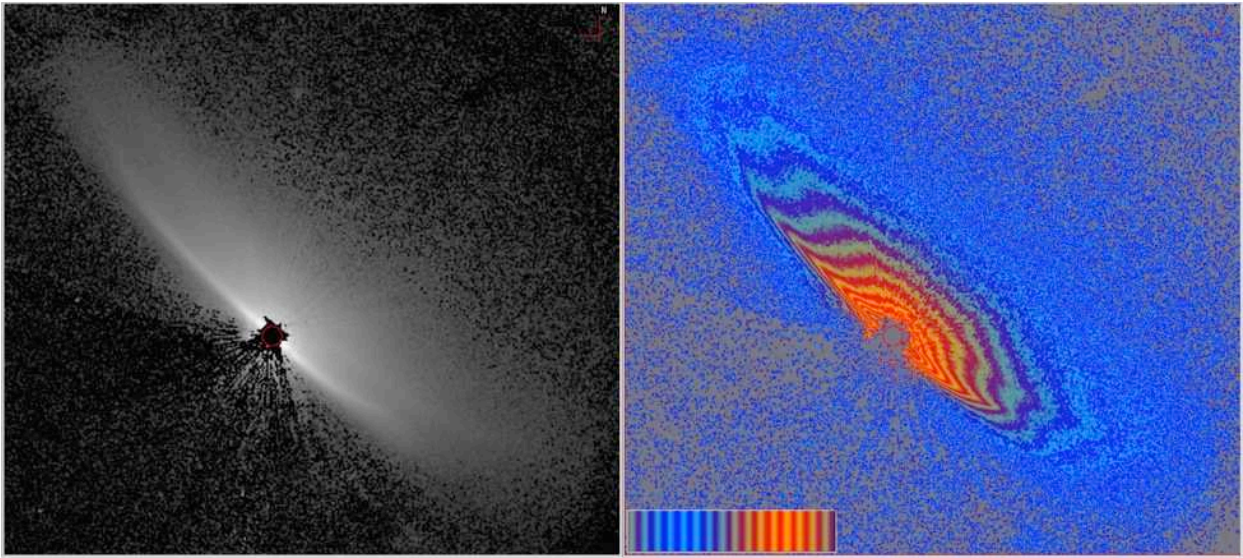

**Figure 18.** HD 32297 six-roll combined PSFTSC imaging (north up, east left) from all WedgeA-0.6 + WedgeA-1.0 observations. Gray scale image (left) and isophotal contours (right) both $\log_{10}$ display from [-3.5] to [+2.15] dex cts s$^{-1}$ pixel$^{-1}$. (app = 0 to 25 mJy arcsec$^{-2}$). Field: 380 x 340 STIS pixels (19.3" x 17.2"), north up, east left. Red circle indicates disk major axis IWA of r = 0.3" (6 STIS pixels = width of WedgeA-0.6). 1-$\sigma$ pixel-to-pixel noise assessed in regions far from disk is app 0.0032 cts s$^{-1}$ pixel$^{-1}$ (0.56 μJy arcsec$^{-2}$).

The STIS image shows a large, highly asymmetric, debris structure with complex sub-structures not previously revealed. Dust-scattered starlight is seen to asymmetrically extend to a distance of ~ 1560 AU from the star with 3 $\sigma$ resel$^{-1}$ detection significance *w.r.t.* the instrumental background noise at the extrema of the "swept back" extensions of the disk major axis. This is, physically, the largest debris system in the GO 12228 ten debris-disk sample. HD 32297 is also the "brightest" of the GO 12228 debris disks in terms of its total starlight scattering fraction with $F_{disk}/F_{star}$ = 0.3%, and is intrinsically bright with a total 0.57 μm flux density of 6 mJy. The concentration of most of this debris-scattered starlight into a narrowly viewed edge-on disk within a few arcseconds of the star (see Fig. 18 logarithmic SB contour image) makes the inner regions of the disk along its mid-plane a comparatively "low contrast" target accessible to ground-based coronagraphy. The most intriguing, and more technically challenging, features of the disk over a very large imaging dynamic (contrast) range are revealed (only) in the STIS image. To facilitate discussion of the debris system features, we present the AQ image, with the disk rotated to the image display horizontal in Fig. 19 with three different dynamic display stretches to fully illustrate the nature of the newly revealed debris system sub-structures.

(1) As previously noted, nearly all of the scattered-light flux from the disk originates from only one side of the star—to the NW, or "above" the plane of the disk, as presented in Fig 18. The large envelope of light-scattering debris above the disk plane has a complex, and asymmetrical morphology, as illustrated by the three different stretches used in panels A, B, and C, to highlight the dust at different stellocentric distances and scale heights above the disk midplane. As suggested for HD 61005 by Hines et al. (2007), and for HD 32297 by Kalas et al. (2005), this too could possibly be due to small particles in the HD 32297 being "ejected" from the debris system as it moves through, and interacts with, the local ISM.

(2) The most prominent (and brightest) part of the HD 32297 debris system are the two diametrically opposed, nearly linear, extensions of the main disk seen in edge-on projection, partially resolved from the ground, and whose relatively narrow width is best seen in Fig. 18 (left). The radial extent of these near-linear features are highly asymmetric, reaching app 6.0" to the SE of the star, but only ≈2.8" to the SW of the star. While Keck AO imaging with comparable spatial resolution suggested diametric co-linearity and equal radial extent of these features in SB and S/N maps obtained with LOCI processing (Currie et al. 2012; *c.f.*, Fig. 1) this appears not to be the case as informed by the STIS images.

(3) At larger stellocentric distances, these features "curve up" significantly at their extrema (best seen in panel B) and bifurcate into very low-contrast "V" shaped (or "wish-bone") features with morphological similarity to those seen more readily (with greater contrast) in HD 61005, and on just one side of HD 15115. The morphological similarity of these features, modulo the perspective and line-of-site dust opacity effects suggest, perhaps, a common causality.

(4) A long "strand" of higher SB scattering material appears "above" the plane of the disk to the NW of the star (only) roughly "parallel" to the plane of the disk, but does not originate at the location of the star. This feature itself appears to have a concavity toward the disk plane, and extends nearly twice as far from the star then the main disk on the same disk of the star. The origin of this feature is uncertain, and may represent a "Pleiades-like" foreground nebulosity rather than a dust enhancement physically associated with the HD 32297 debris disk explicitly.

The causal mechanism(s) for the extreme asymmetries and complex sub-structures seen in the HD 32297 disk pose a challenge to current models of CS debris disks, and may require a combination of both intrinsic and extrinsic forces to replicate and explain the observed structures.

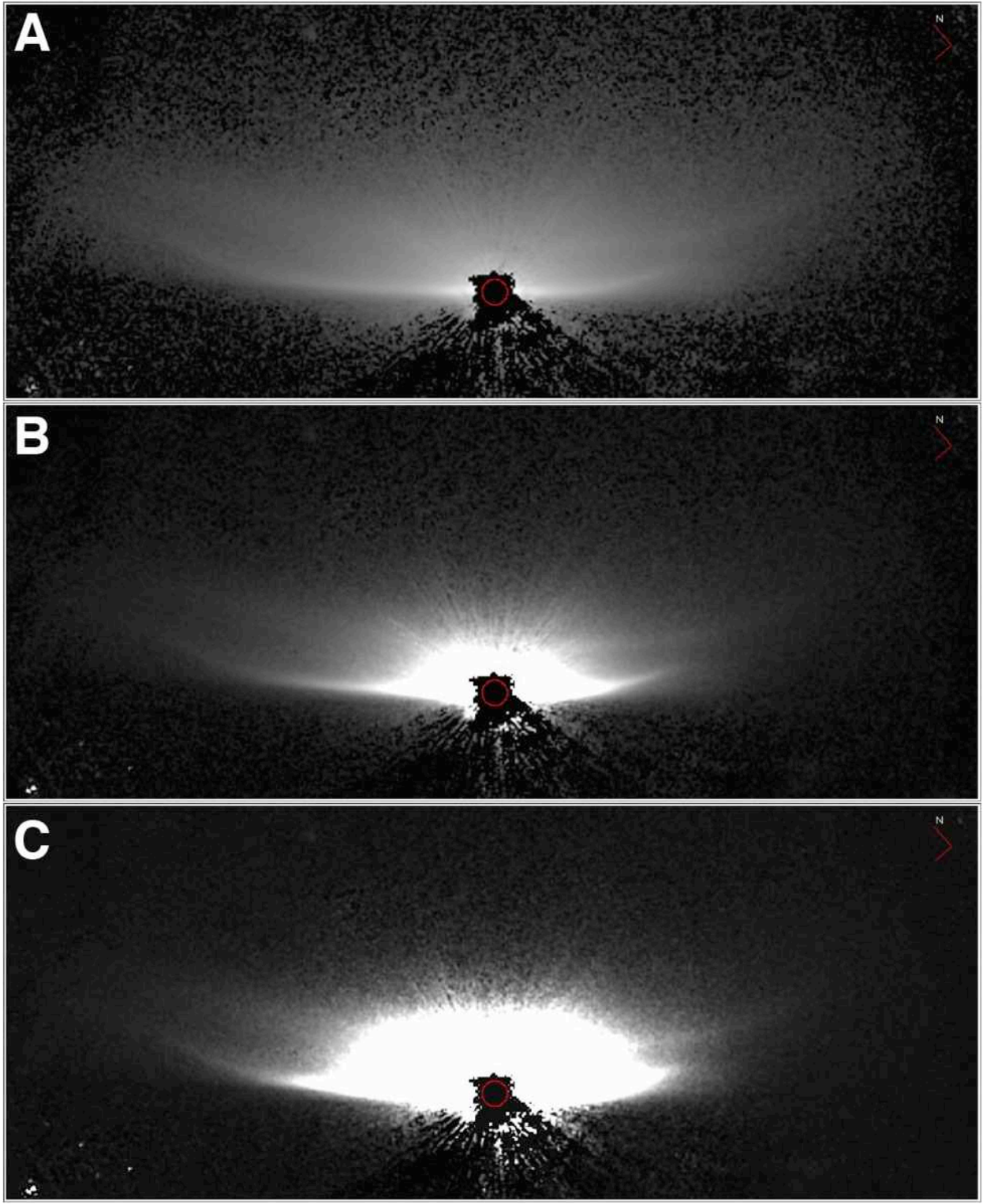

**Figure 19.** HD 32297 6R/PSFTSC imaging (north up, east left) from all WedgeA-0.6 + A1.0 observations rotated

41.5° eastward from north to place the inner disk plane on the image horizontal. Three display stretches to illustrate the complexity and morphology of light-scattering debris structures over a dynamic range of 13 mag arcsec$^{-2}$. Top: Log$_{10}$ display from [-3] to [+2.13] dex cts s$^{-1}$ pixel$^{-1}$ (app 0.002 to 24 mJy arcsec$^{-2}$). Middle: Sqrt display from 0 to 1 cts s$^{-1}$ pixel$^{-1}$ (0 to 175 µJy arcsec$^{-2}$). Bottom: Linear from -0.01 to +0.2 cts s$^{-1}$ pixel$^{-1}$ (app -1.8 to 35 µJy arcsec$^{-2}$). Field: 460 x 186 STIS pixels (23.2" x 9.4"). Red circle r = 0.3" (6 STIS pixels == width of WedgeA-0.6).

## A.4 — HD 53154

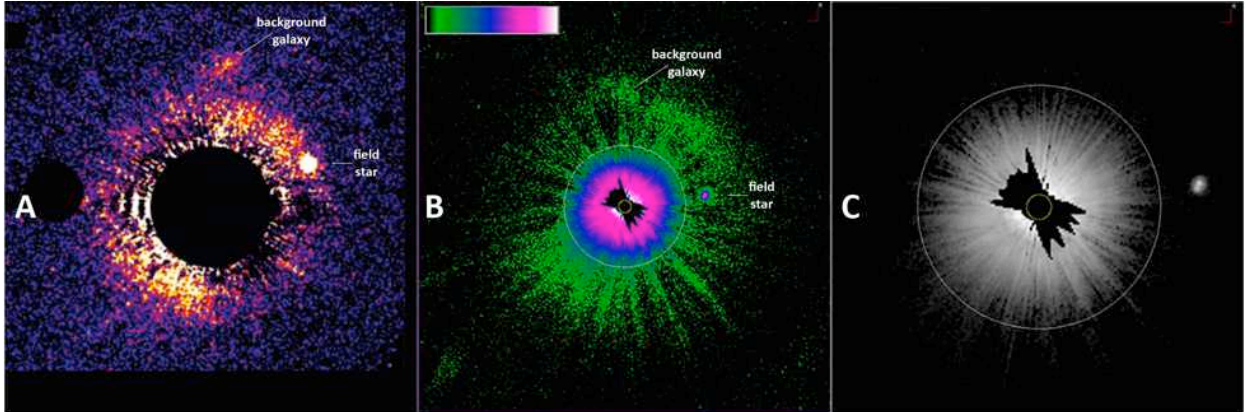

**Figure 20.** A: ACS discovery image (Kalas et al. 2006, *c.f.*, Fig. 1) with IWA$_{effective}$ r = 3.0" digital mask. B: STIS *five*-roll combined PSFTSC image. Same image scale as panel A. Improves IWA$_{effective}$ by x10 to IWA$_{minimum}$ = 0.3" (5.4 AU projected; small yellow circle). 400x400 STIS pixels, log$_{10}$ display [-2.2] to [+1.5] {dex} cts s$^{-1}$ pixel$^{-1}$. C "Inner" disk undetectable with ACS observed with STIS, at 2x spatial scale w.r.t. panel B (same data); 200x200 pixels (11.54"x11.54"), log$_{10}$ display [-1.5] to [+1.5] {dex} cts s$^{-1}$ pixel$^{-1}$. All images north up, east left.

*Introductory Notes.* HD 53154, a G9V solar analog at 18.3 pc with an estimated age of ~ 1 Gy, is the oldest star in the GO 12228 survey and possibly the oldest star (other than the Sun) for which light-scattering CS debris has been imaged. A low SB light-scattering debris disk was discovered by Kalas et al. (2006), seen to a contrast-limited IWA of r $\geq$ 3" (55 AU projected), and to a photon-limited stellocentric angular distance of $\leq$ app 6" with *HST*/ASC F606W (0.6 µm) coronagraphic imaging. These authors inferred an intermediate inclination viewing geometry of the disk of ~ 45° and major axis PA = 147° ± 2° based on the ACS discovery image reproduced herein as originally presented in Fig. 20A. With their determination of "no SB asymmetries detected between (the two) sides of the disk" they then measured the disk's 3" $\leq$ r $\leq$ 6" minor-axis-mirrored SB profile along its mid-plane and found a smooth decline in SB from 22 to 24 mag arcsec$^{-2}$, with a shallow dip between app > 3.5" and < 5" (their Fig. 3).

*Observations and PSF Subtraction.* The first set of four visits (81-84) of HD 53154 and its

(originally chosen) PSF star executed on 10 January 2011, but with a target acquisition failure preceding the planned coronagraphic imaging in Visit 81. The telescope was mis-pointed and no useful data were obtained for that visit. A second problem plagued this first set of visits in that the wrong PSF star (too red) was associated and executed with the disk imaging visits. While the contemporaneous color-mismatched PSF template/calibration data from V83 were obtained, they were not useful due to chromatic residuals strongly polluting the resulting disk images.

The second set of visits (85-89) executed nominally with a replacement PSF star, HD 59780, in V87 on 2011 Apr 11. Serendipitously, this replacement proved a mostly suitable calibration template also for the earlier non-contemporaneous disk-imaging visits. However, because of both "breathing" and differential target centering errors from wedge-deployment non-repeatability, the WedgeA-0.6 (only) attempted PSF-subtractions from V82 and V84 (but not V81) were significantly degraded and not of any utility. The WedgeA-1.0 data from these same visits, after non-contemporaneous PSF-subtraction, were only mildly degraded (w.r.t. the very high quality of the epoch 2 PSF-subtracted imaging using its contemporaneously observed PSF template), but are still quite useful, and in combination better tile the roll space, and overall improve the full data set.

The AQ image for HD 53143 (Fig. 20B/C) thus combines WedgeA-1.0 images from all five roll angles obtained (though mildly degraded, but fully acceptable and still very useful, incorporating visits 82 and 84), and WedgeA-0.6 PSF subtracted images from three roll angles (visits 85, 86, and 88) only. Because of the lack of full Wedge-0.6 roll range planned, the spatial coverage close to the star ( r <~ 1.4") is not compete — i.e., the black "butterfly" region unsampled around the $IWA_{MINIMUM}$ r = 0.3" as illustrated in Fig. 20B/C.

*Principal Results*. The STIS (5/3)-roll PSFTSC imaging of the HD 53143 CS debris disk importantly improves over the ACS discovery imaging in two key aspects. First, by better unveiling an asymmetrical structure in the architecture of the outer (> 55 AU) disk SE vs. NW sectors (see Fig. 20B), and second by revealing a previously unimaged inner, azimuthally symmetric, inner disk component from r ~ 0.3" to 3.0" (~5.5 AU to 55 AU; see Fig. 20C).

(1) The Outer Disk: The ACS and STIS images of the commonly sampled (outer) regions of the HD 53143 CS disk are illustrated in Fig. 20 A/B presented at the same spatial scale and celestial orientation (north up, east left). The STIS image is shown with a $log_{10}$ stretch to permit simultaneous display of both the outer and inner disk components, and to match as closely as possible the dynamic display range of the ACS discovery image as previously published by Kalas et al. (2006). In the STIS image, despite the presence of some low SB residual light from the extended wings of the not-completely-suppressed stellar PSF halo, a scattered light excess in the outer disk peaking at ~ r = 4.7" to the NW and SW of the star is seen, as is also the case in the ACS image. In the STIS image, a background galaxy whose presence in a single epoch image alone would conflate the morphology of the scattered light, is unambiguously identified as a polluter (along with that of a STIS-resolved binary star superimposed within the disk) contributing spatially coincident flux not originating in the disk itself.

The CS scattered light in the outer disk is seen with significance in the two sectors flanking the outer disk major axis on both side of the star to a stellocentric distance < app 5.8" (105 AU). In the STIS image the SB excess to the SE of the star appears spatially contiguous to decreasing radial distances into and interior to the ACS r = 3" IWA$_{effective}$. In the diametrically opposed NW sector, however, a dearth of disk light between about 3.5" – 4.4" (63 to 80 AU) is apparent, which is not seen at the same distance to the SE of the star. Careful, visual, examination of the Kalas et al. (2006) image may implicate also the existence of this previously unnoticed minor-axis centered asymmetry. In Kalas et al. (2006) the major-axis radial profile presented was "mirror averaged" about the disk minor axis. With that, a shallow dip in the profile at the radius of the dearth in SE disk light in the STIS image may be seen (but diluted if due to the NW outer disk alone by mirror averaging). Additionally, beyond the digital mask in the ACS image, to the E/SE of the star, there is some evidence of pollution by PSF diffraction rings that may conflate such a determination, though a relative flux deficit seems present to the NW considered by itself.

The diffuse, elongated, light-scattering structure to the NW of the star in the STIS image, suggests a morphologically description as segment of an arc. This arc seems to be *marginally* at a larger stellocentric distance by a few AU in the 2011 epoch STIS image when compared to 2006 epoch ACS image, though is not well determined or established quantitatively due to its diffuseness (though visually apparent when "blinking" co-registered images). Whether this represents a spatial re-distribution (outward migration) of the scattering dust particles in this region of the disk over this time interval, or a difference due to color-dependent scattering efficiency given the much wider STIS spectral bandpass, or simply from measurement uncertainty, is highly speculative – but noted here in expectation of future (better informing) observations.

(2) The Inner Disk: The STIS AQ image reveals a significantly brighter inner disk component at r < app 2.8″ (blue/pink encoded in Fig. 20B) that (unlike the outer disk component) appears to be circularly symmetric or nearly so (it is *marginally* brighter at equal stellocentric radii on the SE side of the star than on the NW side of the star). Ninety percent of the dust-scattered starlight is contained within 2.03" of the central star. An azimuthally medianed radial surface profile of the inner and immediately flanking regions of the disk from 0.4" ≤ r ≤ 3.6" is very well fit by a single power law of index -3.7 with no significant deviations (goodness of fit R = 0.991).

(3) The Full Disk: The total flux density from the disk, measured to a radius of 5.8" excluding the small region unsampled around the $IWA_{effective} = 0.3$" is $F_{disk} = 6.92$ mJy ± a few percent with then a 0.6 μm disk scattering fraction, $F_{disk} / F_{star} = 0.104\%$.

Contrary to earlier speculation based on the ACS discovery image alone, these new STIS observations may implicate a nearly face-on inner disk, with an unbroken $r^{-3.7}$ power law in its SB to ~ 3.6", with an anisotropic (asymmetric) dust density distribution in the outer disk, giving rise to the localized "arc"-like scattering structure seen to the NW of the star at r ~ 5".

## A.5 — HD 61005 (a.k.a. "The Moth")

*Introductory Notes.* HD 61005 is a nearby (d = 27.5 pc), 40 Myr old (Desidera et al. 2011), solar analog (spectral class G2V) with a strong thermal IR excess ($L_{IR}/L_{star} = 0.27\%$) identified as a high-priority target from the *Spitzer*/FEPS[8] program for *HST* follow-up coronagraphic imaging in *HST* GO 10527. From those NICMOS 1.1 μm data, Hines et. al (2007), discovered a light-scattering debris system associated with HD 61005 of unusual character. The morphology of the debris system was *highly* asymmetric (more-so than any other previously seen) with nearly all of the light-scattering material confined to only one side of the star, earning it suggestively a name in the discovery literature as "the Moth.". This large scattered-light structure had what appeared to be a leading-edge "bow shock" in the direction of the stellar proper motion. The general paradigm suggested for its (then) unique morphology was causal not as a system in isolation, but posited as one interacting with an ISM wind – as we have subsequently seen with some similarities for HD 32297 (§ A.3), HD 15115 (§ A.1), and perhaps HD 15745 (§ A.2).

The morphology of the large debris structure, seen in the wake of HD 61005's proper motion,

---

[8] FEPS: "Formation and Evolution of Planetary Systems" – M. Meyer, PI. This was the first of two light-scattering debris disks discovered with *HST* follow-up from the *Spitzer*/FEPS program. The other was MP Mus, see § A.11.

was confirmed with higher spatial resolution ACS coronagraphic imaging by Maness et al. (2009), but left the inner region of the debris itself at r <~ 1″ unimaged. Those observations, including optical polarimetry at r > 1.4″, suggested from the disk colors and high degree of polarization in the disk plane, large (micron or larger size) particles in the disk plane that may have been differentiated from the larger diffuse debris structure, with the disk undergoing erosion by the ISM. Ground-based AO imaging using aggressive PSF-subtraction methods (Buenzli et al. 2010) revealed the low(er) contrast features associated with the disk itself close to the star, and suggesting a near-edge on ring-like disk (with only one side seen) and diffuse material streaming from the posited disk ansae.

*Observations and PSF Subtraction*. *HST*/STIS 6R/PSFTSC observations of HD 61005 and its PSF star HD 56161 executed per the observing plan at two epochs: 2011 Feb 22 (V25 – V28) and 2011 May 03 (V21 – V24). We followed exactly the procedure described for reduction, calibration, and the creation of a 6-roll combined image using all date from the two occulting wedge positions as described in § 5.7. The 6-roll combined image of the HD 61005 debris system (exquisitely) reveals the edge-on disk, its centrally cleared region interior to a highly-inclined bright ring, as well as the conjectured "blowback" material from its interaction with the ISM from the space motion of the system. However, a low-level stellocentric circularly-symmetric residual halo of light (in this case bright, rather than dark as noted for HD 32297) was seen in the NNW centered sector of the image "above" the star (where the asymmetric debris structure does not appear), closely reminiscent (but to a lesser degree) of the chromatic PSF subtraction residuals in HD 32297 prior to mitigation. This superimposed halo is an image artifact due to (small) chromatic differences in the SED of HD 61005 (which itself may be slightly dust reddened) and the PSF template star HD 56161. The halo is not reproducible through intentional PSF template mis-scaling in intensity; attempts to "null it out" by adjusting the scaling produces VERY deep (non-physical) negative residuals close to the star and along the diffraction spikes. We therefore applied the same method of chromatic correction to mitigate these residuals as we adopted for HD 32297 to produce the AQ image discussed below.

*Principal Results*. The AQ image of the HD 61005 debris system reveals with clarity a highly-inclined debris ring and its (dark) central clearing to a stellocentric IWD approaching ≈10 AU; see Fig. 21 presented in two panels to better illustrate (A) the debris ring itself, and (B) the inner part of the skirt of material "blowing" off the ring and of its front-edge bow enhanced in visibility at the extrema from the dust opacity along the line-of-sight. This "skirt" of (likely small) particles emanating from the previously unresolved inner part of this complex debris structure is likely pushed-back from the disk-plane as the system plows through the ISM.

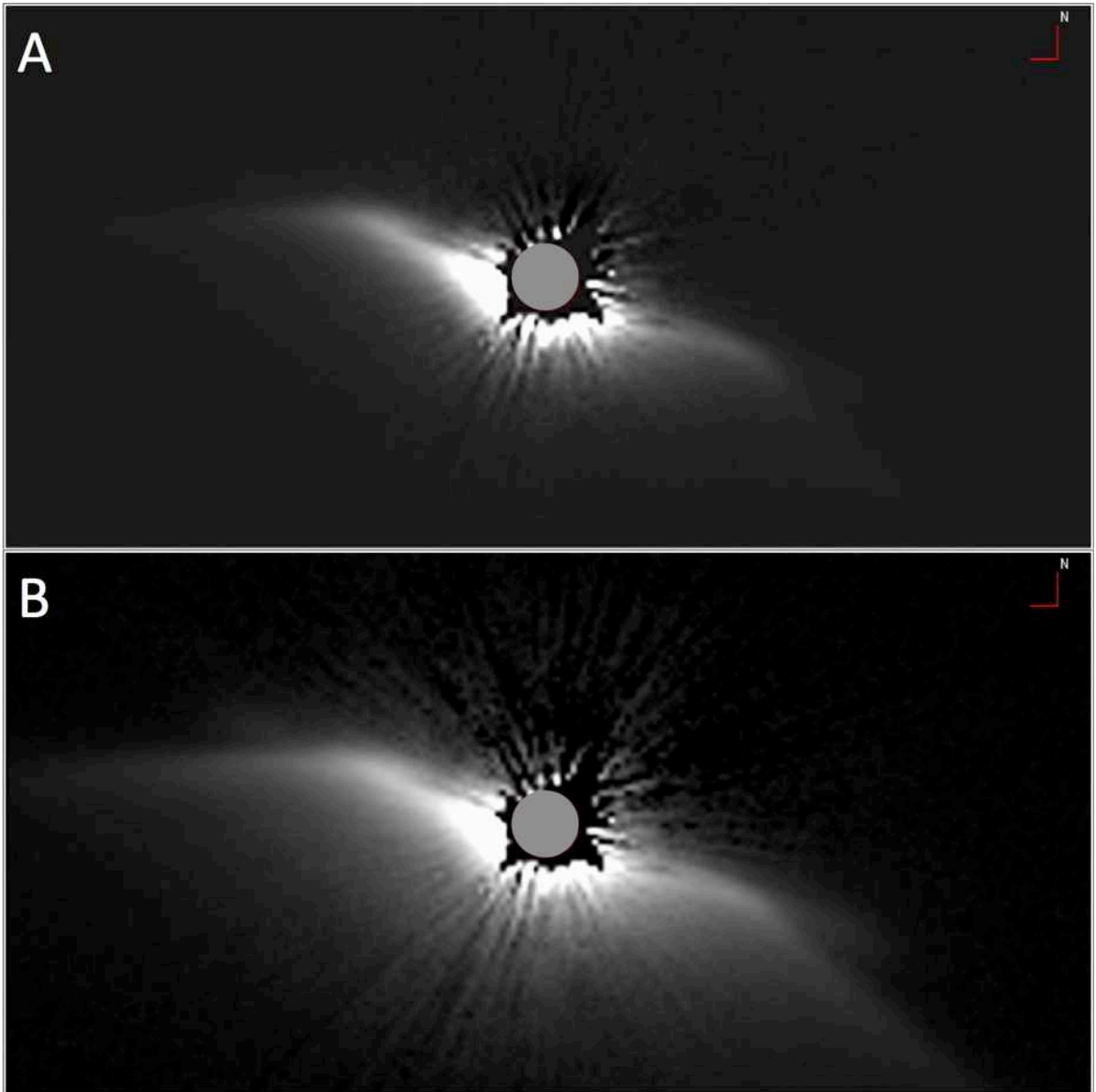

**Figure 21.** A: The HD 61005 debris ring, postulated as interacting with an ISM wind. The highly-inclined, and centrally cleared, debris ring is clearly imaged to a discernable stellocentric IWD of ~ 0.4" with STIS 6R/PSFTSC. Linear display in SB from -1 to +8 cts s$^{-1}$ pixel$^{-1}$ (app -0.177 to 1.462 μJy arcsec$^{-2}$). B: Lower SB material (extending to much further stellocentric distances as shown in Fig. 23D) is seen "streaming" off of the ring in a morphologically similar manner to the "streamers" reported previously for the HD 4976A high-inclination debris ring (Thalmann et al. 2011). Same display range as panel A but with a square-root stretch. FOV: 5.07" x 10.14", north up, east left. Gray circle radius, r = 0.3".

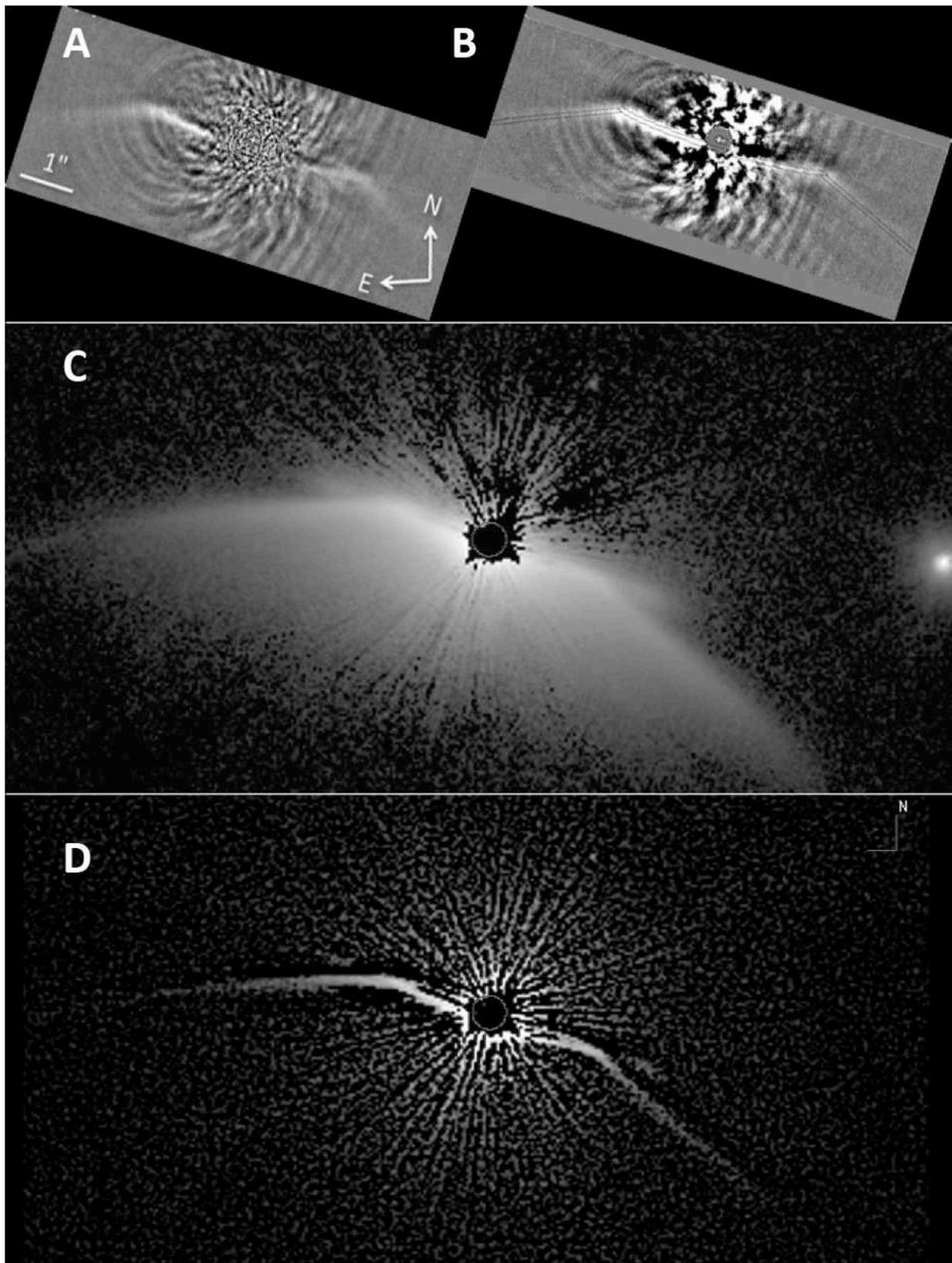

**Figure 22.** Panels A and B from Buenzli et al. (2010). Panel C: $\text{Log}_{10}$ display from [-3] to [+2.13] dex cts s$^{-1}$ pixel$^{-1}$, FOV 400 x 250 STIS pixels (20.28" x 10.14"). Panel D: Gradient-edge enhancement by high-pass spatial

filtering (at the expense of flux conservation)emphasizes the inflection of the high(er) SB ridge where the edge of the skirt joins the ring ansae. Log$_{10}$ display from [-3.5] to [0] dex cts s$^{-1}$ pixel$^{-1}$. Panels C & D. Red circle r = 0.3" (6 STIS pixels = physical IWA angle limit of occulting WedgeA-0.6.

As shown in Fig. 22 panels A & B (reproduced from Buenzli et al 2010), the brighter material in the likely preferentially forward scattering half of the HD 61005 debris ring only, and ansal "streamers" to a distance of ~ 1.5" beyond the ring ansae, had been detected in ground-based imaging with VLT/NACO using LOCI and ADI techniques. The more diffuse, lower spatial frequency, material in the "blow-out skirt" uniquely revealed with STIS 6R/PSFTSC (panel C at the same spatial scale and orientation) was seen prior with lesser efficacy also with ACS and NICMOS imaging, but was undetected from VLT with LOCI and ADI processing that do not conserve the flux density in such features. The structures identified in the ground-based images as processed, however, may be directly compared to the *HST*/STIS image. To do so on a closer footing, the STIS image was high-pass filtered with a < 5 pixel (0.25") spatial-frequency pass-through that rejected low-spatial frequency information in a manner that is similar (by design) to the consequences of LOCI and ADI processing, as shown in Panel D. The ability/utility of ground-based ADI/LOCI techniques to identify such (comparatively) lower-contrast features in CS disks (with the *HST* imaging as a "truth test") is demonstrated with these observations, even though low spatial frequency diffuse structures (as in panel C) are undetectable in the ground-based images. The complexity of the starlight-scattering debris structures in the HD 61005 CS disk is manifested over a very large range of SBs. To better illustrate, in Fig. 23 we present images of the HD 61005 debris system in incrementally increasing, x10 per step shown logarithmically, dynamic display ranges.

In Fig. 23A we show the debris ring and the "wings of the moth" to roughly the same detection sensitivity as provided by the NICMOS discovery, and ACS polarimetry, images. Overlaid on this figure we present a (non-physical) cartoon "model" of the debris ring to scale suggestive of what a complete model (beyond the scope of this paper) might produce, as follows. Using the debris ring ansa to define the disk major axis, we found the celestial orientation of the major axis PA as ~ 70.3°. Assuming the ring ellipticity results simply from its projection onto the sky (i.e., is intrinsically circularly symmetric) by ellipse fitting we estimate the minor:major ellipse axial ratio as 0.0882:1 from which an inclination of ~ 4.9° from edge-on is derived. We measured the stellocentric angular distance to the center of the debris ring along the disk major axis (i.e., to the ansae, i.e., the radius of the debris ring) on average in both directions from the star as r = 1.57" ± 0.07". We then estimate the radius of the inner edge of the debris ring (in convolution with the STIS PSF) as ~ 1.30" along the disk major axis, and assume radial symmetry outward. Finally, we estimate a "front-to-back" scattering efficiency ratio for the dust along the disk minor axis of ~ 3:1. Physical modeling must be done to confirm/optimize these geometrically and photometrically derived parameters, but as is provides observational constraints for image interpretation. Separately, in panel C we call to attention a (first order) axially symmetric scattering feature originating at, and extending beyond, the ring ansae with a "Y" shaped morphological bifurcation of the scattered light flux on the leading edge of the posited bow shock and (more diffusely) along the disk major axis. Morphologically similar features are seen in the (likely) ISM interacting disk of HD 32297 (Fig. 19), and may be related to the bifurcated structure seen only on the east side of the HD 15115 disk (Fig. 11C). Panel D shows the full spatial extent of the "skirt" of (likely small) dust particles emanating from the previously unseen inner part of this complex debris structure that may be escaping the system, pushed-back from the disk-plane as the system plows through the ISM, to the full sensitivity limits imposed by the instrumental and sky background noise.

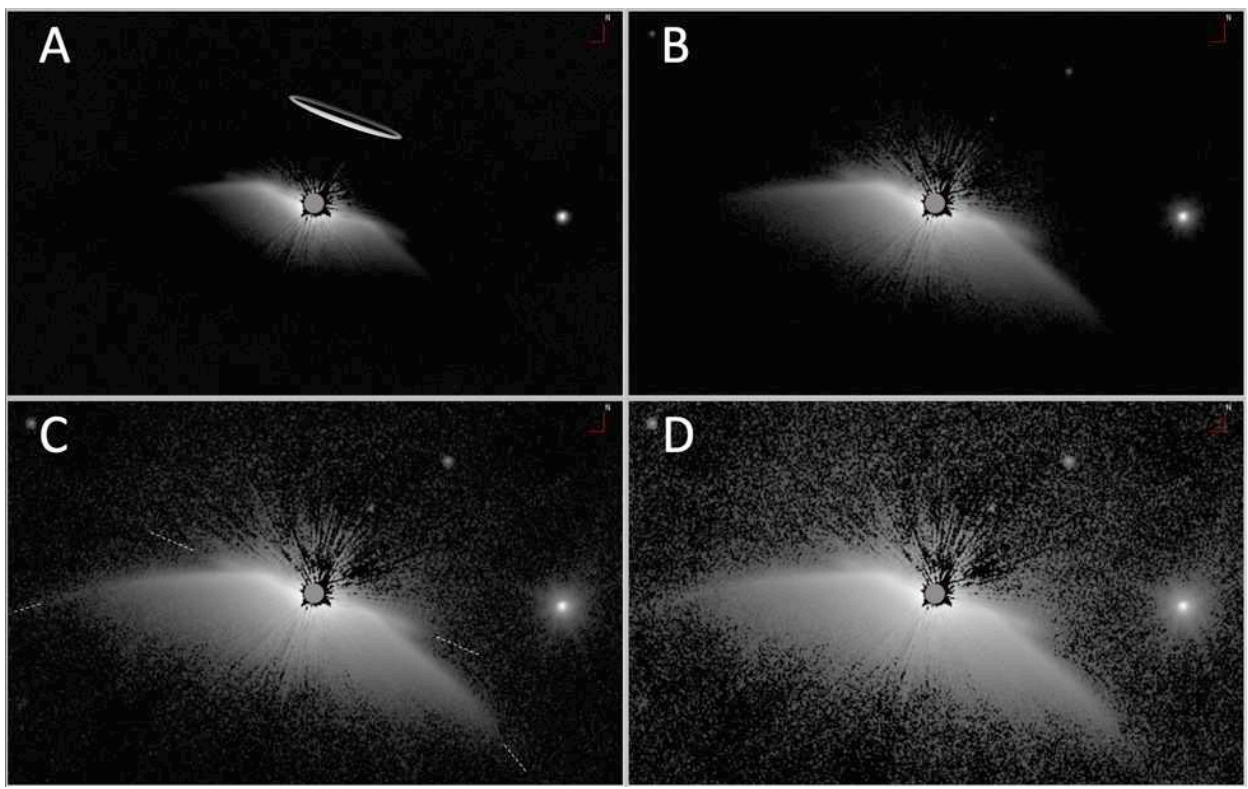

**Figure 23.** HD 61005 six-roll combined PSFTSC imaging (north up, east left) from all WedgeA-0.6 + WedgeA-1.0 observations. $Log_{10}$ display stretches A – D from [-1], [-2], [-3], [-4] to [+1.5] dex (all panels) cts s$^{-1}$ pixel$^{-1}$ (app 18, 1.8, 0.18, 0.018 to 5600 mJy arcsec$^{-2}$). Field: 400 x 250 STIS pixels (20.28" x 10.14") -- North up, east left. Red circle r = 0.3" (6 STIS pixels == width of WedgeA-0.6). The morphology, geometry, and "front-to-back" SB asymmetry of a now well-seen highly inclined debris ring (previously identified in ground based LOCI and ADI imaging by REFS) is suggested by illustration to scale in panel A. The nearly-symmetric observed "ridges" of diffuse scattering grains, morphologically bifurcated in the skirt of lower SB material to the SE of the star are indicated (compare also to HD 11515 panel E).

## A.6 — HD 92945

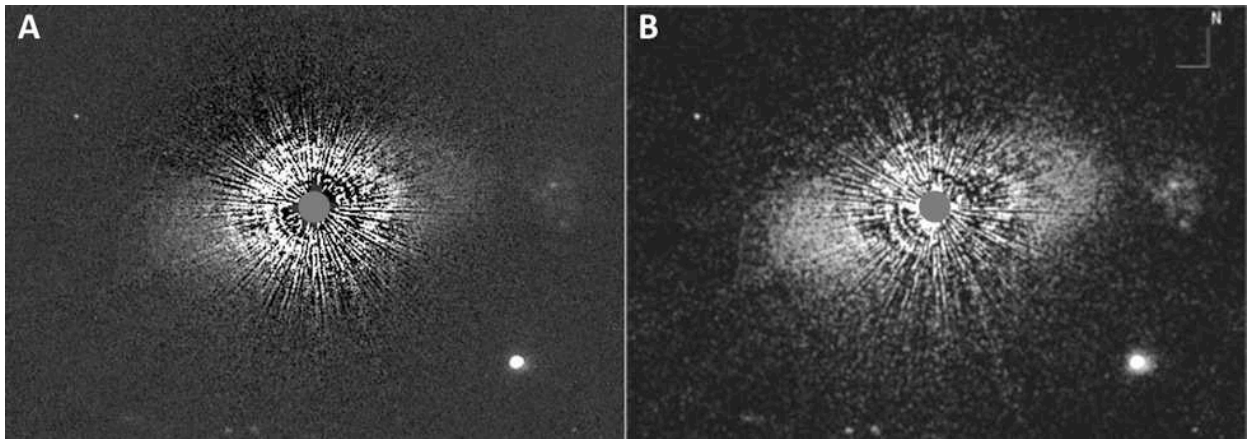

**Figure 24.** HD 92945 6R/PSFTSC imaging using all WedgeA-0.6 + WedgeA-1.0 observations combining eighteen image-artifact masked PSF-subtracted images at six target orientations. A: Linear grayscale display -0.02 (black) to +0.08 (white) cts s$^{-1}$ pixel$^{-1}$. B: Same data post-processed sequentially (1) to reduce chromatic residuals in the inner disk region by disk-masked radial-median image subtraction (see main text) and (2) to additionally improve the visibility of the inner disk by low-pass (0.4 pixel Gaussian kernel) spatial filtering to reduce the amplitude of the high-frequency breathing residuals. B presented with a $log_{10}$ display from [-2.6] to [-1.1] dex cts s$^{-1}$ pixel$^{-1}$. Both: 20.3" x 15.0" FOV with north up, east left. Gray circle represents IWA$_{effective}$ r = 0.46" for these observations.

*Introductory Notes.* The HD 92945 light-scattering debris disk, circumscribing its 80 – 300 My old, 21.4 pc distant, K1V central star, was discovered by Golimowski et al. (2011) with ACS coronagraphy. Their two-band optical imaging informed on the existence of a two-component axisymmetric disk inclined ~ 28° from edge-on seen beyond the r > 2" IWA$_{effective}$ of the ACS observations. The outer disk component was detected to a stellocentric distance of ~ 5.1" (110 AU) along the disk major axis. Golimowski et al. (2011) describe and model a diffuse, featureless, outer disk component declining in SB with stellocentric distance, "truncated" at approximately r = 5.1". Also suggested is a co-inclined higher SB ring extending from inner ring from r = 2" – 3". In their F606W image (exploring the disk to smaller stellocentric angles than with their F814W image), they found the radial SB profile of the disk along its major axis is local peaked (and globally maximum) at the r = 56 AU in the inner ring and also at r = 100 – 110 AU, just interior to the outer disk truncation radius (*c.f.*, their Fig. 4). These peaks were more pronounced on the west side of the disk than on the east side.

*Observations and PSF Subtraction*. STIS 6-roll PSFTSC observations of HD 92945, and its PSF template star HD89585, were obtained on 2011 May 06 (visits 65-68) and on 2012 Feb 29 (visits 61-64). All observations executed nominally, and all data were obtained according to the Phase 2 observation plan. However, the visit 63 WedgeA-0.6 & 1.0 PSF template images were locally corrupted by the presence of a close angular-proximity field star whose existence was previously unknown. That same field star is serendipitously occulted by Wedge A in the visit 67 template images, and with that had no impact. Because of this, the visit 67 WedgeA-0.6 & 1.0 template images (with the offending field star occulted) were tested against the visit 61, 62 & 64 disk images and (fortuitously) were found to be acceptable matches to produce PSF subtracted images for those visits. For this reason, in *addition* to the visit (61, 62, 64)-minus-visit 63, and visit (65, 66, 68)-minus-visit 67 PSF-subtractions, a third set of PSF-subtracted images, visit (61, 62, 64)-minus-Visit 67, was added to the image reduction to provide information on the region otherwise corrupted by the background star appearing in the visit 63 template. In subsequent reduction and analysis of this data set, and the creation of an AQ image after PSF subtractions, a set of 18 PSF-subtracted images (nine WedgeA-1.0 and nine WedgeA-0.6) was used, with the background-star-affected region in the visit 67 template digitally masked (and rejected). The resulting 6R/PSFTSC AQ image of the HD 92935 CS debris disk is shown in Fig. 24A. For completeness only, we note that the inverse was also tested, i.e., using the visit 63 template for disk visits 65, 66 & 68, but was found inferior in all three visits to the visit 67 template due to differences both in breathing and centering, and so was not used in this manner.

The STIS 6R/PSFTSC AQ image reproduces the main features of the disk as described by Golimowski et al. (2011) and probes to smaller stellocentric distances for (potentially brighter) dust in the interior. The inner ring, inside r ≈ 3", is well seen on at least the west side of the star and with lesser visibility on the east side, through the presence of both some low-amplitude chromatic (despite Δ(B-V) = +0.01 target-template color matching), and high spatial frequency radially-directed breathing residuals remaining in this image. The chromatic residuals are manifested primarily as (positive and negative) circularly concentric rings impressed upon the image detectable at approximately r < 3". To partially mitigate their presence, to a simple first order, their azimuthally symmetric brightness was estimated, then subtracted, in the following manner. First, a digital mask was created to exclude flux from the large, diffuse, outer disk regions (and two field objects); see Fig. 25A. Second, a digitally-masked median radial profile derived from image A was transformed to a two-dimensional image; see Fig. 25B. To first order, this primarily represents the largely azimuthally symmetric chromatic residuals only, though other information was included in the initial image. The high spatial frequency radial "tendrils" (positive and negative emanating from the star due to uncompensated breathing errors) are conserved in total flux (or nearly so) circumferentially at any radial distance from the star, so on the azimuthal median are at (or very close to) zero. Flux from the inner disk ring itself is confined to a narrow ellipse, occupying only a small fraction of any annular zone in area, and so is not filtered out with azimuthal medianing. Third, the panel C difference image (A-B) subtracts the empirically determined estimation of the radial biases from chromatic residuals. While the high spatial frequency breathing tendrils are not eliminated, visibility of the inner debris ring with reduced chromatism is improved (compare panels A and C). In Fig. 24B we separately additionally enhanced the visibility of the inner ring by low-pass spatial filtering the image in Fig. 25C with a 0.4 pixel kernel. This conservatively reduces the amplitude of the high-spatial frequency tendrils without a significant effect on the lower spatial frequency ring and outer disk.

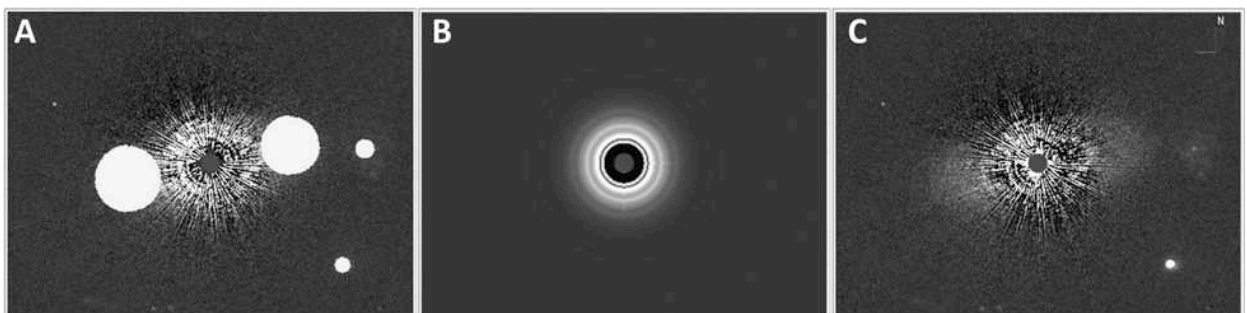

**Figure 25.** (Partially) mitigating chromatic residuals. A: Digital mask applied to AQ image as shown in Fig. 24A. B : Azimuthal median with digital mask applied. C: AQ image with empirical chromatic residual mitigation.

Finally, in Fig. 26 (top), we more aggressively applied a 5x5 pixel (~ ¼ arcsec, similar to the 11x11 kernel used by Golimowski et al. (2011) with the ACS data given its 25 mas pixel$^{-1}$ scale) box-car smoothing kernel to the chromatically post-processed AQ image shown in Fig. 25C. As a result, most of the high-spatial frequency "breathing" noise that otherwise pollutes the image at app 1.7" ≤ r ≤ 3.5" is mitigated.

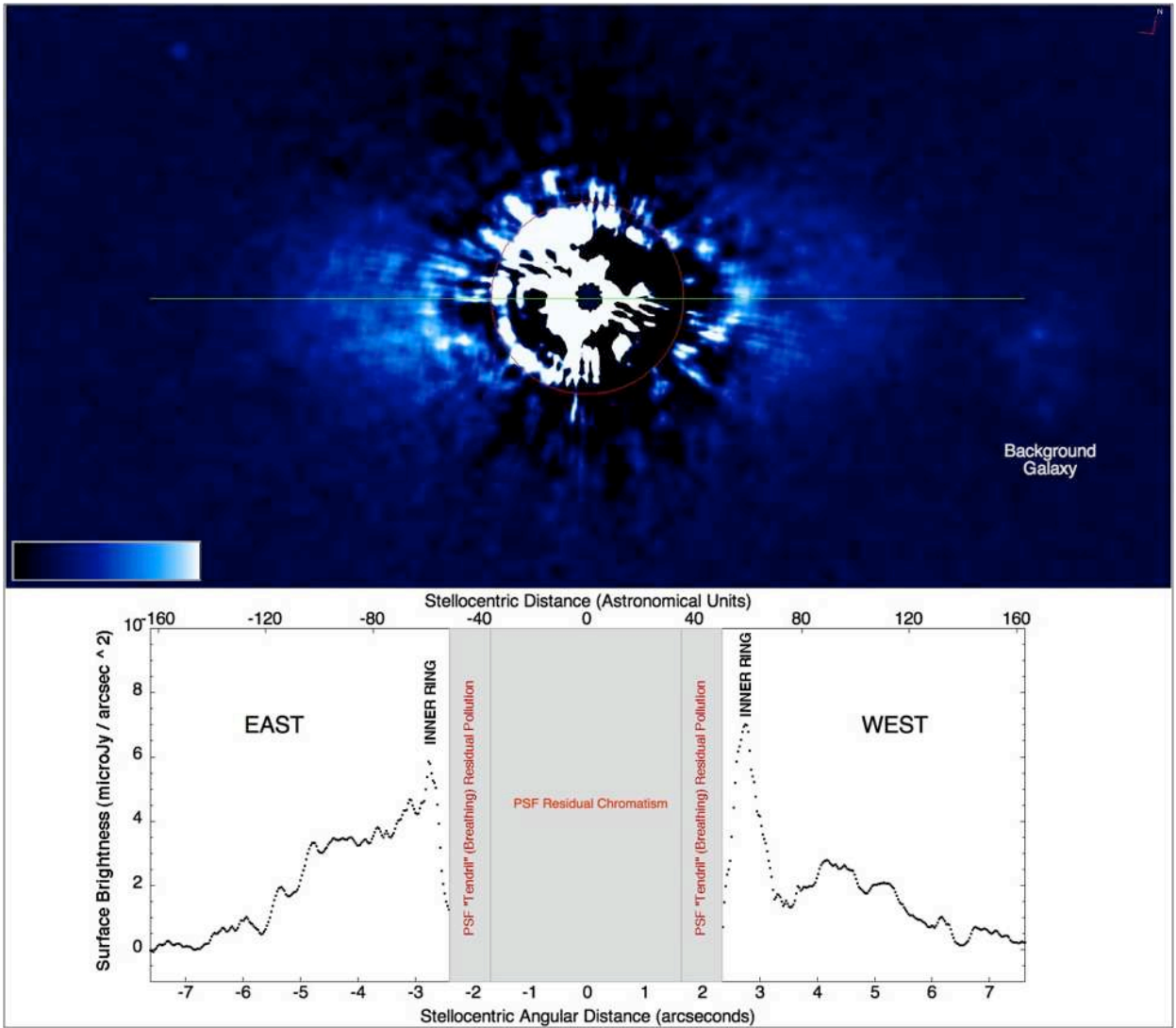

**Figure 26.** Top: STIS 6R/PSFTSC image of the HD 92945 CS debris disk post-processed to remove residual chromatic and high spatial frequency breathing residuals at r > 1.68" (33 pixel radius indicated by the red circle). Linear display stretch from -0.01 to +0.045 cts s$^{-1}$ pixel$^{-1}$. Green line indicates the disk major axis. Bottom: Radial SB profile measured along the disk major axis with the distance scale directly corresponding to the image above.

*Principal Results*. Our new STIS 6R/PSFTSC observations of the HD 92945, obtained at a similar central wavelength to the ACS F606W filtered observations, confirms the previously discussed observational results of Golimowski et al. (2011). The STIS AQ image reveals (at least) the eastern side of the inner ring with somewhat better clarity, and the northern extent of the ring closer to its sky-projected minor axis (e.g., compare also to Golimowski et al. 2011, c.f., Fig. 3). From the STIS data, we find the 50CCD band integrated brightness of the disk from r = 1.67" (~ 36 AU on the disk major axis) outward, largely unaffected by residuals from imperfect PSF subtractions, to be 0.143 mJy, and thus a disk scattering fraction, $F_{disk} / F_{star}$ ~ 0.0051% at r > 1.67". The uncertainty in the absolute calibration both of these measurements is estimated as ± 20% 1-σ. This is in good agreement with what Golimowski et al. (2011) found with $F_{disk} / F_{star}$ = 0.0069% fitting a full disk model to their observations in the F606W band of ACS.

The STIS post-processed AQ image (Fig. 26, top) clearly resolves both the HD 92945 CS debris ring at r ~ 2.77" on the disk major axis, and the circumscribing "featureless" outer disk. The debris inner ring (though somewhat noisy) is well seen in 90° wide sectors flanking the disk major axis on both sides of the star. Assuming intrinsic bi-lateral symmetry, we find the celestial orientation of the disk major axis at PA = 100° ± 2° (in agreement with Golimowski et al. 2011). We then produce a radial SB profile along the disk major axis from the image (Fig. 26, bottom), without quantitative error estimation) to aid in interpretation of the image.

Two significant disk asymmetries are seen. (1) The outer disk appears brighter on the east side than on the west side of the star beyond the inner debris ring. (2) The eastern ansa of the inner debris ring appears inwardly "distorted" in comparison to the western ansa. The latter may be due to a pericentric offset of the debris ring, but this is conflated with remaining PSF-subtraction residuals at the level of measurement uncertainties, and confirmation of this suggestion must await further improved, future, observations.

Modulo uncertainties in stellar ages (see Table 1) HD 92945 may be the oldest debris disk in our GO 12228 sample for which an inner debris ring within a larger debris system is clearly resolved. However, similar large-scale debris has been reported for the 440 ± 40 Myr old Fomalhaut system (Kalas et al. 2008, Kalas et al. 2013, Mamajek et al. 2013) and may be a routine feature, at low SB levels, of many debris disks.

### A.7 — HD 107146

*Introductory Notes.* HD 107146 is a 27.5 pc distant, 80 – 200 Myr old, close-solar analog star (spectral type G2V) with a large infrared excess, $L_{IR}/L_{star}$ = 0.12%, attributed to several tenths of an Earth mass of re-radiating dust orbiting the star. The presence of a broad starlight-scattering CS debris ring about the star was reported, and discussed in detail, by Ardila et al.

(2004, 2005). The disk, then seen in ACS/HRC PSF template-subtracted coronagraphic observations at 0.6 and 0.8 μm, was suggested to be inclined ≈ 25° from pole on and broadly peaking in radial SB in both bands ~ 4.7" (130 AU) from the central star. While the ACS observations have been successfully used to ascertain the primary characteristics of the disk, the above authors have stated "*{PSF} subtraction errors, caused by mismatches in the colors of the stars or PSF time variability, dominate the emission within ~2" from the star and contribute light at large distances. With only one reference star, we cannot quantify very precisely the magnitudes of these two error sources.*" The debris belt was initially detected in the millimeter (Williams et al. 2004) and has been imaged in the submillimeter (Corder et al. 2009), and modeled (Ertel et al. 2011). The availability of these data plus HD 107146's status as a solar analog, made HD 107146 a natural candidate for 6R/PSFTSC.

*Observations and PSF Subtraction*. All observations of HD 107146 and its PSF reference star HD 120066 executed nominally on 2011 Feb 22 and 2011 May 30. After visit-level PSF subtraction, it was obvious that the V24 PSF-subtracted images for both WedgeA-0.6 and 1.0 were clearly inferior to the other visits with "breathing" driven PSF-subtraction residuals dominating the inner regions of the disk. The (initial) inclusion of the V24 data into a six-roll combined AQ image, in a systematic inter-visit manner as described in § 5.7, marginally degraded the final product rather than improving it compared to a five-roll combination excluding V24. Fortunately, this was not critical to the metrical analysis of this target's angularly large, broad, and nearly face-on ring-like disk that is spatially well sampled in the five other, nearly equally superior, visit-level images included in then a five-roll combination. The 5-roll combined AQ image is presented in Fig. 27 (left panel). The interior region of the disk, along the direction of and flanking its major axis (SE to NW) is imaged to an $IWA_{efective}$ = 0.41", about 2 resels beyond the edge of the occulting wedge. The r ≤ 1" CS region in the 5R/PSFTSC combined image roughly in the direction of the disk minor axis only (depicted as the contiguously hard-black stellocentric region in Fig. 27) is unsampled due to the rejection of the V24 imaging data that uniquely tiled that space. Separately, two 3-roll combined images, divided by epoch inclusive of the V24 data in the second epoch, were created for the purpose of investigating potential companionship by common proper motion for point sources identified in close proximity to the disk not adversely affected by breathing degradation of the V24 disk image itself.

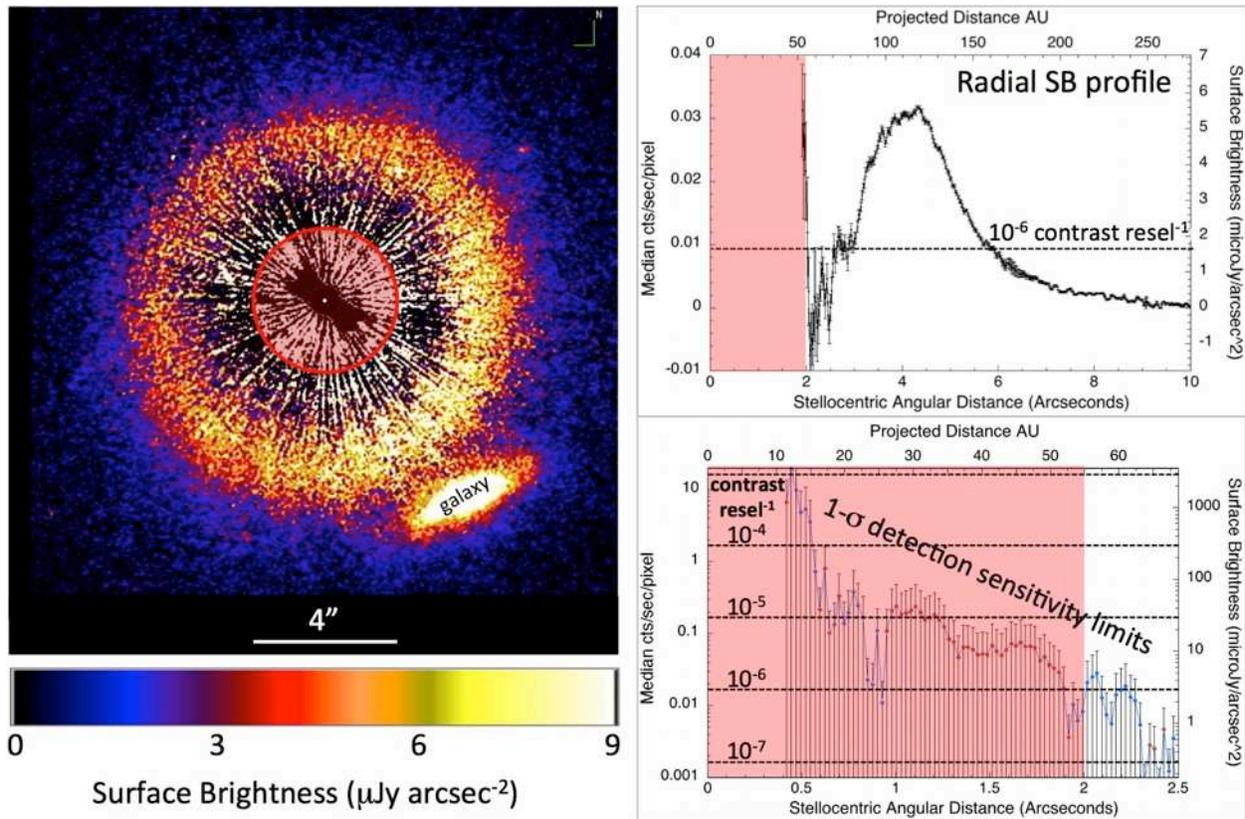

**Figure 27.** Left: STIS 5R/PSFTSC image of the nearly face-on HD 107146 debris disk. 15.23" x 15/23" FOV, north up, east left. The red circle indicates the region where the brightness of PSF-subtraction residuals dominate over disk-light declining toward smaller stellocentric distances interior to radius of peak brightness at $10^{-6}$ resel$^{-1}$ and lesser contrasts. Top Right: Azimuthal median radial SB profile in the non-contrast limited regions of the disk. Error bars plotted on the radial SB profile represent the 1-$\sigma$ standard deviations about the median brightness measured in 1-pixel radial increments fully around the star. Absolute calibration uncertainties due to possible under or over subtractions of the template PSF of ~ few percent at most are not included. Bottom Right: 1-$\sigma$ contrast-limited disk detection sensitivities as a function of stellocentrc distance in the presence of PSF-subtraction residuals as measured from the median absolute deviations in flux density per resel in 1 pixel wide annuli. (Red/blue points indicate positive vs. negative median residuals).

*Principal Results*. The 5R/PSFTSC AQ image of the HD 107146 disk improves significantly in several regards over the discovery imaging obtained by Ardila et al. (2005: *c.f.* Fig 1). The new STIS image is presented in Fig. 27 with a linear display stretch over a dynamic range and with a color table matching as closely as possible that of the ACS F606W discovery image to enable a direct visual comparison. A background galaxy, seen equally well in the Ardilla et al. discovery imaging, now marginally intrudes into the outer periphery of the broad ring-like disk, as it has been "closing" on HD 107146 due to the stellar (reflex) proper motion since the 2004 epoch discovery imaging. Circa 2020 the galaxy will be superimposed upon the intrinsically brightest part of the disk, radically altering its apparent SB distribution then and for a decade to follow (a warning to future observers).

Improving on earlier ACS and NICMOS data, the STIS AQ image sensitively traces dust depletion in the inner regions of the broad HD 107146 debris ring to within ≈ 60 AU of the star, half the stellocentric distance from where the debris ring is brightest. The HD 107146 CS debris disk is intrinsically quite faint both in terms of its STIS 50CCD band integrated flux density, $F_{disk}$ = 0.40 mJy at r ≥ 2.0" (~ 55 AU; radius of the red "here be dragons" circle in Fig. 27), and its visible-light scattering fraction $F_{disk} / F_{disk}$ = 7.7 x $10^{-5}$ (±4.0 x $10^{-6}$), and in comparison to its 16x larger infrared excess. An azimuthal median radial SB profile (measured with the background galaxy masked) is shown along side of the disk image in Fig. 27 with $SB_{peak}$ = 5.6 µJy arcsec$^{-2}$ at a stellocentric median distance r = 4.3".

While dust depletion interior to the radius of peak SB was inferred from the prior ACS data, the extent and depth of the clearing was uncertain (e.g., see Fig. 2 of Ardilla et al. 2004). The STIS SB image and its azimuthal median radial profile (Fig. 27), clearly show a steep decrement in SB interior to the broad ring with the region cleared of light-scattering debris to a 1-σ per resel sensitivity limit of ~ 0.2 µJy arcsec$^{-2}$ (image contrast ~ 2 x $10^{-7}$) at 2.3".

The ring SB declines more rapidly interior of $r_{peak}$ toward the star and less steeply radially outward from $r_{peak}$. The median photometric FWHM of the ring, as one characterizing metric of the ring width, $W_{FHWM}$ = 2.2" (≈ 60 AU), with $W_{FHWM} /D_{peak}$ = 0.34 (significantly broader than the HR 4796A [Schneider et al. 2009] and HD 181327 ring-like debris belts).

We note here that the debris belt, and its inner clearing revealed with STIS 5R/PSFTSC exhibit a high level of bi-axial symmetry, with no obvious sub-structures or distortions of detectable significance around the ring, such as have been seen in sub-millimeter interferometry (Corder et al. 2009). Conversely, though not explicitly spoken to by Ardila et al. (2004), the ACS F606W discovery image, *c.f.* Fig. 1 (at similar wavelength as the STIS image), on visual examination appears asymmetrically extended to greater peak radial distance along the SW disk major axis. This is also reflected (when examined in detail) in their F606W SB profile along the disk major axis, (*c.f.* Fig 2) and corresponding Tau x Omega plot (*c.f.* Fig 3) derived from that image, but is not (obviously) seen in their F817W discovery image. With the higher-fidelity

STIS image, and lack of correlation in the asymmetry implied only in the ACS F606W image, we suggest this was an optical artifact from pollution by incomplete PSF subtraction beyond r > 2" and/or imperfect geometrical distortion correction in the ACS F606W image.

From the STIS data we can also quantitatively assess SB limits on any dust undetected between 0.4" to 2.3" (Fig. 27, bottom right) where dust detection by scattered-light becomes contrast limited. Therein we also cast this in terms of image contrast. With such dimensionless quantification, we suggest HD 107146 would make an ideal "test case" for more aggressive state-of-the-art ground- and space-based coronagraphs now under design and development.

Two point-like objects (putative low-mass companion candidates) were detected in close proximity exterior to, and superimposed upon, the broad HD 107146 debris ring at r = 6.40", PA = 300.3° (Fig. 28 left panel), and r = 3.81", PA=307.05° (Fig. 28 right panel); two stretches of the same 5R/PSFTSC image. Thanks to the relatively large proper motion of HD 107146 ([-174.16, -148.90] mas yr$^{-1}$), and its annual parallax, in concert with an uncertainty in the location of the occulted star of only ~ 5 mas in the SIAF (see § 5.4), both objects were found to be non-commoving background objects over the three month interval between the two visit sets.

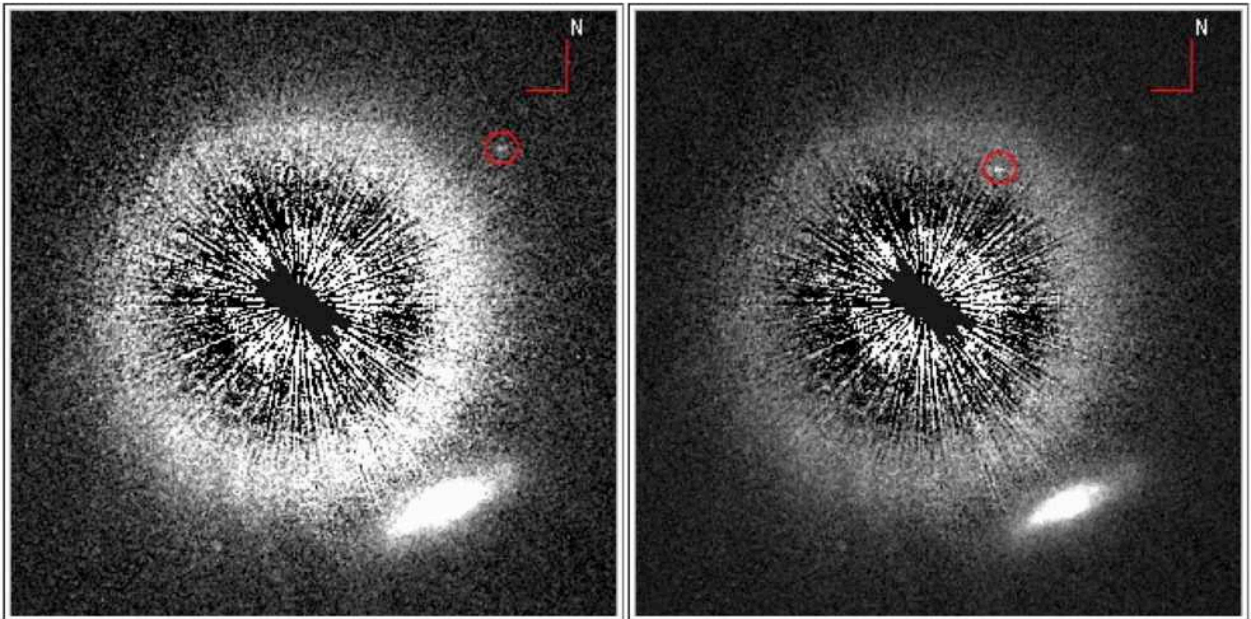

**Figure 28.** Identified, but rejected, companion candidates found to be background stars from their reflex motions relative to the location of HD 107146 ascertained with "X marks the spot" differential astrometry (see § 6.4).

## A.8 — HD 139664

*Introductory Notes.* HD 139664 is a 17.4 pc distant F5 main sequence star with a starlight-scattering CS disk that was revealed at stellocentric distances > 3" (52 AU) with *HST*/ACS coronagraphic imaging (Kalas et al. 2006). Its estimated age of 300 (-200, +700) Myr, comparable to Fomalhaut and possibly as old as HD 51354 (1 Gy, see § A.4), makes it one of the oldest stars for which a CS debris disk has been imaged in scattered light. HD 139664 also has the smallest IR excess, $L_{IR}/L_{star}$ = 0.009%, of any of the stars on the GO 12228 sample and the general population of ~ two dozen CS debris disks that have been imaged in scattered light to date, and follows the general trend of IR excess declining with stellar age (Fig. 9). The ACS scattered-light discovery image revealed the ansal regions of a faint, apparently edge-on, debris disk from a contrast limited IWA ≈ 3" (54 AU) to a stellocentric distance of ≈ 6" (104 AU) suggested by Kalas et al. (2006) to be the outer boundary of the disk (see Fig. 30A). No SB asymmetries were detected between the two diametrically opposed sides of the disk. A minor-axis mirrored (to improve S/N) radial SB profile suggested a peak in the dust density at r ≈ 83 AU, the radius of a postulated inclination-obscured planetesimal belt.

*Observations and PSF Subtraction.* The STIS observations of HD 139664 and its PSF star HD 99353 executed nominally at two epochs, 2011 May 23 (V95 – V98) and 31 July 2011 (V91 – V94), and would prove to be the most challenging of all the GO 12228 targets to improve over previous observations. HD 139664 itself at V = 4.64 is significantly brighter than our other disk target stars (see Table 1). Thus, to permit sufficient imaging dynamic range across the inner to outer disk regions in the face of image saturation in the region adjacent to the STIS occulting WedgeA, for HD 139664 and its commensurately bright PSF star, we imaged using WedgeA-0.6 with both very short and intermediate duration exposure times, in addition to the deeper (centrally saturated as expected) WedgeA-1.0 imaging; see Table 3 for summary details.

Following our homogeneous multiple-orientation observing paradigm across the target set, even with foreknowledge of the edge-on disk major axis PA from the ACS observations, we did not constrain the absolute (celestial) orientation of the disk in any of the STIS visits to enable us to probe for, and study, any out-of-plane material that, a priori unknowingly, might have existed (e.g., as seen in the cases of the HD 61005 (§ A.5) and HD 32297 (§ A.3) nearly edge-on disks).

All of the PSF-subtracted WedgeA-1.0 images, as well as those of intermediate exposure depth at the WedgeA-0.6 occulter position, suffered from rotationally invariant chromatic residuals from a small color mismatch (Δ(B-V) = -0.10) in the PSF star (shown similarly, by example, in the case of HD 32297 in Fig. 17) causing a zonal under-subtraction of the disk closer to the star. These were (separately) corrected for the data acquired at the two different wedge positions by same method described in § A.3.

With the inter-visit relative orientations unconstrained against absolute celestial PA, the disk major axis itself was unobscured by the STIS wedge, OTA diffraction spikes, and IWA saturation region beyond the mask edge, in only two of the six STIS visits (V96 & V98) constrained by a combination of guide star and target visibilities. This limited, in aggregate, the sensitivity depth of detection in the outer regions of the disk along the disk major axis explored with the (deeper) WedgeA-1.0 exposures. V95 was significantly degraded by breathing residuals at both wedge occulter positions after PSF subtraction, and hence was rejected in the creation of the final multi-roll (chromatism corrected) AQ image shown in 29.

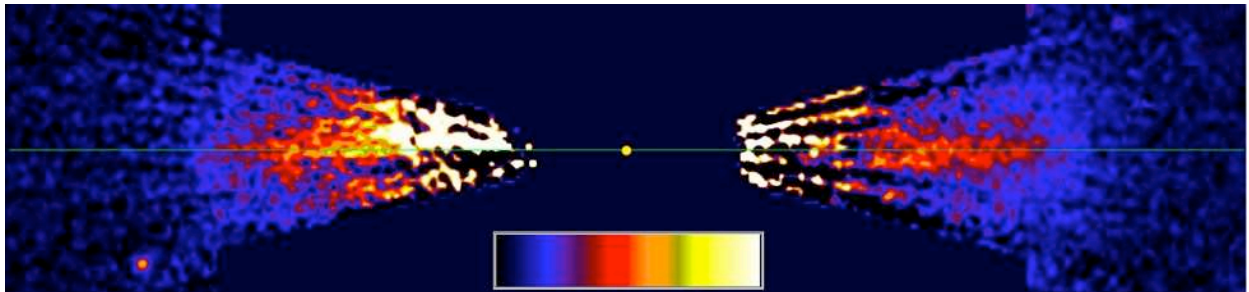

**Figure 29.** The STIS AQ image of the HD 139664 CS edge-on debris disk oriented with the morphological major axis (green line) on the image horizontal (celestial PA of the disk major axis estimated from this image as 75.5° ± 0.5°). Linear display stretch from -0.02 to +0.44 cts s$^{-1}$ pixel$^{-1}$. FOV 15.23″ x 3.55″. IWA = 1.2″.

*Principal Results.* The Fig. 29 AQ image reproduces, and confirms, the main features of the debris disk observable with ACS reported by Kalas et al. (2009), with the following differences deviating from their determination that "no surface brightness asymmetries were detected between each side of {the} disk". To facilitate visual comparison, we additionally suppress still remaining high-spatial frequency breathing residuals in the fully circum-azimuthal FOV by applying a 3x3-pixel boxcar kernel as a low-pass spatial filter. The result is shown in Fig. 30B. An examination also of a radial SB profile along the disk major axis (prior to smoothing, i.e., derived from the image in Fig. 29) in Fig. 31 confirms an "east/west" SB asymmetry.

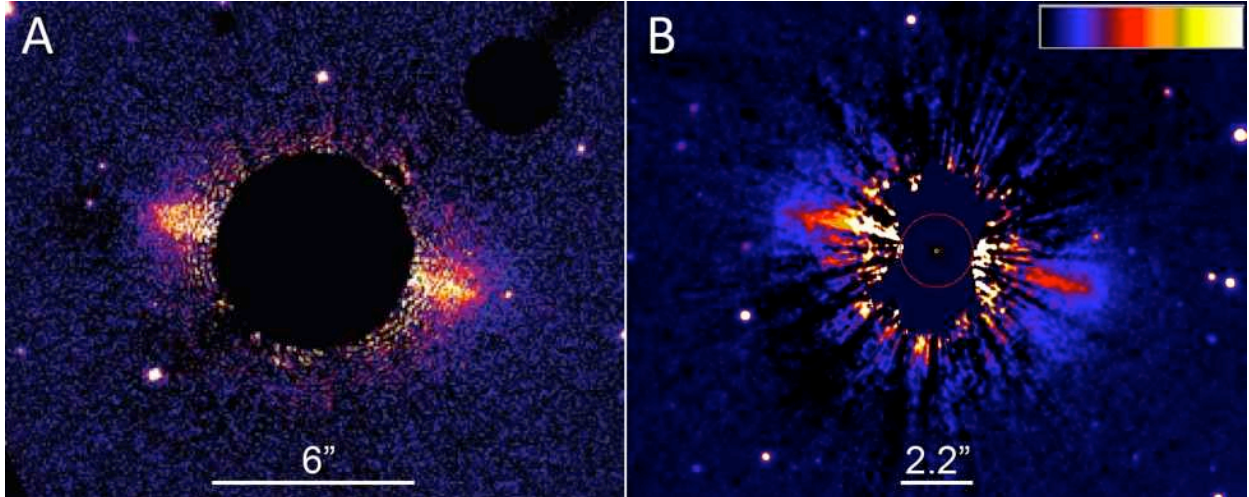

**Figure 30.** A: ACS discovery image of the HD 139664 CS debris disk, r = 3″ IWA indicated by the black digital mask (reproduced from Kalas et al. 2006). B: STIS 3x3 boxcar smoothed AQ image with -0.02 to +0.44 cts s$^{-1}$ pixel$^{-1}$ display stretch similar to the ACS image for ease of comparison. The STIS image probes closer to the star (red circle, r = 1.2″), though in the presence of remaining PSF subtraction residuals in region masked in the ACS image.

Though somewhat polluted with PSF-subtraction artifacts, starlight-scattering dust on the east side of the edge-on disk is traceable in the STIS AQ image to a (masked, but saturation limit obscured) inner working distance of r ~ 1.2" (21 AU), with a SB ~ 50 μJy arcsec$^{-2}$ at r = 2.7" and declines inwardly by a factor of two at r = 1.3" just beyond where the disk is obscured in the STIS image. On the opposite (west) side of the star, the disk SB peaks further from the star at r = 3.7", reaching a maximum SB of ~ 30 μJy arcsec$^{-2}$ and declines linearly. Directly opposite the east side peak at r = 2.7" the west side only ~ 17 μJy arcsec$^{-2}$. See Fig. 31.

Additionally, the vertical (out of plane) morphology of the disk to the north, "above" the disk plane as shown in Fig. 29, appears different on the east and west sides of the disk and, on close examination, correlates with the ACS image. On the west side, the bright part of the edge-on disk (color scaled as red in the STIS image) is a narrow linear feature defining the disk plane surrounded above, below, and exterior by dust scattered starlight at lower SB (encoded light blue). On the east side only, dust-scattered starlight extends at higher SB to a greater distance "above" the disk plane, with greatest height in the STIS image at r ~ 2.8" to 3.7" where the SB peaks. A similar morphology is seen in the r > 3" region in the ACS image (giving a "wedge", rather than linear, shape to the outer part of the eastern side of the disk).

Together, these differences implicate a non-axi-symmetric disk SB, unknown to Kalas et al.

(2006) in their minor-axis mirrored reflected treatment of the posited dust density distribution. Whether this results from a stellocentric offset of a geometry-obscured debris ring, or asymmetric differences in the dust density distribution, is an area for future investigation.

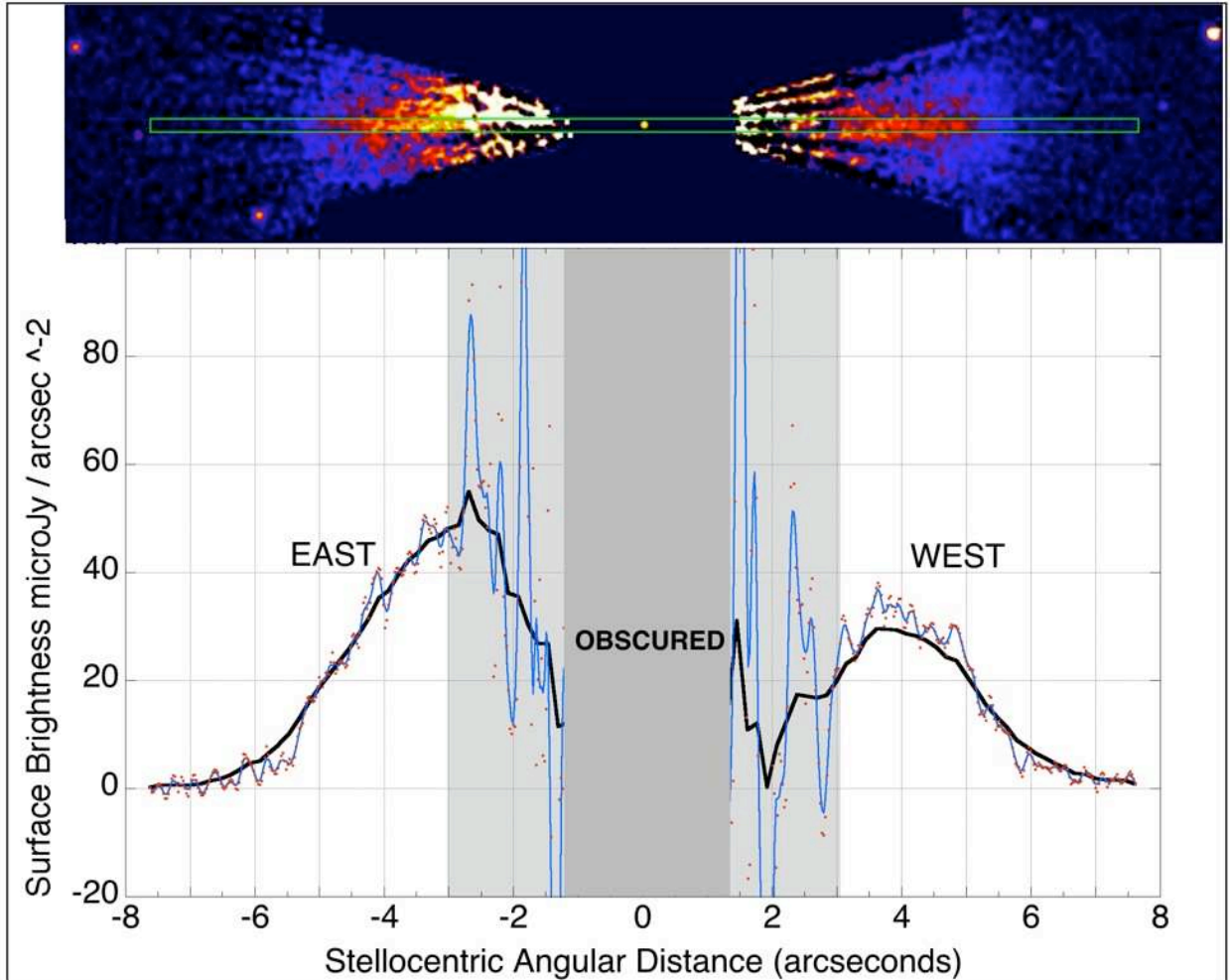

**Figure 31.** Radial SB profile along the disk major axis in 1-pixel radial increments (red points) in a 2-pixel (~ 1 resel) wide strip ± 8″ long centered on the star (green strip superimposed on disk image at the same horizontal angular scale). Blue line is with 3-pixel wide boxcar smoothing (corresponding to spatial filtering in Fig. 30B). Black line further adaptively filters out the high spatial frequency noise. The region in light gray was inaccessible to ACS. The dispersion in the measures over small spatial scales is indicative of the relative measurement uncertainties that become breathing-dominated, when unfiltered, close to the star.

We measure the total flux density of the disk, via background (far from the star) subtracted aperture photometry conforming to the FOV shown in Fig 30 that captures > 98% of the scattered from the disk, as 260 μJy. Given the stellar V magnitude of HD 139664, we find the

0.57 µm disk scattering fraction, $F_{disk}/F_{star} = 0.0005\%$, more than an order of magnitude smaller than any of the other disks in the GO 12228 sample. This excludes the region of the disk at r < 1.2" inaccessible in the STIS AQ image.

In V95 the edge-on disk (mostly obscured by one of the *HST* diffraction spikes) was sufficiently "off rolled" (by 48°) to use separately as an independent PSF subtraction template for the V98 disk image, mitigating residual chromatism near the star as previously discussed. This was possible to a limiting non-saturating distance of r = 1.2" (similar to the two-roll combined PSF-template subtracted AQ image) in the intermediate depth exposures. This worked well, but only for the wedge-unobscured V98 image of the disk due both the large roll angle differential and serendipitously very low amplitude breathing residuals in comparison to other visits. The resulting image is shown in Fig. 32 (left, wherein a negative imprint of the disk in the V95 template is seen flanking one of the diffraction spike masks and partially biases the disk flux from 1.2" to ~ 1.7". This image (which is combined also with the WedgeA-1.0 deeper imaging) does not improve over the AQ image at r > 1.2", but reveals the relative absence of bright dust close to the star between 1.2" and 1.7" (locally against the biased background) providing an additional constraint for later modeling of this older, and very faint, CS debris system. A radial SB profile along the disk plane from this "off-rolled" subtraction (Fig. 32, right) is highly correlated and in good agreement with the identically-measured two-roll subtraction (Fig. 31) where both are photometrically reliable at r > 1.7".

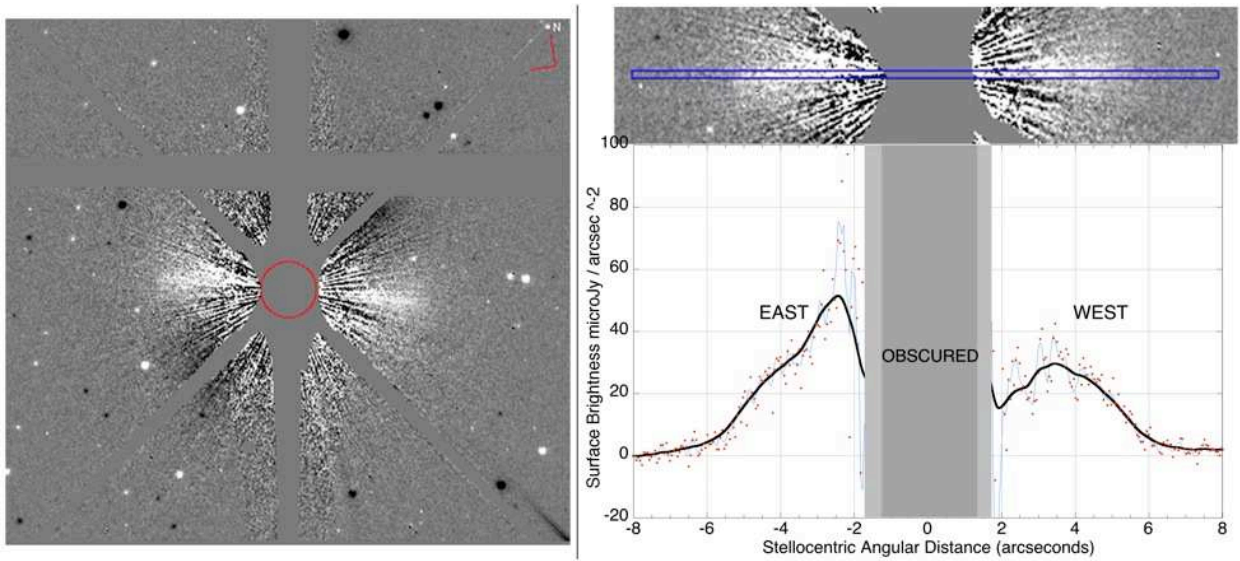

**Figure 32.** Left: Using the off-rolled V95 images as a PSF template for the Visit 98 images of the disk to intermediate exposure depth probes the CS region at a non-saturating IWA of 1.2" (radius of red circle) for the presence (or absence) of bright dust. Linear display stretch: -0.2 to +0.2 cts s$^{-1}$ pixel$^{-1}$. Right: Radial SB profile along the plane of the disk using the achromatic V95 "off-rolled" target PSF. The 1.2" ≤ r ≤ 1.7" region masked in light gray is photometrically unreliable due to background biases from the off-rolled imprint of the disk at these stellocentric distances.

## A.9 — HD 181327

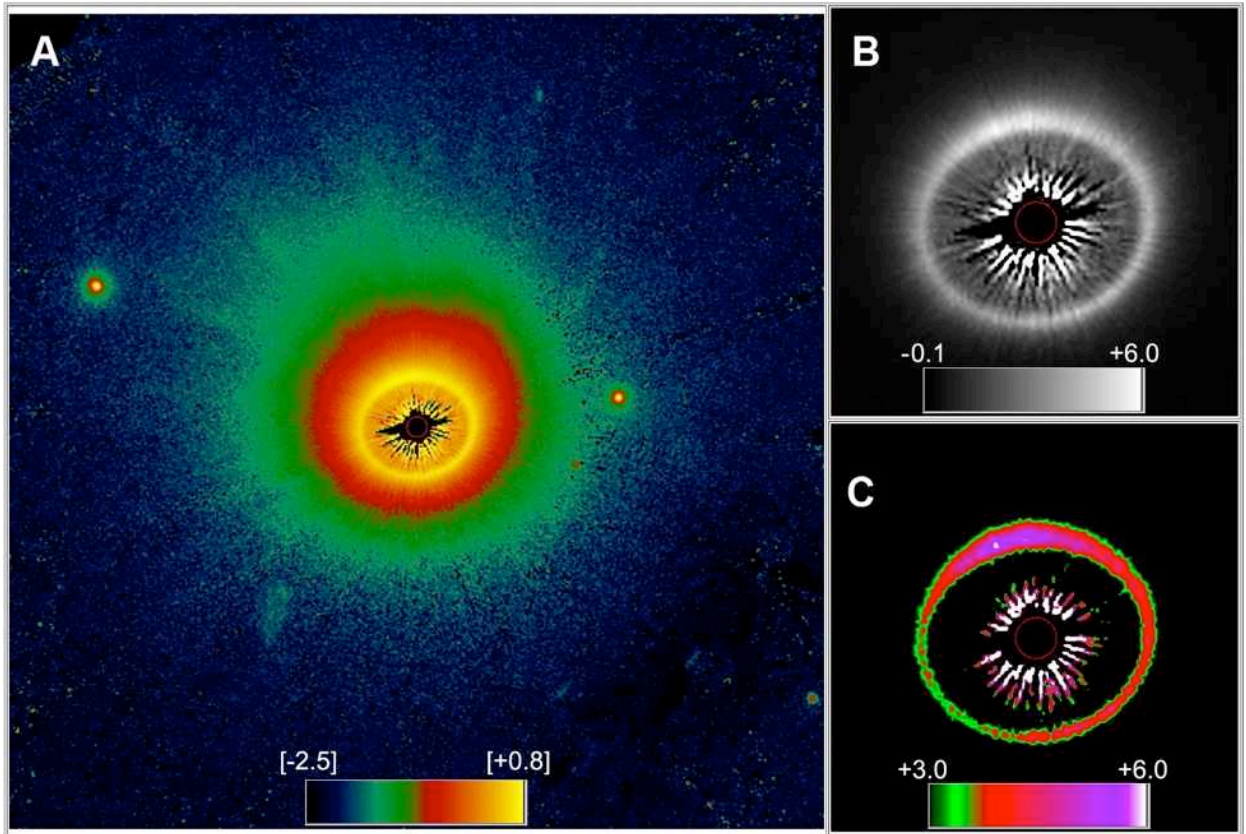

**Figure 33.** STIS six-roll AQ image of the HD 181327 CS debris disk (celestial north "up") utilizing all twelve WedgeA-0.6 + WedgeA-1.0 images Visit-level PSF subtracted images in artifact-masked median combination. (A): 24.36" x 24.36" FOV centered on the occulted star. Log$_{10}$ display stretch from [-2.5] to [+0.8] {dex} cts s$^{-1}$ pixel$^{-1}$ (app 0.6 to 1100 μJy arcsec$^{-2}$) simultaneously revealing the debris ring and its large, asymmetric, diffuse halo. (B, C): 6.09" x 6.09" FOV centered on the occulted star in two linear display stretches (range in cts s$^{-1}$ pixel$^{-1}$ ranges indicated) show the narrow width and "sharpness" of the bright debris ring (B), and its asymmetric SB distribution (C). Central red circle (all images): r = 0.3" (15 AU).

*Introductory Notes.* HD 181327 is a 51.8 pc distant F5/6 main sequence star with a strong thermal infrared excess ($L_{IR}/L_{star}$ = 0.25%). HD 181327 is a member of the β Pic moving group (canonical age ∼ 12 My), and a likely common proper motion companion to the 0.1 pc distant HR 7329 (a likely coeval star with close brown dwarf companion, see Lowrance et al. 2005).

These characteristics made it a high-priority target for prior *HST* high-contrast (coronagraphic) imaging investigations at both near-IR (*HST*/GO 10177, 10847) and visible-light (*HST*/GTO 9987) wavelengths. An optically bright light-scattering debris disk circumscribing HD 181327 ($F_{disk}$/ $F_{star}$ = 0.17% at 1.1 μm) was first reported from *HST*/NICMOS and ACS observations by Schneider et al. (2006), while mid-IR imagery along with IR spectroscopy was discussed by Chen et al. (2008). Together, those observations revealed the existence of a relatively narrow ring-like debris belt responsible for ~ 70% of the total disk flux, peaking in SB at a stellocentric distance of ≈ 86 AU. The ~ 31.7° from face-on inclined debris system appeared to be largely cleared of light-scattering material closely interior to the bright ring as determined from the NICMOS images and, exterior, surrounded by a much larger, asymmetric, halo of diffusely scattering low SB particles further from the star as detected in deeper ACS optical imaging.

*Observations and PSF Subtraction*. Our STIS GO 12228 6R/PSFTSC observations of HD 181327 are detailed in §5.5 as archetypical of the observation and data reduction/processing methodologies as applied to all GO 12228 targets. In particular, for HD 181327, an ensemble of 168 images targeting the inner regions of the debris ring to a WedgeA-0.6 limiting stellocentric angle of r = 0.3", and 64 deeper WedgeA-1.0 images to probe the outer regions of the debris system, were obtained at six different spacecraft roll (field orientation) angles with up to ~ 12.8 ksec of total integration time in most pixels (see Table 3 for details). An intermediate set of twelve "visit level" combined images (derived from multiple images, separately at each wedge position in each orbit) were created; see Figs. 3 and 4. Though there is some (anticipated) small dispersion in the quality (occurrence and amplitude of PSF-subtraction residuals) between these images in the regions unaffected by the *HST*/OTA diffraction spikes and the STIS occulting wedges (e.g., qualitatively compare the panel A "worst case" panel of Fig. 3 to the panel E "best") none were excluded as outliers in subsequent multi-roll image combination. An AQ data image (with error estimate propagation), combining all twelve astrometrically registered, artifact-masked, PSF-subtracted visit-level images, each oriented in the celestial ("north up") frame, was then created with up to ~ 12.8 ksec of total integration time in most pixels. The HD 181327 AQ image discussed here is presented in Fig. 33.

*Principal Results*. The new *HST*/STIS 6R-PSFTSC imaging of the HD 181327 debris system improves dramatically over the previous NICMOS (Fig. 1) and ACS (Fig. 6) co-discovery imaging in the range of stellocentric regions of the disk previously imaged, enabling novel science. A newly discovered radial variation in the apparent scattering phase function may imply the presence of a yet unimaged planetary-mass perturber (Stark et al. 2014). The new 6R/PSFTSC imaging of the HD 181327 debris disk (Figs. 33A/B) also reveals previously unseen sub-structures and brightness asymmetries along the ring (Fig. 33C). The bright narrow ring of dust at r = 1.86" (88.5 AU projected distance), exhibits non-axisymmetric SB asymmetries (Fig. 33C) that cannot readily be explained by simple directionally preferential (e.g., Henyey & Greenstein [H-G] 1941) scattering by the disk grains alone (Stark et al. *ibid*).

Morphologically, the HD 181327 debris disk appears as an asymmetrically bright narrow elliptical ring cleared of scattering material in the interior, and exterior surrounded by a larger, fainter diffuse region. Assuming intrinsic circular symmetry, after elliptical isophote fitting of the flux density in the ring, and finding a celestial PA of 102°±4° for the major axis of the bright ring inclined 30.1°±1.2° from face-one, we deproject the HD 181327 disk to a face-on viewing geometry as shown in Fig. 34 with debris ring major axis horizontal. Photometric SB contours external to the bright ring deviate from perfect ellipses (circles in de-projection) and exhibit some azimuthal "skewing" with stellocentric distance about the presumed axially concentric bright debris ring.

The most immediately noticeable (and striking) features about the bright ring itself are: (1) the narrowness of the ring; (2) the steeply declining SB gradients exterior, and particularly interior to the radius of (deprojected) peak SB; with (3) a dearth of dust-scattered starlight in the interior region. Additionally, (4) the diametrically opposed sides of the ring about minor axis (vertical in Fig. 34) at mirror-symmetric deprojected azimuthal phases angles (theta in Fig. 34B/C; theta = 0° coincident with the azimuth angle of the peak SB along the ring), are of significantly different brightness and, (5) with a minimum in the SB around the ring not coincident with deprojected theta = 180° (i.e., not H-G scattering) and, (6) is also not diametrically opposed to the direction of the "extension" of the outer disk halo to the north.

Compensating for the $r^{-2}$ decline of the stellar radiation field with increasing stellocentric distance, we transform the deprojected SB image of the debris ring (Fig. 35A) into a proxy for the product of the optical depth and the scattering phase function of light-scattering particles (Fig. 35B). The radial sharpness of this approximation for the radial dust density distribution, its abrupt clearing interior to the ring peak radius, and axial asymmetry with mirror-symmetric theta angles are illustrated (and readily apparent) in Fig. 35C.

The combination of previously unresolved sub-structures in the bi-axially asymmetric HD 181327 r = 88 AU debris ring, with a very narrow width, sharp inner edge, little or no light-scattering material interior, and large, skewed, diffuse outer halo, all implicate the possible presence of yet unimaged planetary-mass perturbers.

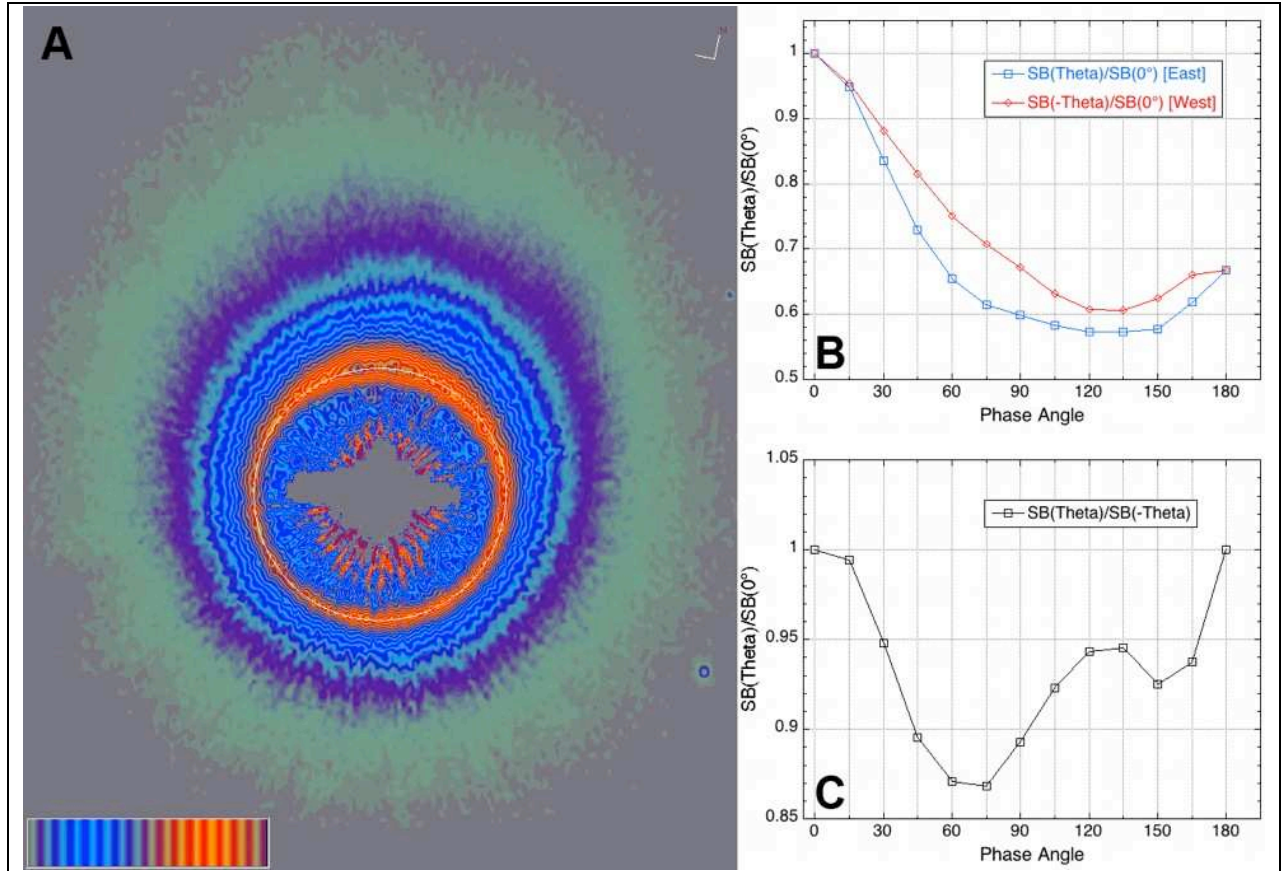

**Figure 34.** (A) Face-on deprojected isophotal view of the HR 181327 CS disk. 0 to 6 cts s$^{-1}$ pixel$^{-1}$ linear display stretch with 3.125% intensity contour intervals. Image orientation of the peak SB along ring with stellocentric deprojected azimuth angle (theta) = 0° at a celestial PA = 12° ("up") in this representation. White circle is best circular fit to circumferential radial SB peaks. (B) Azimuthal SB profile (normalized to the theta = 0° ring-peak SB) asymmetry about the disk minor axis. (C) "mirror symmetric" SB ratios, as a function of azimuthal (theta) angle about deprojected minor axis.

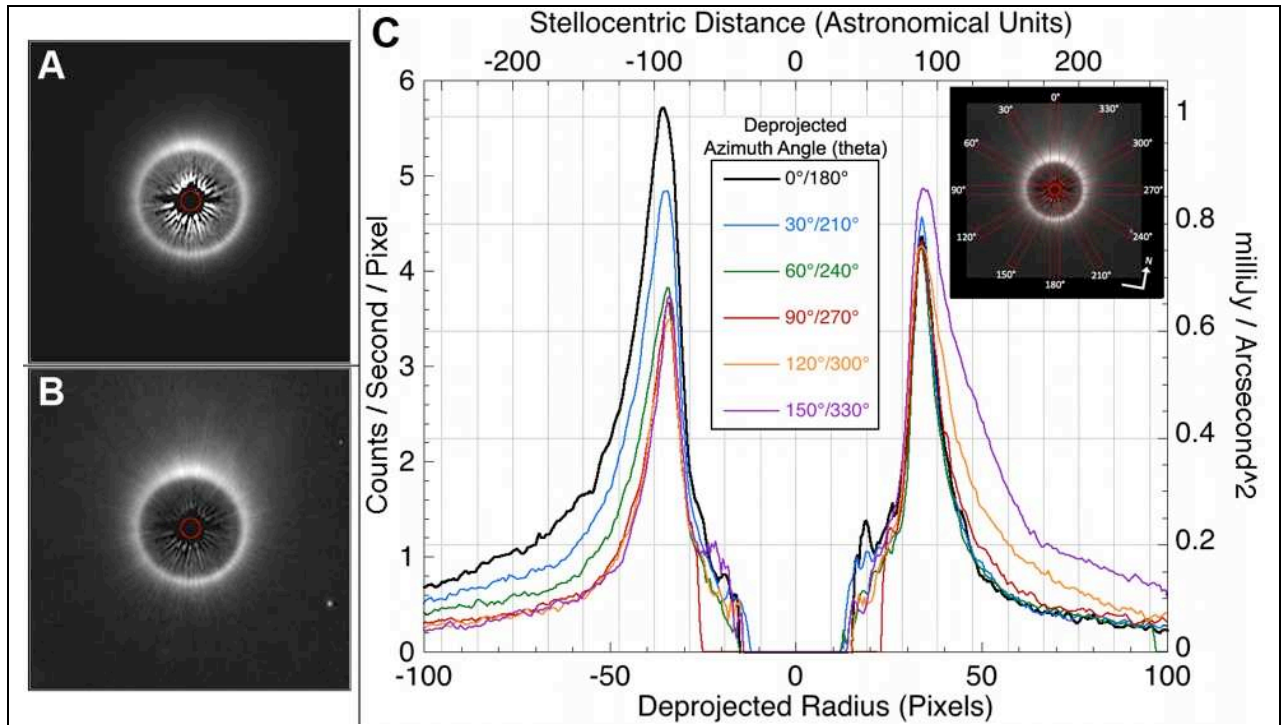

**Figure 35.** Face-on deprojected images of the HR 181327 debris ring (A) before, and (B) after compensating for the $r^{-2}$ dilution of the stellar radiation field (liner display stretches 0 to 6 cts s$^{-1}$ pixel$^{-1}$; red circle r = 0.3"). (C): 0.5"-wide radial $r^{-2}$ compensated brightness profile "cuts" in 30° deprojected azimuthal increments from theta = 0° (direction of peak brightness along the ring). Physical units based on *HST*/STIS photometric calibration per STScI MAST/OPUS pipeline, SIMBAD parallax, and STIS scale per instrument handbook; see Table 4.

### A.10 – AU MIC

*Introductory Notes*. AU Mic is an ~ 12 Myr old member of the β Pic moving group and the spectrally latest (M0V), and lowest mass, star for which a CS debris disk has been imaged directly. Thanks to its only 9.9 pc fortuitous proximity to Earth, large angular extent, and $F_{disk}$/$F_{star}$ = 0.2% flux ratio with natural contrast enhancement owed to the edge-on viewing geometry of its debris system, the AU Mic disk was discovered with ground-based near-IR coronagraphy by Kalas et al. 2004. Follow-on Keck II NIRC2/H-band images (Liu 2004) revealed not only the disk itself, but suggested some sub-structures along the disk major axis at stellocentric distances beyond about 25 AU, suggestive (perhaps) as influenced by co-orbiting, unseen, planets. Subsequent *HST*/ACS imaging by Krist et al. (2005), in three broad optical bands, revealed (diffraction polluted) light from the disk as close as 0.75" (interior to the r = 0.9″ focal plane mask at the *HST* spherically aberrated focal plane) from the central star (e.g., see Fig. 1 top left panel reproduced from Krist et al. 2005) and cleanly traced the disk along its edge-on major axis from app r = 2" – 15". The ACS F606W (0.61 μm) image reproduces some of the sub-structure suggested in the Keck II image in this radial range, though with a "*relatively large discrepancy in position of feature C {that} ... may be due to subtraction errors in either or both data sets.*" Krist et al. (2005) state that a three component fit to the disk radial SB profile along the disk major axis, decreases in steepness (in two breaks) with stellocentric distance. Subsequently the disk has been modeled using multiwavelength and polarized light data (Graham et al. 2007; Fitzgerald et al. 2007) and the signatures of grain growth probed.

*Observations and PSF Subtraction - Details*. STIS 6R/PSFTSC observations of AU Mic and its contemporaneously observed PSF star, HD 191847, were conducted at two epochs nearly a year apart: 2010 Aug 09 (V35 – V38), and 2011 July 16 (V31 – V34). In V35, the edge-on debris disk was aligned with, and obscured by, STIS occulting WedgeA. Thus, only five rolls in the two epochs combined contribute to the imaging of the disk along its mid-plane. Despite the excellent $\Delta(B-V) = -0.02$ color match of its PSF template star, PSF-subtractions from all six visits revealed bright residuals in the form of a chromatic ring at r = 0.8" and a large diffuse chromatic halo beyond. These residuals are illustrated in Fig. 36 with two representative visits each from WedgeA-0.6 (panels A & B) and WedgeA-1.0 (panels E & F). These residuals mostly arise from mis-match in target:template photospheric SED on the red end of the very broad, unfiltered, STIS bandpass of greater impact for this reddest (B-V = +1.46) of all our disk hosting stars. In panels C & D and E & F we respectively use the disk-obscured images of AU Mic itself obtained in V35 as its own PSF template, in which the self-mitigating effects to chromatism are apparent.

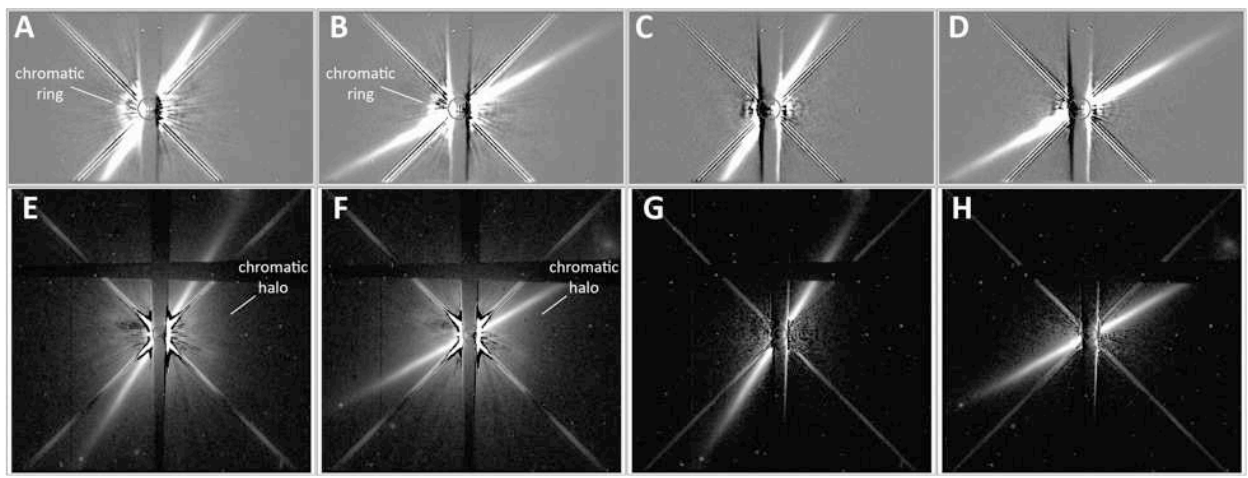

**Figure 36.** Top – WedgeA-0.6, and Bottom – WedgeA-1.0 PSF subtracted images of the HD Au Mic CD disk (V36 & V38 progressing left to right) with the HD 191847 PSF template star (left four panels) and mitigating residual chromatic effects, using the disk-obscured V35 image of AU Mic itself as its own PSF template.

Given the above, we used the V35 disk-occulted images of AU Mic itself as its own self-referential PSF template that, fortuitously, exhibited no significant breathing excursions w.r.t. the non-contemporaneous disk imaging visits 31 – 34. In doing so, we first tested whether or not this could over-subtract with significance any disk flux from out-of-plane light due to the disk in the template image possibly unobscured by Wedge A. In Fig. 37A we show the large chromatic halo that emerges when subtracting the HD 191847 PSF star (V37) from the disk-obscured V35 image (similar to panels E & F in Fig. 36 with the disk unobscured by Wedge A). No negative over-subtraction due to light from out of the disk plane, along the edges of the A wedge, is seen at this same display stretch in which the images of the disk in Fig. 36 panels E – H are shown.

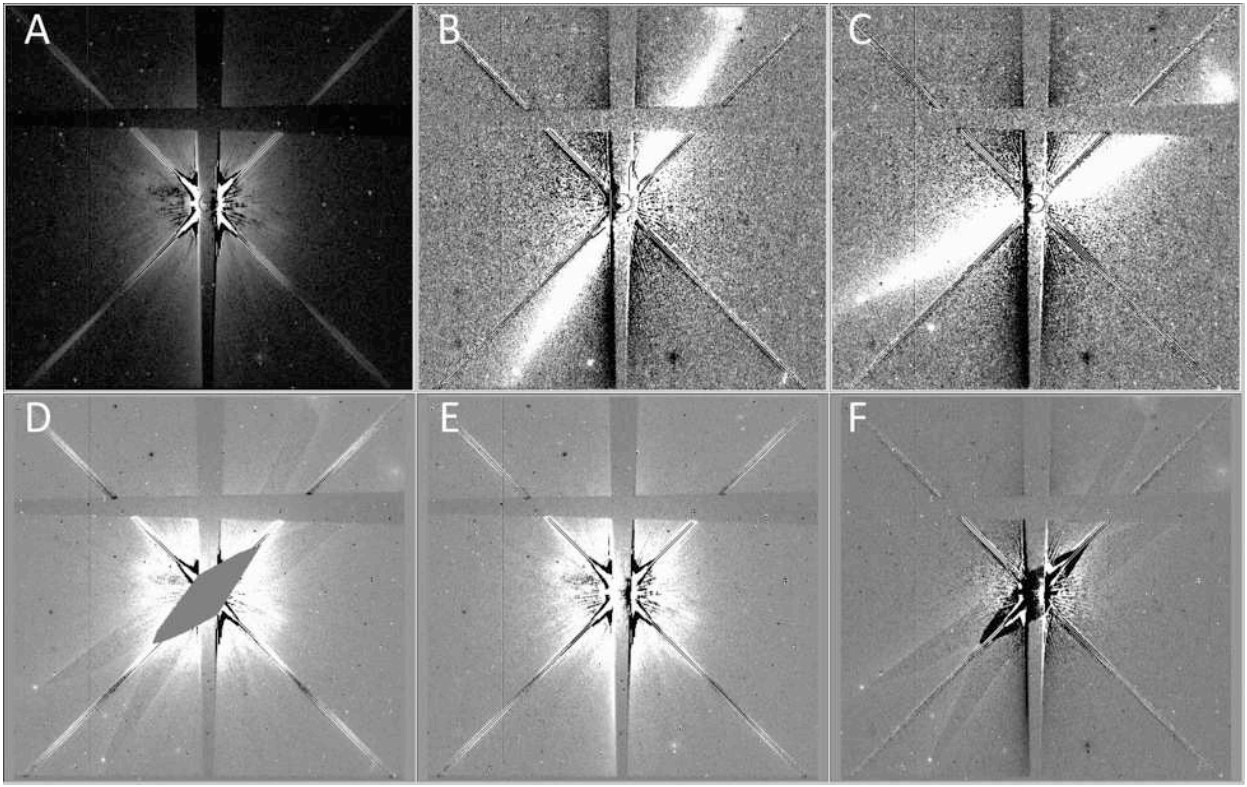

**Figure 37.** (A): Chromatic halo from disk-obscured V35 image minus V37 PSF template HD 19847, same stretch as Fig. 36 panels E – H. (B & C): V36 & V38 images, respectively, minus disk-obscured V35 image stretched ± 0.02 cts s$^{-1}$ pixel$^{-1}$. A small amount of lux from the disk in the V35 template over subtracts the sky background close to the wedge edges (especially where the wedge taper is narrowest) but has little effect in the off-rolled plane of the disk itself. (D): Disk-masked image combination by averaging of V36 and V38 images in the SIAF frame reproduces the halo as a rotationally invariant artifact (E): Same as panel A, but stretched as Panel D. (F) Difference demonstration: (D) – (E) eliminates nearly all of the chromatic halo light from the disk-obscured image.

In Fig. 37B/C, we subtract the V35 AU Mic template from the V36 & V38 disk images as in Fig. 36G/H, but here shown with a very hard bi-directional linear stretch of ± 0.02 cts s$^{-1}$ pixel$^{-1}$, from which we can see at a very low level, and quantify (nearly inconsequential) out-of-plane over-subtraction of background light far (in the vertical plane) from the disk itself. This is very significantly smaller than the level of the chromatic residuals using the HD 191847 PSF star, and thus we proceed using the AU Mic V35 image as a PSF subtraction template.

Cautiously, as a further test illustrated in Fig. 37, we confirm the chromatic halo that appears when subtracting the PSF star is indeed an artifact (and not due to an extended disk halo) by its rotational invariance in V36-PSF and V38-PSF images (as well as in the V35-PSF image where the disk is behind the wedge). Here we combine in the detector (SIAF) frame the two PSF-subtracted images shown in panels B & C (each containing the disk visible at two different orientations, not rotated to a common celestial orientation) after separately digitally masking the visible portions of disk. This gives a slightly imperfect image of only the halo as shown in panel D. In detail: (a) regions unmasked in both images, where the disk does not appear, were averaged; (b) regions masked in one image where the disk appears but not in the other were rejected; and (c) regions where the disk appears in both images (i.e., in uniform gray near image center) were zeroed. The resulting combined image in panel D "eliminates" the disk (based upon the empirical mask of where the disk appears in each image) and shows only the posited rotationally-invariant chromatic "halo." Panel E then shows the chromatic halo in the V35 image of AU Mic (after subtracting the HD 191847 color mis-matched template), as shown in panel A but here in a symmetric ± linear display stretch. If the halo seen in panel D two-visit combined difference image is a rotationally invariant artifact, subtracting the V35 (panel E) image difference should (in double differencing) null the halo. This is seen to be the case as shown in panel F with no significant azimuthal sub-structure (just noise) after differencing, indicative of a rotationally invariant image artifact in the two-roll detector-frame combined image in panel D. I.e., the "halo" is indeed an optical artifact, and not intrinsic to the CS environment. Therefore, the AQ data images we present, and discuss, here all use the V35 disk-occulted image of AU Mic itself as its own self-referential PSF template.

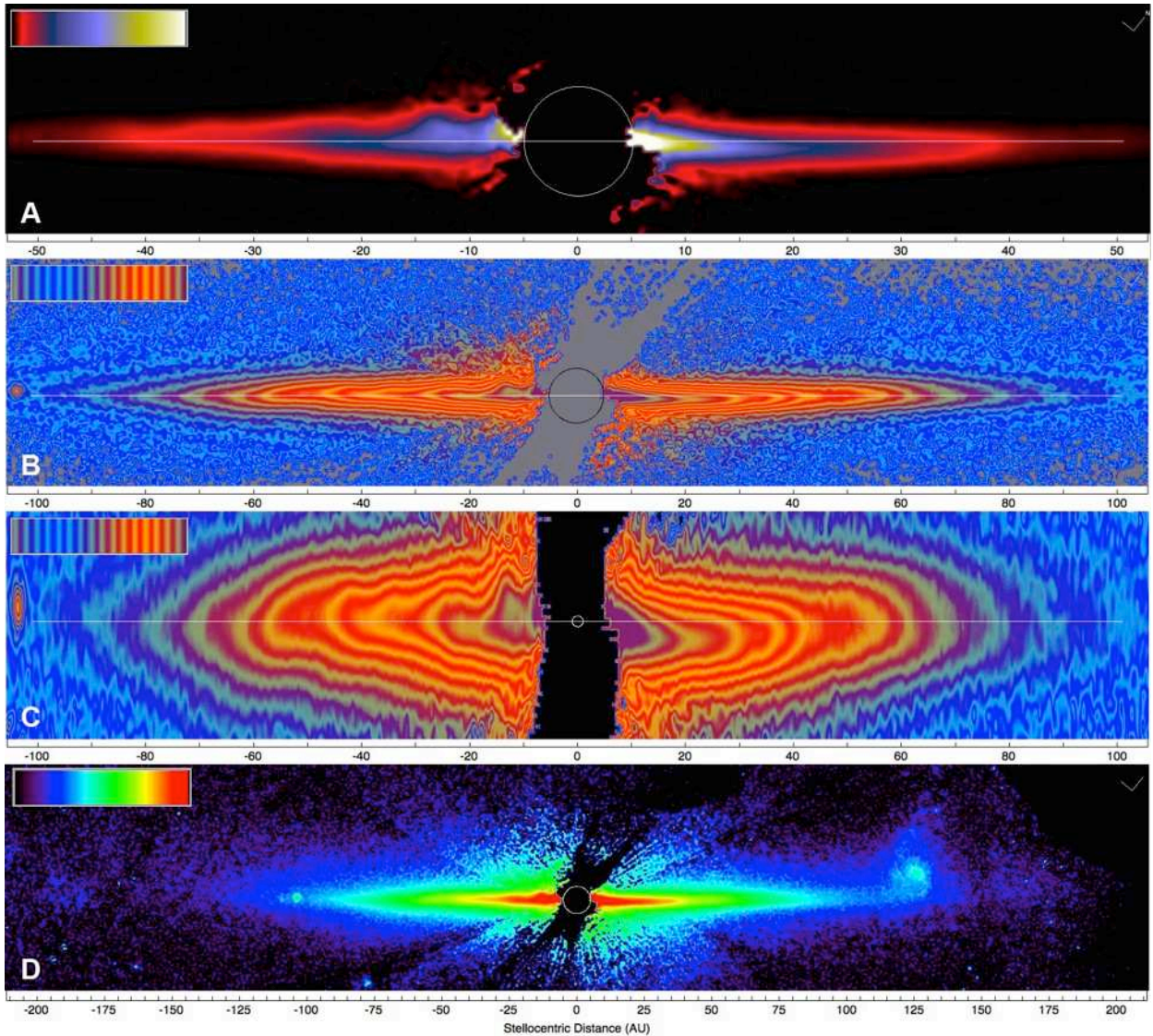

**Figure 38.** STIS five-roll (10 image) combined PSF-subtracted images of the AU Mic CS disk (celestial north 38.7° CW of image vertical). (A): "Inner" region $5 \leq d \leq 50$ AU in 10.62" x 2.67" with a linear display stretch 0 to 15 cts s$^{-1}$ pixel$^{-1}$ (0 to app 2650 µJy arcsec$^{-2}$ pixel$^{-1}$). (B, C): Isophotal contour images $5 \leq d \leq 100$ AU in 21.24" x 5.34" FOV. (C): with 4x vertical scale expansion. (D): Full spatial extent in 42.64" x 16.66" FOV. (B–D) $\log_{10}$ display stretch [-3.0] to [+1.0] cts s$^{-1}$ pixel$^{-1}$ {dex} (app 0.177 to 1770 µJy arcsec$^{-2}$ pixel$^{-1}$). White line (A–C) is disk mid-plane defined by panel (B) isophotes from $50 \leq d \leq 100$ AU. Central circle r = 0.5" except panel (B) marking star position. A background (non-commoving) star and galaxy, at this epoch, flank the opposite side of the disk along its major axis.

*Principal Results*. The AU MIC edge-on disk is cleanly imaged with STIS 5R/PSFTSC using the disk-obscured AU Mic image as its own PSF template to a stellocentric inner working distance of 5 AU along the disk mid-plane. Here, we photometrically define the disk mid-plane

(illustrated by the horizontal white line in panels A-C) by fitting the central ridge of isophotal SB maximum at stellocentric distances 50 – 100 AU on either side of, and passing through, the star (finding a celestial position angle of 37.8 ± 0.2°).

In the inner ( r < 50 AU) region of the disk, the new image reveals:

(1) a significant NW (brighter) to SE (fainter) brightness asymmetry between the two "sides" of the disk at equal stellocentric distances interior to ~ 15–20 AU inner working distances (see Fig. 38 A) that had not been cleanly probed with earlier scattered-light imaging.

(2) a prominent brightness enhancement ("bump") on the SE side of the star "above" (to the NE) of the disk mid-plane at 13 AU (see also panel A). We have readily confirmed the "bump" as intrinsic to the disk (not due to background contamination) as established by its common proper motion with AU Mic itself (+380, -361 mas yr$^{-1}$) over the two observational epochs nearly a year apart.

(3) significant out-of-plane asymmetries on the NW side of the disk "below" (to the SW) of the mid-plane, opposite the SE-side "bump", with a "warp" below the disk mid-plane to a stellocentric distance of ~ 45 AU *only* on the NW side of the disk. This is shown in the isophotal contour maps in panel B and with a 4x vertical (only) scale expansion in panel C.

The outer regions of the disk are sensitively detected to ~ 130 AU on both sides of the star (panel D). In the outer regions of the disk image, on the SE a background star (confirmed from the two observational epochs as not co-moving with the disk) coincidentally falls on the disk mid-plane at disk-projected distance of 105 AU from the star. On the NW side, at 125 AU a background galaxy to the just to the NE of, and blending with, the disk mid-plane pollutes the image.

The disk SB along the mid-plane, as measured from earlier ACS images, had been fit and discussed by Krist et al. (2005) with a three-component power-law of increasing index in segregated zones with increasing distance from the star. Here, without invoking physical arguments, we separately fit the STIS derived SB profile on both sides of the disk each to a simple exponential. We did so to better examine the apparent local deviations from a smoothly continuous radial SB profile with stellocentric distance to identify small spatial scale sub-structures in the disk above the local measurement uncertainties. The as-measured profiles, and the two resulting fits, are illustrated in Fig. 39.

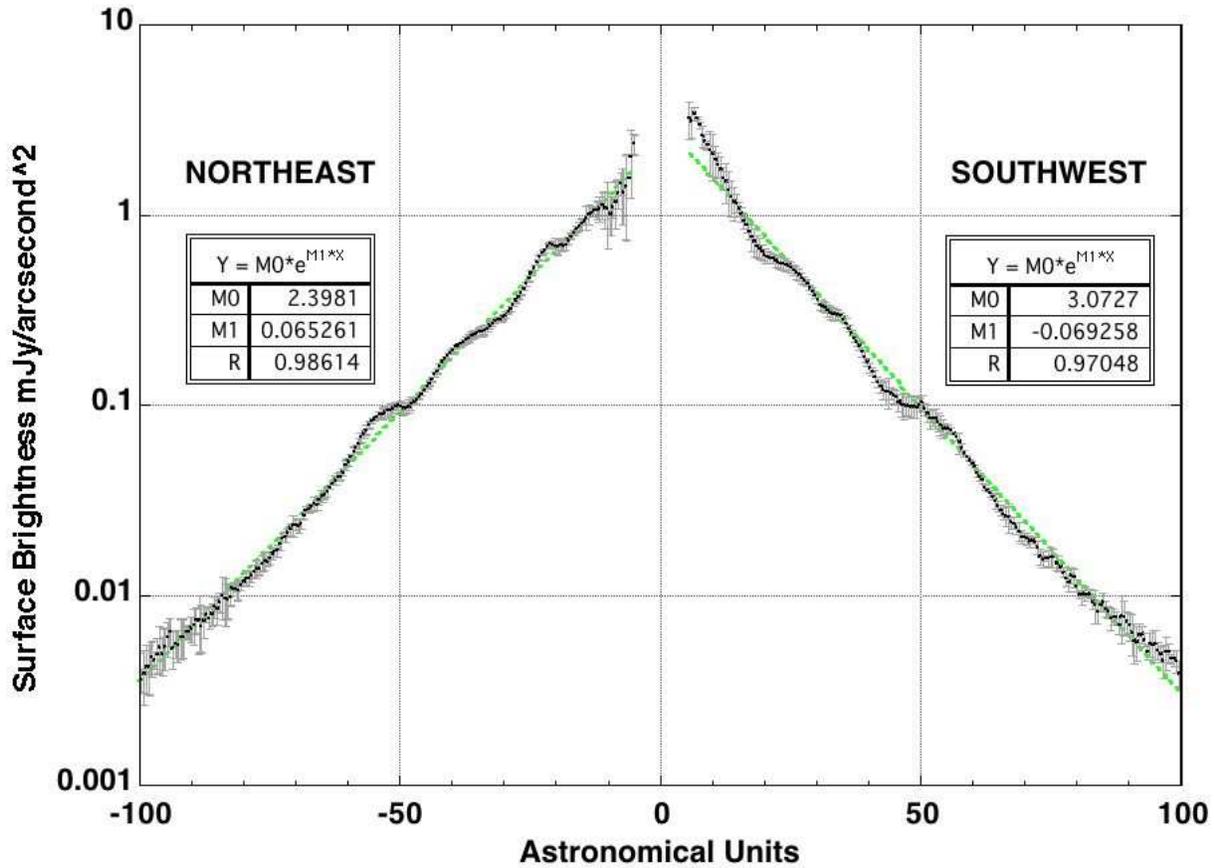

**Figure 39.** Radial SB profile of the AU Mic CS debris disk, corresponding to Fig. 38B, measured in a 50 mas wide central strip along the disk mid-plane ± 100 AU from the central star, separately fit on both sides of the star to single exponentials. The 1-sigma error bars on the observed data are computed from the dispersion in measures from the individual PSF-subtracted images about the roll-combined median image (up to 10 input images for any single pixel) and do not include likely < few percent uncertainty in the absolute flux density calibration. Statistically significant deviations on both sides of the star, both asymmetric (e.g. rise at ~ 20 AU to the NE and depression same distance to the SW) and symmetric (e.g., the significant "break" on both sides) about the minor axis, and other deviations elsewhere in the inner (< 50 AU) part of the disk.

In the vertical dimension, the aforementioned "bump" at 13 AU to the north of the mid-plane on the east side of the disk is a most prominent feature. Lower SB material "above" the disk mid-plane at and beyond the radial distance of the bump on the east side of the disk; E.g. in green in Fig. 38D from -45 AU to -10 AU. This scattered light in excess of the exponential fit does not appear either south of the disk on the east side, nor north of the disk on the ("no bump") west side – though visibility there is precluded at $r <\sim 12$ AU due to the remaining obscuration by the STIS occulting wedge. The "bump" and excess to the NE of the star, and "warp" in the photometric mid-plane of the disk to the diametrically opposed SW *may* have a common causality with dust in the inner part of the system orbiting non-coplanar (slightly inclined) to the outer part of the disk, e.g., like β Pictoris, possibly (here) to a yet unseen inner planet.

### A.11 — MP MUS (PDS 66)

*Introductory Notes.* Unlike the other GO 12228 targets, MP Mus (a.k.a. PDS 66, Hen 892) is a pre-main sequence, classical T-Tauri star (cTTS) hosting a late stage starlight-scattering protoplanetary/transition disk discovered by Cortes at al. (2009) with *HST*/NICMOS as part of a *Spitzer*/FEPS-selected coronagraph imaging survey of Sun-like stars. Its age was estimated by Cortes et al. (2009) as ~ 13 Myr, as a member of the Sco Cen OB association (Lower Centaurus-Crux moving group) and by fitting stellar evolutionary tracks. Despite its advanced age for a CTTS (having been described in the literature as "mature"), CS material is still accreting onto the star, as evidenced by variability in the thermal SED and other accretion signatures typical for much younger stars. MP Mus is known to be optically variable with a 5.75 day period and peak-to-peak amplitude of ≈ 0.1 mag in V-band; see Batalha et al. (1998); the GCVS 2007–2011 gives $V_{mag}$ range 10.30 to 10.42. In combination with ACTA interferometric observations in four continuum bands from 3 mm to 6 cm, and an IR SED derived from 2MASS, IRAS, and *Spitzer* IRAC & MIPS photometry and IRS spectroscopy, the grain and physical properties of the disk were partially constrained by Cortes et al. (2009), with its geometry informed by the NICMOS imaging to assess the potential for planet formation in the disk. Kastner et al. (2010) detected molecular gas (CO) orbiting MP Mus as had been found previously for the prototypical "old" cTTS TW Hya, and for V 4046 Sgr (both, like MP Mus, within 100 pc of the Earth).

The primary focus of the GO 12228 investigation and its target sample is on CS debris disks. MP Mus, however, was importantly included as a "mature" protoplanetary/transition disk to directly bridge those data spanning two decades in stellar ages to a nearby (d < 100 pc) protoplanetary/transition disk system, spatially resolvable with *HST*, at this earlier phase of disk (and exoplanetary system) evolution, coeval with the youngest of our debris disk targets.

*Observations and PSF Subtraction*. STIS 6/PSFTSC observations of MP Mus and its PSF template star executed per the observation plan at two epochs three months apart with no issues on 2011 Mar 03 (VA1 – VA4) and 2011 Jun 06 (VA5 – VA8). Unlike our other targets, in the case of MP Mus and its PSF star, our observing strategy included, by design, only WedgeA-0.6 imaging. The combination of the relative faintness of the star (V = 10.44) and efficiently scattering disk dust particles, as known from the NICMOS imaging, allowed us to expose more deeply with WedgeA-0.6 (only) without risk of saturation close to the star. This, and the anticipated smaller angular size of this 86 pc distant protoplanetary disk, obviated the need for WedgeA-1.0 imaging (and the additional, and longer, overheads incurred from its utilization).

MP Mus's intrinsic optical variability informed of the possibility of significant/detectable stellar brightness changes at some light-curve phases even over the ≈ 7-hour period of each of our two sets of visits, in addition to the 3-month timescale between the two observational epochs. Hence, in multi-roll combination, the contributing single-visit PSF subtracted images were re-normalized to the stellar brightness in Visit A1 (when the star was found to be brightest); I.e., so inter-visit PSF-subtracted images of the disk are constant in $F_{disk} / F_{star}$ as $F_{star}$ varies[9]. The inter-visit brightness of the occulted star was determined by ratioing the flux from the unocculted OTA diffraction spikes, before PSF-subtraction, beyond the WedgeA-0.6 mask. In detail we found the following relative brightnesses, by Visit: (Epoch 1): A1 = x 1.0000, A2 = x 1.0041, A4 = x 1.0081; (Epoch 2): A5 = x 1.0784 , A6 = x 1.0714, A8 = x 1.0554.

In the thusly-weighted six-roll combined image, the disk is very cleanly resolved to an instrumentally asymmetric IWA of as close as 0.305" from the disk-hosting star. Additionally, AQ images partitioned from the first and second observational epochs were separately created to permit investigation of spatially resolved temporal changes in the disk surface illumination over time (and thus SB) in regions commonly sampled at both epochs.

*Principal Results*. The improvement in image quality from the NICMOS discovery image reproduced in Fig. 40A to the STIS 6R/PSFTSC AQ image in panel B is striking, and bespeaks the utility of this observational technique not only for studying optically thin debris disks, but for optically thick protoplanetary disks as well. From this image we have, more accurately than from the NICMOS image, redetermined the principal characterizing attributes of the disk.

The disk geometry was explored by elliptical isophote fitting in the stellocentric region r ≥ 0.863" (17 STIS pixels) where the disk isophotes are elliptically concentric (see Fig. 42B). At this inner distance the disk brightness, in instrumental units, is ~ 2.0 cts s$^{-1}$ pixel$^{-1}$ on the eastern semi-minor axis and ~ 1.5 ct s$^{-1}$ pixel$^{-1}$ on the western semi-minor axis. With an assumption of apparent ellipticity due to a circularly symmetric disk seen in inclined projection, the disk isophotes were found offset from the location of the star (by the "X marks the spot" method, § 5.4) by +0.66 (farther from the star) ± 0.2 pixels along the brighter disk semi-minor axis. This photometric (isophotal) offset does not necessarily imply a physical offset in the particle density distribution as the large scattering fraction ($F_{disk}/F_{star}$, see below) is consistent with a disk that is at least partially optically thick and non-isotropic scattering may account for this offset (as later modeling can explore).

---

[9] MP Mus is optically variable with ΔV ~ 0.1 mag and a 5.75 day period. In combining, and comparing, the image data obtained at the two observational epochs, the disk flux density was first re-normalized on a per-visit basis to a compensate for a ± 3.5% stellar in stellar brightness between the two observational epochs.

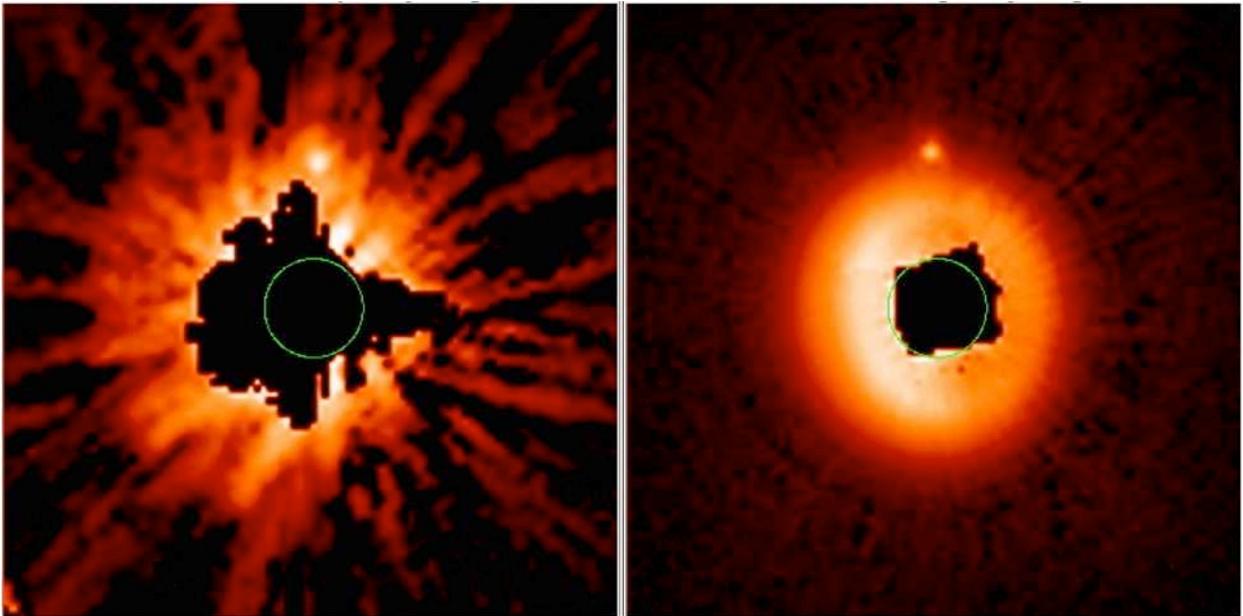

**Figure 40.** Left: NICMOS 1.1 μm 2-roll scattered-light discovery image of the MP Mus CS disk (as discussed by Cortes et al, 2009). Right: STIS 6R/PSFTSC AQ image reproduced at the same spatial scale and orientation (north up, east left). Both $\log_{10}$ display stretch from [-2] to [+1] {dex} cts s$^{-1}$ pixel$^{-1}$. NICMOS 1 cts s$^{-1}$ pixel$^{-1}$ = 218 μJy arcsec$^{-2}$, STIS 1 cts s$^{-1}$ pixel$^{-1}$ = 177 μJy arcsec$^{-2}$. Green circle is r = 0.41". The prominent point-like object to the north of MP Mus, superimposed on the outer periphery of the disk, was unequivocally found to be a background star from differential proper motion measures between the NICMOS (circa 2007) and STIS (circa 2011) epochs, as previously suggested comparatively from VLT/NACO imaging (circa 2003) of the stars (but not the disk).

We revisited the disk geometry, applying this observed isophotal stellocentric offset to define the disk center. We find: (1) the celestial orientation (position angle) of the disk major axis is 10° +/- 2° east of north; (2) the minor:major axial ratio is 0.889 ± 0.026 implying; (3) an inclination of 27.3 ± 3.3° from a face-on viewing geometry (in statistical agreement with, but improving upon the earlier determination of i = 32° ± ~5° by Cortes et al. (2009) from the NICMOS discovery image.

Photometrically, the disk is truncated along the major axis at r ~ 1.3" (see axial radial profiles in Fig. 41, left panel), which may be roughly considered the outer edge of the main disk. But, only 90% of the total starlight scattered by the MP Mus CS disk (excluding the central region obscured by the STIS wedge in the STIS 6-roll combine image) is within 2.0" of the star (see encircled energy plot in Fig. 41, right panel). A diffuse low-SB halo surrounds the main disk to r ≤ 6" (see Fig. 41 right panel inset). The flux density of disk measured in an annulus 0.36" ≤ r ≤ 6.09" (120 STIS pixels), normalized to the temporally-variable stellar brightness in Visit A1, is 1.86 mJy. Beyond this stellocentric distance any posited additional contribution to total disk flux density is at or below the level of the noise from the sky-plus-instrumental background background. *N.B.*: A similar measure in the F110W band found a total disk 1.1 μm disk flux density of 2.4 ± 0.4 mJy (as reported by Cortes et al. 2009). Given the V magnitude of the star, we find the total optical disk plus diffuse halo scattering fraction, $F_{disk}/F_{star}$ (r < 6.09") ~ 0.73%. To a conservatively inclusive distance beyond the outer boundary of the optically thicker main disk of r = 2.5", $F_{disk}/F_{star}$ (r < 2.5") ~ 0.68%.

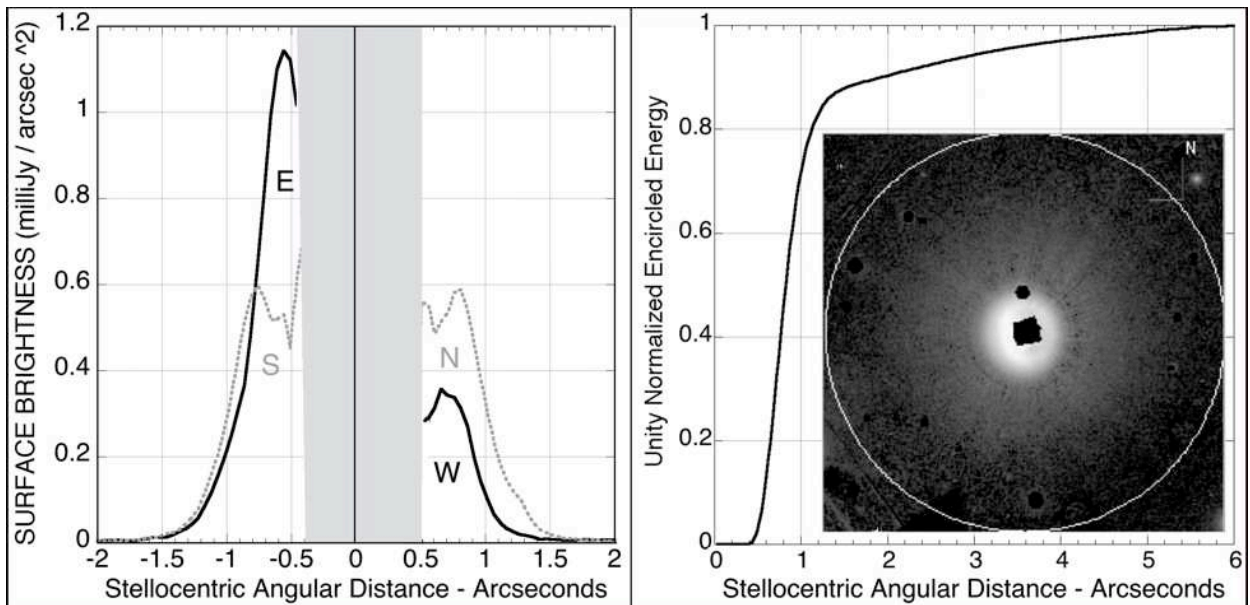

**Figure 41**. Left: Radial SB profiles of the MP Mus CS disk, measured in two pixel wide strips along its major (dotted gray) and minor (solid black) axes, ± 2" from the star. Right: Encircled energy plot to r ≤ 6" corresponding to the AQ image inset (white circle radius = 6.09") with stellar background sources masked reveals the MP Mus CS disk is surrounded by large halo of low SB light-scattering dust. Inset image displayed with a $\log_{10}$ stretch from [-3]

to [+1] {dex} cts s$^{-1}$ pixel$^{-1}$.

The detailed spatially resolved morphology of the disk is better revealed in a series of three representations of the 6R/PSFTSC AQ image as shown in Fig. 42. The disk apparently exhibits strong directionally preferential scattering centered at a celestial PA ~ 92°, that is offset from the disk's brighter semi-minor axis by −8°. E.g., at r = 0.6" along the disk minor axis, the disk SB is ~ 3.5x brighter on the eastern side of the disk compared to its diametrically opposed western side. The 6R image is composed of images from two epochs with strong evidence, discussed below, for time-variable illumination resulting in spatially localized fluctuations of disk SB at levels of tens of percent that must be considered in analytic modeling of the disk.

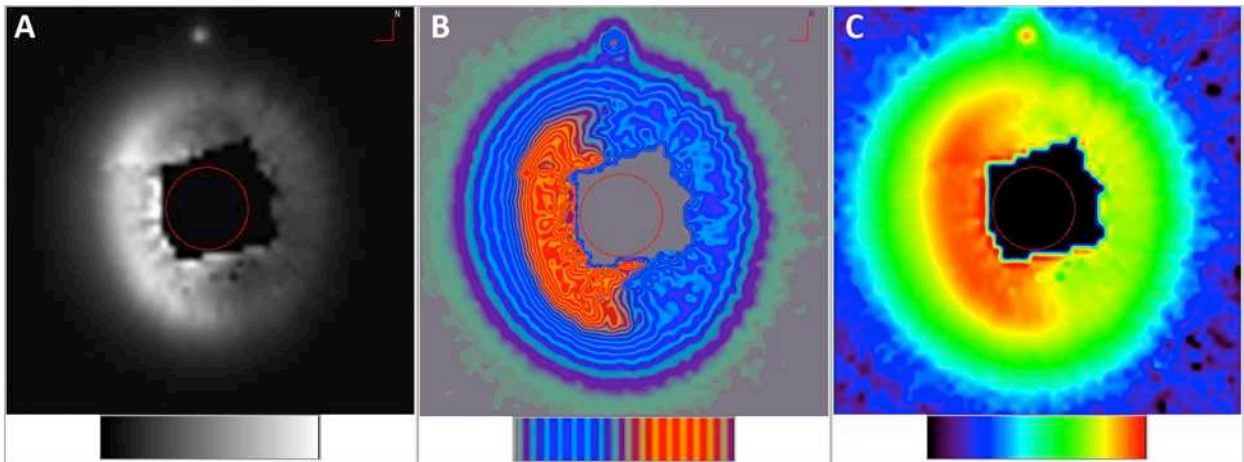

**Figure 42.** The MP Mus disk. Field: 3.05" x 3.05" (60 x 60 pixels) centered on occulted star (celestial North up). Left: Linear display image from -0.5 (black) to +8.0 (white) counts/sec/pixel. Middle: Linear display isophote map from 0 to 8 counts/sec/pixel with 3.125% intensity contour intervals. Right: Log$_{10}$ image display from -2.0 to +1.0 {dex} counts/sec/pixel. Red circle: r = 6 pixels (0.305"). *N.B.*: The point-object to the north is (unfortunately) unequivocally a background star from non-common proper motions as established with 2006 epoch NICMOS disk discovery images.

Our STIS 6-roll PSFTSC observations were obtained at two epochs, in two sets of three-target orbit, three months apart. In the spatial regions commonly sampled at both epochs, a significant change in the disk SB with stellocentric azimuth angle was seen, separate from an anticipated correlation of the global disk brightness with stellar variability. By ratioing first epoch to (globally renormalized) second epoch AQ images, a strong signature of azimuthally differentiated SB variability unassociated with the (global) stellar variability is seen. Specifically, as illustrated in Fig. 43, in an annulus from 0.6" ≤ r < 1.4" in regions commonly sampled at both epochs, the first epoch to second epoch globally renormalized ratio of median disk SB varied significantly on opposite sides of the disk. In particular, in sector from approximately celestial PA ≈ 20° to 230° (roughly centered on the presumed forward scattering direction to the SE of the star; green dashed lines in Fig. 43) the median disk SB increased by +11% relative to the star. Contemporaneously, in near diametric opposition from celestial PA approximately 240° to 360° (blue dashed lines) to the NW, the medianed disk SB declined by -24% relative to the star.

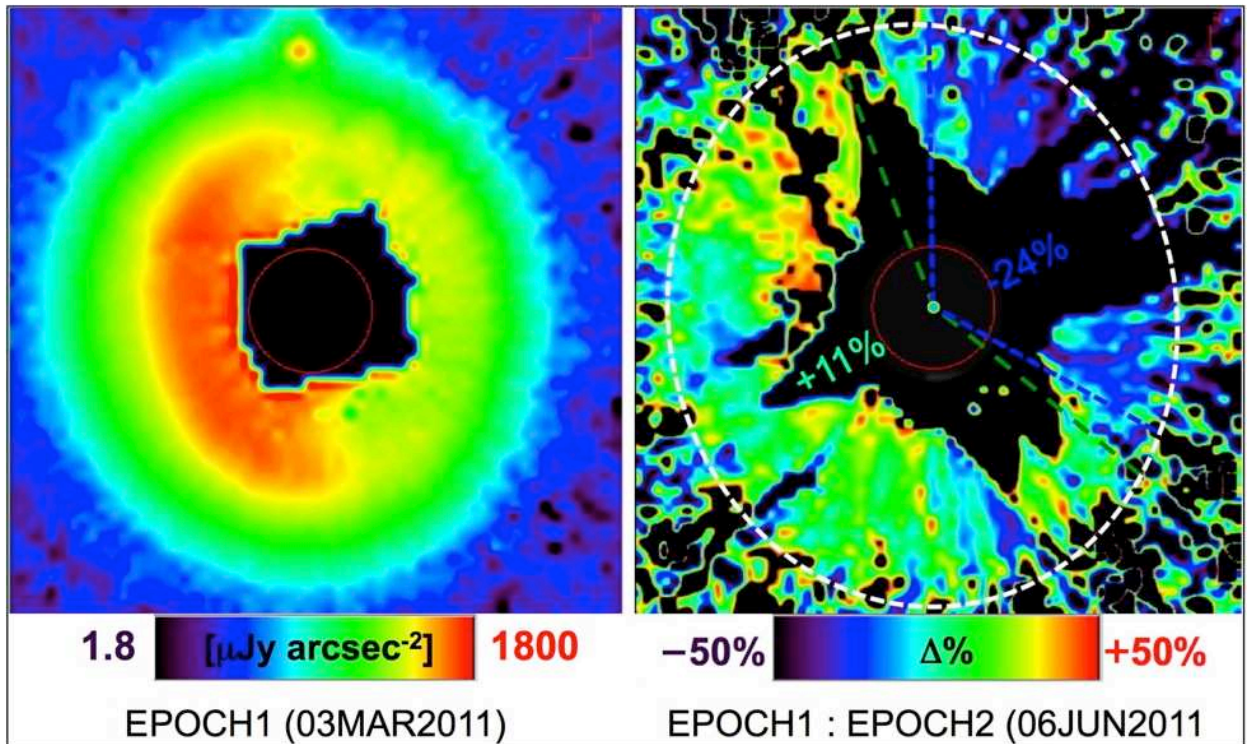

**Figure 43.** Left: $Log_{10}$-scaled SB image of the MP Mus disk using all data combined, normalized to the 1$^{st}$ epoch (Visit A1) stellar brightness. Right : Spatially resolved, temporal, variability in the MP Mus CS disk relative to the 1st epoch image. The median changes in SB in the two diametrically opposed regions considered within the ellipse representing the outer boundary of the disk are noted. The central regions in black are excluded due to mutually exclusive spatial coverage between the two epochs, and in the image periphery due to insufficient S/N from rapidly declining disk flux density. Red circle radius (both panels), r = 0.35".

Given the orbital time scales, these changes cannot be from a change in the surface density distribution of light-scattering grains. We interpret this, possibly as due to time-variable shadowing (or "beaming") of starlight to the outer disk beyond, either by opaque material orbiting close to the star, or the "puffing" (change in vertical height) of a unseen inner cavity wall, interior to the inner working distance (~35 AU) that could probed by the STIS coronagraph. Alternatively, some variation in the illumination pattern onto the disk might arise from differentiated brightness or localized on the stellar photosphere from accretion "hotspots." These possibilities might be arbitrated with properly sampled time-resolved imaging on the stellar rotation versus longer timescales, that our two-epoch data inform, but themselves are insufficient to differentiate between the possibilities.